 \documentclass[preprint,review,12pt]{elsarticle}

\usepackage{amssymb}

\usepackage{lineno}
\usepackage{algorithm}
\usepackage{algorithmic}
\usepackage{booktabs}
\usepackage{threeparttable} 
\usepackage{subfigure}
\usepackage{ulem}

\usepackage{comment}

\usepackage{graphicx,amsmath,amsthm,amsfonts,xcolor,bm,url,fullpage}

\biboptions{square,comma,compress}

\newcommand*\patchAmsMathEnvironmentForLineno[1]{%
  \expandafter\let\csname old#1\expandafter\endcsname\csname #1\endcsname
  \expandafter\let\csname oldend#1\expandafter\endcsname\csname end#1\endcsname
  \renewenvironment{#1}%
     {\linenomath\csname old#1\endcsname}%
     {\csname oldend#1\endcsname\endlinenomath}}%
\newcommand*\patchBothAmsMathEnvironmentsForLineno[1]{%
  \patchAmsMathEnvironmentForLineno{#1}%
  \patchAmsMathEnvironmentForLineno{#1*}}%
\AtBeginDocument{%
\patchBothAmsMathEnvironmentsForLineno{equation}%
\patchBothAmsMathEnvironmentsForLineno{align}%
\patchBothAmsMathEnvironmentsForLineno{flalign}%
\patchBothAmsMathEnvironmentsForLineno{alignat}%
\patchBothAmsMathEnvironmentsForLineno{gather}%
\patchBothAmsMathEnvironmentsForLineno{multline}%
}

\newcommand{\tensor}[1]{\bm{\mathsf{#1}}}

\def\diff{\mathrm{d}}

\def\CFLmax{\mbox{CFL}_{\max}}

\def\Ma{\mbox{Ma}}

\def\eg{\textit{e.g.}}

\def\ie{\textit{i.e.}}

\journal{Journal of Computational Physics}

\begin{document}

\begin{frontmatter}

\title{An Exponential Time-Integrator Scheme\\
   for Steady and Unsteady Inviscid Flows}


\author[csrc]{Shu-Jie Li}
\ead{shujie@csrc.ac.cn}
%
\author[csrc,odu]{Li-Shi Luo}
\ead{lluo@odu.edu}
\author[uk]{Z.J. Wang}
\ead{zjw@ku.edu}
\author[usc]{Lili Ju}
\ead{ju@math.sc.edu}

\address[csrc]{Beijing Computational Science Research Center,  Beijing 100193,
  China}
\address[odu]{Department of Mathematics and Statistics, 
Old Dominion University, Norfolk, VA 23529, USA}

\address[uk]{Department of Aerospace Engineering, University of
  Kansas, Lawrence, KS 66045, USA}
\address[usc]{Department of Mathematics, University of South Carolina, 
Columbia, SC 29208, USA}

\begin{abstract}
An exponential time-integrator scheme of second-order 
accuracy based on the predictor-corrector methodology, denoted PCEXP,
is developed to solve multi-dimensional nonlinear partial differential
equations pertaining to fluid dynamics.  The effective and efficient
implementation of PCEXP is realized by means of the Krylov
method. The linear stability and truncation error are analyzed through
a one-dimensional model equation.  
{ The proposed PCEXP scheme is applied to the Euler
  equations discretized with a discontinuous Galerkin method in both
  two and three dimensions. The effectiveness and efficiency of the
  PCEXP scheme are demonstrated for both steady and unsteady inviscid
  flows. The accuracy and efficiency of the PCEXP scheme are verified
  and validated through comparisons with the explicit third-order
  total variation diminishing Runge-Kutta scheme (TVDRK3), the
  implicit backward Euler (BE) and the implicit second-order backward
  difference formula (BDF2).  For unsteady flows, the PCEXP scheme
  generates a temporal error much smaller than the BDF2 scheme does,
  while maintaining the expected acceleration at the same time.  
 Moreover, the PCEXP scheme is also shown to achieve the computational
  efficiency comparable to the implicit schemes  for steady flows.}
\end{abstract}


\end{frontmatter}


\section{Introduction}
\label{sec:intro}

Significant progress has been made recently in the development of
high-order spatial discretization methods in computational fluid
dynamics (CFD), such as the discontinuous Galerkin (DG)
\cite{Reed1973lanl, Lesaint1974, Cockburn1989cm, Bey1991aiaa1575,
  Bassi1997jcp, Cockburn1998jcp, Cockburn1998siamjna, Arnold2000,
  Cockbur2000, Hesthaven2007}, multi-moment constrained finite-volume
(MCV) \cite{Ii2009jcp}, flux reconstruction (FR) or correction
procedure \textit{via} reconstruction (CPR) method
\cite{Huy2012aiaa4079,WangZJ2009jcp,Huynh2014caf}, and others
\cite{ShuCW2002, WangJZ2013ijnmf, WangZJ2016mer}.  These high-order
techniques have exhibited a great potential as effective numerical
solution methods amenable for efficient implementation on massively
parallel high-performance computers.
For complex geometries, an efficient solution, however, also depends
on the availability of a fast time advancement solver.  In contrast to
a relative ubiquity of efficient techniques for spatial
discretizations, efficient time-marching approaches for both steady
and unsteady flows seem to be limited.  Efficient time-integration
approaches are thus the focus of the present work.

For unsteady flows, explicit methods, such as Runge-Kutta (RK)
approaches are prevalent for their simplicity.  However, with highly
clustered nonuniform meshes, the Courant-Friedrichs-Lewy (CFL)
condition can severely limit the local time-step size.  The
restriction due to the CFL condition is particularly acute for direct
numerical simulation (DNS) and large-eddy simulation (LES) of
turbulent flows, which usually require very fine grids of high aspect
ratios in near-wall regions. Thus, the restriction due to the CFL
condition becomes a critical bottleneck in computational efficiency
for explicit time-marching schemes.

To enhance the computational efficiency of explicit time-marching
schemes, it is desirable to relax or to remove the limitation of the
CFL condition. To this end, a class of schemes based on the
exponential time integration shows a great potential \cite{Cox2002jcp,
  Hochbruch1997siamjna, Hochbruck1998siamjsc, Ostermann2006bitnm,
  Tokman2006jcp, Nie2006jcp, Nie2008jcp, Chen2011jcp, Caliari2009anm,
  Hochbruch2009siamjna, Tokman2011jcp, Loffeld2013jcam,
  JuLL2015siamjsc, ZhuLY2016siamjsc}.  In contrast to usual explicit
time-marching schemes, these schemes allow much larger time-step sizes
while maintaining excellent numerical stability.

In explicit time-marching methods, information cannot propagate beyond
one element in each time step, which is the physical significance of
the CFL condition.  In exponential time-marching methods, however,
information is propagated to the entire computational domain
instantaneously through the global Jacobian, similar to implicit
methods, thus significantly alleviating the restriction on time-step
size imposed by the CFL condition, if not eliminating it
altogether. As mentioned previously, a variety of schemes based on the
exponential integration have been developed already (cf., \eg,
\cite{Cox2002jcp, Hochbruch1997siamjna, Hochbruck1998siamjsc,
  Ostermann2006bitnm, Tokman2006jcp, Nie2006jcp, Nie2008jcp,
  Chen2011jcp,Caliari2009anm, Hochbruch2009siamjna, Tokman2011jcp,
  Loffeld2013jcam, JuLL2015siamjsc, ZhuLY2016siamjsc}).
{ While the basic idea of exponential integration has
  been adopted in the aforementioned methods, the existing algorithms
  differ from each other in some aspects.  There are two types of
  exponential schemes depending on the treatment of the nonlinear
  term, \ie, explicit and implicit.  The ETD scheme is a typical
  implicit scheme (cf., \eg, \cite{Cox2002jcp}), while the
  semi-implicit integrator factor method is of an implicit one, which can
  alleviate the stiffness due to the nonlinear term (cf., \eg,
  \cite{Nie2006jcp, Nie2008jcp, Chen2011jcp}).

While most of the exponential schemes are applied to specialized
equations \cite{Cox2002jcp, Hochbruch1997siamjna,
  Hochbruck1998siamjsc, Ostermann2006bitnm, Tokman2006jcp, Nie2006jcp,
  Nie2008jcp, Chen2011jcp,Caliari2009anm, Hochbruch2009siamjna,
  Tokman2011jcp, Loffeld2013jcam} with either scalar exponentials or
constant matrix exponentials, such as the applications to semilinear
parabolic equations \cite{JuLL2015siamjsc, ZhuLY2016siamjsc}, and
relatively few are applied to practical CFD problems (cf., \eg,
\cite{Edwards1994jcp, JC2009, Clancy2013}) with time-dependent full
matrix exponential computations.  } There are some key issues, such as
the computational efficiency for steady problems and the temporal
accuracy for unsteady problems, have yet to be fully investigated.
The overarching goal of the present work is to develop an efficient
and time-accurate exponential scheme to solve multi-dimensional fluid
dynamic equations.
Specifically, we  develop a second-order exponential time-integrator scheme 
to solve the Euler equations for steady and unsteady problems in both two 
and three dimensions, and assess its accuracy and computational efficiency
 by comparing with several well-known explicit and implicit approaches.

The remainder of this paper is organized as follows.
Section~\ref{sec:exp-integrator} discusses the construction of a
second-order exponential scheme based on the predictor-corrector
methodology, denoted as PCEXP, and its efficient implementation
through the Krylov method.
Section~\ref{sec:stability} describes a linear stability and error
analysis of PCEXP for a simple model equation in one dimension.
Section~\ref{sec:DG-Euler} presents the application of PCEXP to the
Euler equations discretized with a high-order DG method
in space.
Section~\ref{sec:results} presents the numerical results of this work
including three inviscid flow problems: (a) the transportation of an
isentropic vortex in 2D with a constant velocity; (b) subsonic flow
over a \mbox{NACA0012} airfoil with a Mach number $\Ma = 0.63$; and
(c) subsonic flow over a sphere in 3D with $\Ma = 0.3$.
The numerical results obtained with PCEXP are compared with
third-order Total Variation Diminishing Runge-Kutta scheme (TVDRK3),
implicit backward Euler (BE), and second-order backward difference
formula (BDF2).
Finally, Section~\ref{sec:finale} summarizes and concludes this work.
The Appendix provides the details of the Jacobian matrices.

\section{Exponential time-integrator schemes}
\label{sec:exp-integrator}

In this section, we first develop a predictor-corrector based the
second-order exponential time-integrator scheme, and then discuss the
efficient implementation through the Krylov method.  We also carry out
a linear stability analysis of the proposed scheme applied to a model
equation in 1D to demonstrate its feasibility of time marching with
large time steps.

\subsection{Predictor-Corrector EXPonential time-integrator scheme (PCEXP)}
\label{sec:PCEXP}

We start with the following semi-discrete system of autonomous
ordinary differential equations which may be obtained from a spatial
discretization:
\begin{equation}
  \frac{\diff \mathbf{u}}{\diff t} 
  = 
  \mathbf{R}(\mathbf{u}) 
  ,
\label{eqn:starteq}
\end{equation}
where { $\mathbf{u} = \mathbf{u}(t) \in \mathbb{R}^K$
  denotes the vector of the solution variables and
  $\mathbf{R}(\mathbf{u}) \in \mathbb{R}^K$ the right-hand-side term
  which may be the spatially discretized residual terms of the
  discontinuous Galerkin method used in this work.  The dimension $K$
  is the degrees of freedom which can be very large for 3D problems.
}
Without loss of generality, we consider { $\mathbf{u}(t)$}
in the interval of one time step, \ie, $t \in [t_n,\, t_{n+1}]$.

We apply the term splitting method \cite{Caliari2009anm} to 
treat Eq.~\eqref{eqn:starteq}:
\begin{equation}
  \frac{\diff \mathbf{u}}{\diff t} 
  =
  \tensor{J}_{n} \mathbf{u}
  + \mathbf{N}(\mathbf{u})
  ,
\label{eqn:starteq2}
\end{equation}
where the subscript  $n$ indicates the value evaluated at $t =
t_n$, $\tensor{J}_{n}$ denotes the Jacobian matrix 
$
\tensor{J}_{n} 
:=
\frac{\partial \mathbf{R}(\mathbf{u})}{\partial \mathbf{u}}
\Big|_{t = t_n}
$
and
$
\mathbf{N}(\mathbf{u}) 
:= 
\mathbf{R}(\mathbf{u}) 
- \tensor{J}_{n}{\mathbf{u}}
$
denotes the reminder, which in general is nonlinear.
Equation~\eqref{eqn:starteq2} admits the following formal solution:
\begin{equation}
  \mathbf{u}_{n+1}  
  = 
  \exp (\Delta t \tensor{J}_n)  \mathbf{u}_n 
  +
  \int_{0}^{\Delta t} 
  \exp \left( (\Delta t-\tau) \tensor{J}_n \right)
 { \mathbf{N} ( \mathbf{u}(t_n + \tau) ) }
  \,
  \diff \tau 
  ,
\label{eqn:s3}
\end{equation}
where $\Delta t := t_{n+1}-t_n$ and
\begin{equation}
  \exp(- t \tensor{J}_n) 
  := 
  \sum_{m=0}^\infty  
  \frac{ \left(- t \tensor{J}_n \right)^m}{m!}
\end{equation}
is the integrating factor.
The formal solution \eqref{eqn:s3} is the starting point to derive the
proposed exponential scheme in which the stiff part is computed
analytically whereas the nonlinear term is approximated numerically.

By substituting the nonlinear term { $\mathbf{N}(t_n +
  \tau)$} with its Taylor expansion about $t_n$
\begin{equation}
  {\mathbf{N}(\mathbf{u}(t_n+\tau))}
  = 
  \sum_{k=1}^\infty 
  \frac{\tau^{k-1}}{(k-1)!} 
  \frac{\partial^{k-1} \mathbf{N}(\mathbf{u}_n)}{\partial \tau^{k-1}}
  ,
\label{eqn:s4}
\end{equation}
the solution~\eqref{eqn:s3} becomes
\begin{align}
  \mathbf{u}_{n+1}  
   = 
  &
  \exp (\Delta t \tensor{J}_n) \,  
  \mathbf{u}_n 
  + 
  \exp (\Delta t \tensor{J}_n)
  \sum_{{k}=1}^\infty 
  \frac{1}{{(k-1)!}}
  \left[
  \int_{0}^{\Delta t}
  \hspace*{-2ex}
  \exp (-\tau \tensor{J}_n) \,
  \tau^{k-1} \,
  \diff \tau 
  \right]
  \frac{\partial^{k-1} \mathbf{N}(\mathbf{u}_n)}
       {\partial \tau^{k-1}}    
  \nonumber
  \\
  =
  &
  \exp (\Delta t \tensor{J}_n) \, \mathbf{u}_n 
  + \sum_{k=1}^{\infty}{\Delta t}^k \,
  \tensor{\Phi}_k(\Delta t \tensor{J}_n) 
  \frac{\partial^{k-1} \mathbf{N} (\mathbf{u}_n) }{\partial  \tau^{k-1} } 
  ,
\label{s5}
\end{align}
where the tensorial function $\tensor{\Phi}_k(\Delta t \tensor{J}_n)$
is defined as the following:
\begin{equation}
  \tensor{\Phi}_{k}(\Delta t \tensor{J}_n) 
  := 
  \frac{\exp( \Delta t \tensor{J}_n )}{{\Delta t}^{k} (k-1)! }  
  \int_{0}^{\Delta t} 
  \hspace{-2.0ex}
  \exp ( - \tau \tensor{J}_n ) \tau^{k-1} \, 
  \diff \tau ,
  \quad
  k \geq 1,
\end{equation}
and it satisfies the following recursion relationship:
%
\begin{subequations}
\begin{align}
  &
  \tensor{\Phi}_{k+1}(\Delta t \tensor{J}) 
  =
  \frac{\tensor{J}^{-1}}{\Delta t \, k! }
  \left[
  k! \, \tensor{\Phi}_k ( \Delta t \tensor{J} ) 
  -
  \tensor{I}
  \right]
  ,
  \quad
  k \geq 1,
\label{s6}
  \\
  &
  \tensor{\Phi}_1( \Delta t \tensor{J} )
  :=
  \frac{\tensor{J}^{-1}}{\Delta t}
  \left[
  \exp\left( \Delta t \tensor{J} \right)
  - \tensor{I}
  \right]
  ,
\label{eqn:Phi1}
\end{align}
\end{subequations}
where $\tensor{I}$ denotes the { $K \times K$} identity matrix.
Thus, an approximation of the integral in \eqref{eqn:s3} by a
truncated Taylor expansion of the nonlinear term $\mathbf{N}$ leads to
an exponential scheme consisting of linear combinations  of
functions $\tensor{\Phi}_k$.
Specifically, with $k=1$, the nonlinear term $\mathbf{N}$ is
approximated by a constant, \ie, its left-end value $\mathbf{N}_n$ on
the interval $[t_n,\, t_{n+1}]$, hence,
\begin{equation}
  {\mathbf{N}(\mathbf{u}(t_n+\tau)) }
  \approx 
  \mathbf{N}_n
  =
  \mathbf{R}_n 
  - 
  \tensor{J}_n 
  \mathbf{u}_n 
  , 
\end{equation}
leading to a simple exponential scheme
\begin{equation}
  \mathbf{u}_{n+1}
  =
  \exp (\Delta t \tensor{J}_n) 
  \, \mathbf{u}_n + {\Delta t} \, \tensor{\Phi}_1(\Delta t \tensor{J}_n) 
  \, \mathbf{N}_n 
  =
  \mathbf{u}_n + {\Delta t} \, \tensor{\Phi}_1(\Delta t \tensor{J}_n) 
  \, \mathbf{R}_n 
  ,
\label{s8}
\end{equation}
which is the first-order exponential-time differencing scheme, ETD1
\cite{Cox2002jcp}, also referred as exponential Rosenbrock-Euler
method \cite{Caliari2009anm}.

With $k=2$, a first-order finite-difference approximation to the
derivative of $\mathbf{N}$, \ie,
$$
\partial_{\tau} \mathbf{N}_n 
\approx 
\left(
\mathbf{N}_n - \mathbf{N}_{n-1} 
\right) 
/ \Delta t 
,
$$ 
leads to the second-order scheme ETD2
\cite{JuLL2015siamjsc}:
\begin{subequations}
\begin{align}
  &
  \mathbf{u}_{n+1}  
  = 
  \underbrace{
  \exp ( \Delta t \tensor{J}_n ) 
  \, \mathbf{u}_n + \Delta t 
  \, \tensor{\Phi}_1( \Delta t \tensor{J}_n )
  \, \mathbf{N}_n 
  }_{\mbox{\scriptsize ETD1 defined by \eqref{s8}}}
  +
  {\Delta t} 
  \, \tensor{\Phi}_2(\Delta t \tensor{J}_n) 
  \, \left( \mathbf{N}_n - \mathbf{N}_{n-1} \right) 
  ,
\label{s7}
  \\
  &
  {
  \tensor{\Phi}_2( \Delta t \tensor{J} )
  := 
  \frac{\tensor{J}^{-2}}{{\Delta t}^2}
  \left[
    \exp\left( \Delta t \tensor{J} \right)
    -{\Delta t}\tensor{J}- \tensor{I}
    \right]
  }
  .
\label{eqn:Phi2}
\end{align}
\end{subequations}
{ Clearly the ETD2 scheme requires an extra term $\tensor{\Phi}_2(\Delta t \tensor{J}_n) ( \mathbf{N}_n -
  \mathbf{N}_{n-1} )$ compared to the ETD1 scheme.
It should be stressed that the calculation of $\tensor{\Phi}_2$ is
computationally more demanding than that of $\tensor{\Phi}_1$ for a
system with large degrees of freedom $K$.  Hence, the ETD2 scheme might
not be practical for large systems.}

The Runge-Kutta or multi-step approximations for the nonlinear term
$\mathbf{N}(\mathbf{u})$ can also be used to construct high-order
schemes (cf., \eg, \cite{Cox2002jcp, Tokman2011jcp, ZhuLY2016siamjsc}).
However, the objective of this work
is to construct an effective and efficient second-order ETD scheme
which only requires $\tensor{\Phi}_1$.  To this end, we design a
scheme based on the idea of the predictor-corrector methodology
consisting of two stages.
First, the solution is advanced with the first-order ETD scheme
\eqref{s8} to obtain a predicted solution $\mathbf{u}_*$.
{ Next, the solution $\mathbf{u}_{n+1}$ is corrected by
  replacing the nonlinear term ${\mathbf N}_n :=
  \mathbf{N}(\mathbf{u}_n)$ by the algebraic average of itself and its
  predicted solution $\mathbf{N}_* := \mathbf{N}(\mathbf{u}_*)$, which
  is a standard second-order Gaussian quadrature or midpoint
  approximation.  This simple procedure enhances the accuracy of the
  scheme from first order to second order.  The two-stage scheme can
  be summarized as below:
  }
\begin{subequations}
\begin{align}
  &
  \mathbf{u}_* 
  = 
  \mathbf{u}_n 
  + {\Delta t} \, \tensor{\Phi}_1(\Delta t \tensor{J}_n)  
  \mathbf{R}_n
  ,
\label{s9}
  \\
  &
  \mathbf{u}_{n+1} 
  =  
  \mathbf{u}_* 
  + \frac{1}{2} \Delta t 
  \, \tensor{\Phi}_1({\Delta t \tensor{J}_n})
  \left( \mathbf{N}_* - \mathbf{N}_n \right)
  .
\label{s10}
\end{align}
\label{eqn:PCEXP}
\end{subequations}
The above two-stage scheme is designated as the predictor-corrector
exponential (PCEXP) scheme.  The first stage of PCEXP is designated as
EXP1, which is only used for steady problems.  Note that the PCEXP
scheme is in fact a one-step scheme, \ie, only the solution
$\mathbf{u}$ at the current time $t=t_n$ is required.

\subsection{Realization of PCEXP with the Krylov method}

The exponential time-integrator schemes require evaluations of
matrix-vector products, and in particular, the product of the
exponential functions of the Jacobian and a vector, \eg,
$\tensor{\Phi}_1 ( \Delta t \tensor{J}_n ) \mathbf{N}$ in \eqref{s10}.
If the inverse of the Jacobian $\tensor{J}$ exists, then it is
possible to use $\tensor{J}^{-1}$ to compute $\tensor{\Phi}_1 (\Delta
t \tensor{J})$ defined by \eqref{eqn:Phi1}.
However, $\tensor{J}$ { may be singular, \eg, in the
  presence of periodic boundary conditions, thus $\tensor{J}^{-1}$ may
  have to be computed with an operator restriction treatment for
  generalized matrix inversion, such as the singular value
  decomposition (SVD).}  In addition, for a problem with a very large
number of degrees of freedom, { direct} inversion of
$\tensor{J}$ can be prohibitively expensive to compute. These
impediments could be the reason why the exponential schemes have yet
to gain much traction.

The matrix-vector products in \eqref{s10} can be approximated
efficiently using the Krylov method \cite{Tokman2010pcs,
  Saad1992siamjna}, which can also treat a singular $\tensor{J}$.  The
basic idea of the Krylov method is to approximate the product of
$\exp(\Delta t \tensor{J})$ and a vector, { such as
  $\mathbf{N}$ in \eqref{s10}}, by projecting it onto a small Krylov
subspace, resulting in a much smaller matrix thus cheaper in
computational effort.  The algorithm will be discussed in detail next.

With the Taylor expansion of $\exp(\Delta t \tensor{J})$, the product
$\tensor{\Phi}_1 \mathbf{N}$ can be written as:
\begin{equation}
  \tensor{J}^{-1} 
  \frac{\exp(\Delta t \tensor{J}) - \tensor{I}}{\Delta t} \mathbf{N} 
  = 
  \sum_{k=0}^\infty
  \frac{ (\Delta t \tensor{J})^k}{(k+1)!} \mathbf{N}
  =
  \left( 
  \tensor{I} 
  + \frac{(\Delta t \tensor{J})}{2!} 
  + \frac {(\Delta t \tensor{J})^2}{3!} 
  + \cdots 
  \right)  
  \mathbf{N}
  .
\end{equation}
It can be approximated by the following function projection onto the
Krylov subspace of dimension $m$:
\begin{equation}
  \mathbb{K}_m(\tensor{J}, \mathbf{N}) 
  = 
  \mbox{span} 
  \{ \mathbf{N},\, 
  \tensor{J} \mathbf{N},\, 
  \tensor{J}^2 \mathbf{N},\, 
  \ldots,\, 
  \tensor{J}^{m-1} \mathbf{N} \}
  .
\end{equation}
%

The orthogonal basis matrix $\tensor{V}_m := ( \mathbf{v}_1,\,
\mathbf{v}_2,\, \cdots,\, \mathbf{v}_m ) \in \mathbb{R}^{K \times m}$
satisfies the so-called Arnoldi decomposition \cite{Saad1992siamjna}:
\begin{equation}
  \tensor{J} \tensor{V}_m 
  =
  \tensor{V}_{m+1}  \widetilde{\tensor{H}}_m,
\label{eqn:vm}
\end{equation}
where $ \tensor{V}_{m+1} := ( \mathbf{v}_1,\,
\mathbf{v}_2,\, \cdots,\, \mathbf{v}_m,\, \mathbf{v}_{m+1} ) = (
\tensor{V}_{m},\, \mathbf{v}_{m+1} ) \in \mathbb{R}^{K \times (m+1)}$
and $\widetilde{\tensor{H}}_m$ is the following $(m+1) \times m$
upper-{ Hessenberg} matrix:
\begin{equation}
\widetilde{\tensor{H}}_m  
= 
\begin{bmatrix}
  h_{1,1} & h_{1,2} & h_{1,3} & h_{1,4} & \cdots & h_{1,m} 
  \\
  h_{2,1} & h_{2,2} & h_{2,3} & h_{2,4} & \cdots & h_{2,m} 
  \\
  0   & h_{3,2} & h_{3,3} &h_{3,4} & \cdots &  h_{3,m} 
  \\
  & 0  & h_{4,3} & \ddots & \ddots & \vdots 
  \\
  \vdots &  & 0 &  \ddots  & h_{m-1,m-1} & h_{m-1,m}
  \\
  & & & \ddots & h_{m,m-1} & h_{m,m} 
  \\           
  0 & & \cdots & & 0 & h_{m+1,m}
\end{bmatrix}
  .
\end{equation}
The matrix $\widetilde{\tensor{H}}_m$ can be written as the following:
\begin{equation}
  \widetilde{\tensor{H}}_m 
  =
  \begin{bmatrix}
    \tensor{H}_{m}
  \\
  h_{m+1,m} \mathbf{e}_m^{\tensor{T}} 
  \end{bmatrix}
  ,
\end{equation}
where $\tensor{H}_{m}$ is the matrix composed of the first $m$ rows of
$\widetilde{\tensor{H}}_m$ and $\mathbf{e}_m := (0,\, \cdots,\, 0,\,
1)^{\tensor{T}} \in \mathbb{R}^m$ is the $m$-th canonical basis vector
in $\mathbb{R}^m$, then Eq.~\eqref{eqn:vm} becomes
\begin{equation}
  \tensor{J} \tensor{V}_m
  =
  \tensor{V}_m \tensor{H}_m 
  + h_{m+1,m} \mathbf{v}_{m+1} \mathbf{e}_m^{\tensor{T}} 
  . 
\label{vm1}
\end{equation}
{ Because $\tensor{V}_m^{\tensor{T}} \tensor{V}_m =
  \tensor{I}$,
}
the upper-Hessenberg matrix $\widetilde{\tensor{H}}_m$ is given by:
\begin{equation}
  \tensor{H}_m 
  =
  \tensor{V}^{\tensor{T}}_m \tensor{J} \tensor{V}_m 
  .
\label{vm2}
\end{equation}
Therefore $\tensor{H}_m$ is the projection of the linear
transformation of $\tensor{J}$ onto the subspace $\mathbb{K}_m$ with
the basis $\mathbb{V}_m$.
Because $\tensor{V}_m \tensor{V}_m^{\tensor{T}} \neq \tensor{I}$,
\eqref{vm2} leads to the following approximation:
\begin{equation}
  \tensor{J}
  \approx
  \tensor{V}_m \tensor{V}^{\tensor{T}}_m 
  \tensor{J} 
  \tensor{V}_m \tensor{V}^{\tensor{T}}_m
  =
  \tensor{V}_m \tensor{H}_m \tensor{V}^{\tensor{T}}_m 
  ,
\end{equation}
and $\exp (\tensor{J})$ can be approximated by $\exp ( \tensor{V}_m
\tensor{H}_m \tensor{V}^{\tensor{T}}_m )$ as the following:
\begin{equation}
  \exp (\tensor{J}) \mathbf{N}   
  \approx   
  \exp ( \tensor{V}_m \tensor{H}_m  \tensor{V}^{\tensor{T}}_m)  
  \mathbf{N} 
  = 
  \tensor{V}_m  
  \exp  ( \tensor{H}_m )  \tensor{V}^{\tensor{T}}_m   \mathbf{N}
  .
\label{evv}
\end{equation}
The first column vector of $\tensor{V}_m$ is $\mathbf{v}_1 =
\mathbf{N}/ \| \mathbf{N} \|_2$ and $\tensor{V}^{\tensor{T}}_m
\mathbf{N} = \| \mathbf{N} \|_2 \, \mathbf{e}_1$,
 thus \eqref{evv} becomes:
\begin{equation}
  \exp (\tensor{J}) \mathbf{N} 
  \approx   
  \| \mathbf{N} \|_2  
  \tensor{V}_m \exp(\tensor{H}_m) \, {\mathbf{e}_1} 
  .
\end{equation}
Consequently $\tensor{\Phi}_1$ can be approximated by:
\begin{equation}
  \tensor{\Phi}_1(\Delta t  \tensor{J})  \mathbf{N} 
  = 
  \frac {1}{{\Delta t}}
  \int_{0}^{\Delta t} \hspace*{-2ex} \exp ((\Delta t-\tau)
  \tensor{J})   
  \mathbf{N} \, 
  \diff \tau 
  \approx 
  \frac {1}{{\Delta t}}  
  \int_{0}^{\Delta t} \hspace*{-2ex}  \| \mathbf{N} \|_2  \tensor{V}_m  
  \exp \left((\Delta t-\tau) \tensor{H}_m \right) 
  \, {\mathbf{e}_1} \, \diff \tau.
\label{final1}
\end{equation}
In general, the dimension of the Krylov subspace, $m$, is chosen to be
much smaller than the dimension of $\tensor{J}$, $K$, thus
$\tensor{H}_m \in \mathbb{R}^{m\times m}$ can be inverted easily,
so $\tensor{\Phi}_1$ can be easily computed as the following:
\begin{align}
  \tensor{\Phi}_1(\Delta t \tensor{J})  \mathbf{N}  
  &\approx 
  \frac {1}{\Delta t}  
  \| \mathbf{N} \|_2 \tensor{V}_m  
  \int_{0}^{\Delta t}
  \exp \left((\Delta t-\tau) \tensor{H}_m \right) \, 
  \mathbf{e}_1 \, \diff \tau
  \nonumber 
  \\
  &= 
  \frac {1}{{\Delta t}}  \|\mathbf{N}\|_2  
  \tensor{V}_m   \tensor{H}_m^{-1}  
  \left[ \exp (\Delta{t}   \tensor{H}_m) - \tensor{I} \right] 
  \mathbf{e}_1
  ,
\label{final}
\end{align}
where the term $\exp(\Delta t \tensor{H}_m)$ can be computed
efficiently by the Chebyshev rational approximation (cf., \eg,
\cite{Saad1992siamjna, Moler2003siamjna}) due to the small size of
$\tensor{H}_m$.

\section{Linear stability analysis and local truncation error}
\label{sec:stability}

In this section, we carry out a linear stability analysis of the
proposed PCEXP scheme by considering a scalar equation for which
analytic results can be obtained. This example is instructive because
the growth rate in the scalar equation, $\exp( \Delta t J )$ with a
constant $J$, is the degenerated case of $\exp( \Delta t \tensor{J}
)$. The stability of the PCEXP scheme is compared with the TVDRK3
scheme. 
In addition, we will also analyze the local truncation error of the
PCEXP scheme applied to the scalar equation.

\subsection{Linear stability analysis}

We shall analyze the stability of the following scalar equation with a
constant $J$:
\begin{equation}
  u^\prime = J u + N (u).
\label{ff1}
\end{equation}
Linearization of equation \eqref{ff1} about a fixed point $u_0$, such
that $ J u_0 + N(u_0)=0$ leads to the following simple linear equation
\begin{equation}
  u^\prime  
  = 
  J \,u + \lambda \,u
  :=
  \zeta \, u
  ,
\label{model}
\end{equation}
where $u$ is now the perturbation to $u_0$, $\lambda = N^\prime(u_0)$,
and $\zeta := J + \lambda$. Obviously, the fixed point $u_0$ is stable
if and only if
\begin{equation}
  \mbox{Re} (\zeta) 
  \equiv
  \mbox{Re}(J + \lambda) 
  \leq 0.
\label{eqn:stabilitycondition}
\end{equation}
To analyze the dependence of the stability region on the finite-term
Krylov basis approximation,
we apply the PCEXP scheme \eqref{eqn:PCEXP} to the model equation
\eqref{model} and we obtain the following very simple solution:
\begin{equation}
  u_{n+1} = \exp (\Delta t \zeta) \, u_n := G \, u_n,
  \quad
  G 
  :=
  \exp (\Delta t \zeta)
  .
\label{eqn:growthfactor}
\end{equation}
{ Note that the solution of the linear equation
  produced by the PCEXP scheme is the exact solution
  of the same linear equation in a standard exponential form.  This
  capability of producing the exact solution of a linear equation is
  an important feature of exponential schemes.}
The function $G$ can be approximated by its Taylor expansion up to $m$-th
order:
\begin{equation}
  G_m 
  = 
  1 
  + (\Delta t \zeta) 
  + \frac {(\Delta t \zeta)^2}{2!} 
  + \frac {(\Delta t \zeta)^3}{3!} 
  + \cdots 
  + \frac{(\Delta t \zeta)^{m}}{m!} 
  .
\label{eqn:G_m}
\end{equation}
The polynomial $G_m$ approximates the growth rate $G$,
and the stability criterion \eqref{eqn:stabilitycondition} requires
that $| G_m | \leq 1$.

\begin{figure}[htb!]
\centering
\subfigure[$m$-dependence of stability curves]{%
\label{m1}
\includegraphics[width=0.45\textwidth]{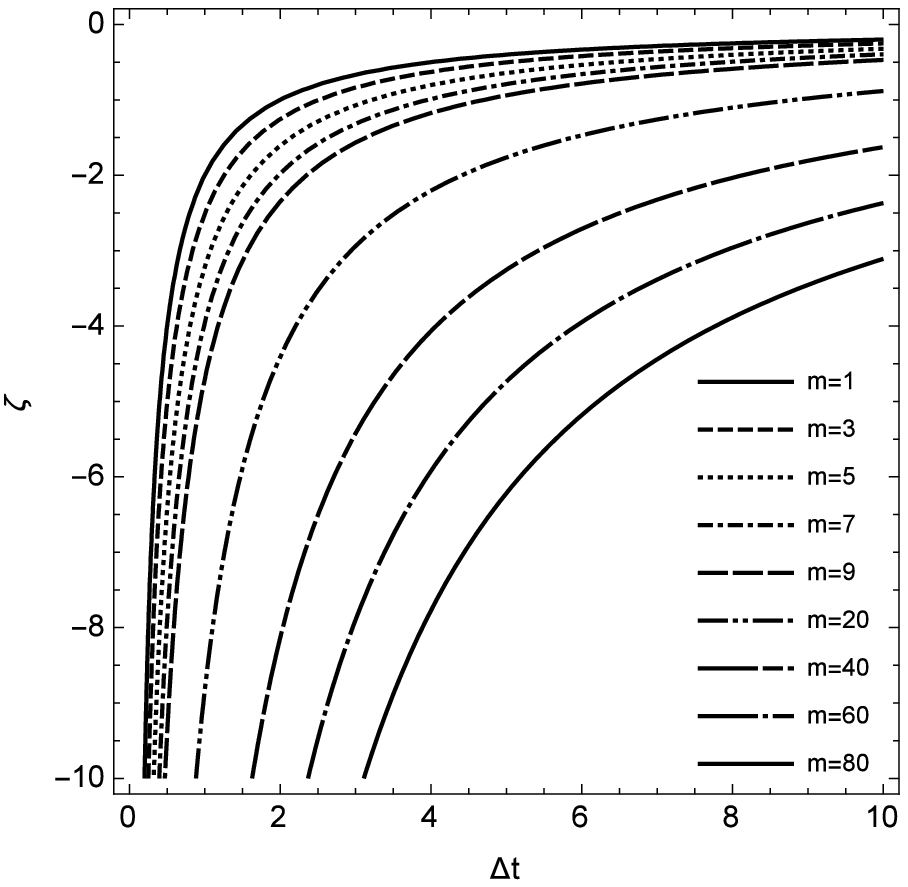}}
\subfigure[$m$-dependence of $|\Delta t \zeta |$]{%
\label{m2}
\includegraphics[width=0.44\textwidth]{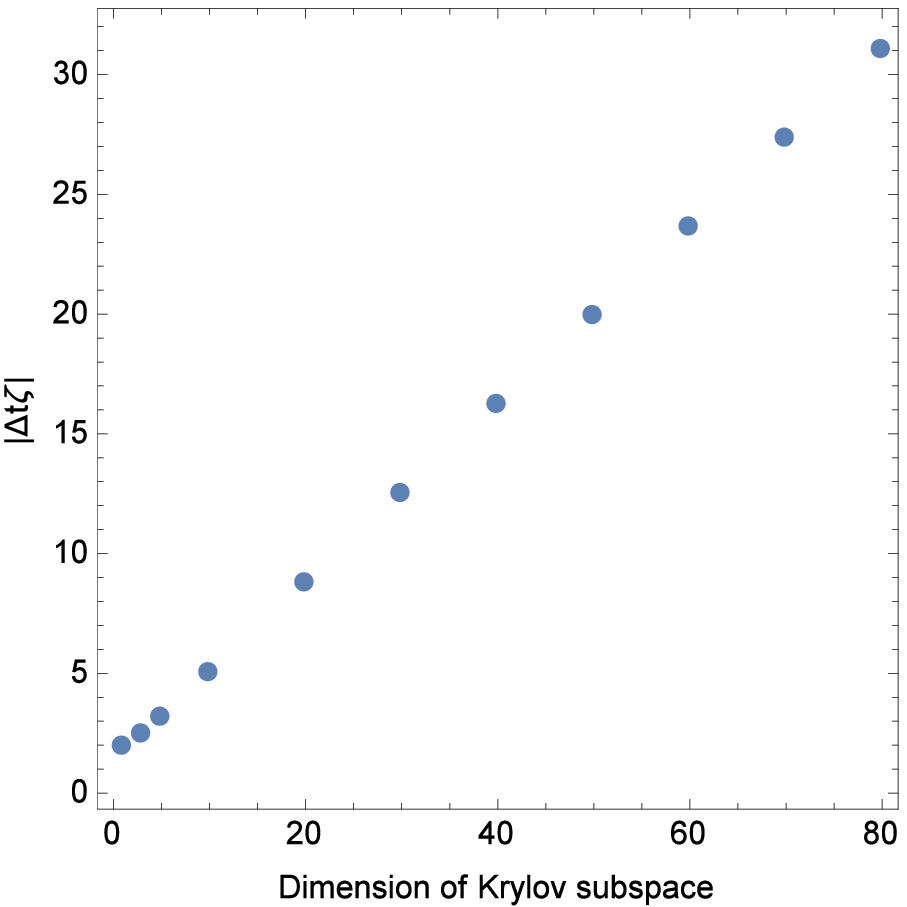}}
\caption{The $m$-dependence of the stability boundary in the parameter
  space $(\Delta t, \, \zeta)$ (left) and the stability constant
  $|\Delta t \zeta |$ (right).  { The stable region in
    $(\Delta t, \, \zeta)$ is bounded by the both axes of $\Delta t$
    and $\zeta$ and the $m$-dependent curve.}}
\label{fig:m80}
\end{figure}

We compute the dependence of the stability region determined by $| G_m
| \leq 1$ in the parameter space $(\Delta t,\, \zeta)$ on the order of
the polynomial $G_m$, $m$. First, for real $\zeta$, we compute the
boundary of the stability region with $1 \leq m \leq 80$. The
$m$-dependence of the stability boundary in the parameter space
$(\Delta t,\, \zeta)$ is shown in Fig.~\ref{fig:m80}(a);
{ the stability region is bounded by both $\Delta t$ and
  $\zeta$ axes and the $m$-dependent boundary. Clearly, the the
  stability region expands as $m$ increases.
}
We also compute the $m$-dependence of the maximum stable value of
$|\Delta t \zeta|$, as shown in Fig.~\ref{fig:m80}(b).  It can be seen
that the maximum stable value of $|\Delta t \zeta|$ increases with $m$
linearly, for the linear problem considered. The result of
Fig.~\ref{fig:m80}(b) is also tabulated in Table~\ref{timesize}, which
also gives the $m$-dependent time-step sizes normalized by $\Delta
t_1$ corresponding to $m=1$.

\begin{table}[htb!] 
\setlength{\belowcaptionskip}{10pt}
\begin{center} 
 \caption{Time step with various number of Krylov basis}
  \label{timesize}
\begin{tabular}{rrr | rrr} 
\toprule
\ $m$ & \multicolumn{1}{c}{$\Delta t$}  & { $ \Delta t / \Delta t_1$ } &
\ $m$ & \multicolumn{1}{c}{$\Delta t$}  & {$ \Delta t / \Delta t_1$} \\
\midrule
  1 &  $2.00000/|\lambda|$ &  1.00 &  40 & $16.2705/|\lambda|$&  8.14 \\
 5 & $3.21705/|\lambda|$ &  1.61 & 50 & $19.9819/|\lambda|$&  9.99 \\
10 & $5.06952/|\lambda|$&  2.53  & 60 & $23.6883/|\lambda|$&  11.84\\
 20 &$8.82143/|\lambda|$ &  4.41 & 70 & $27.3910/|\lambda|$&  13.70 \\
 30 & $12.55170/|\lambda|$&  6.28 & 80 & $31.0908/|\lambda|$&  15.54\\
\bottomrule
\end{tabular}
\end{center}
\end{table}

\begin{figure}[htb!]
\centering
\includegraphics[height=0.4\textheight]{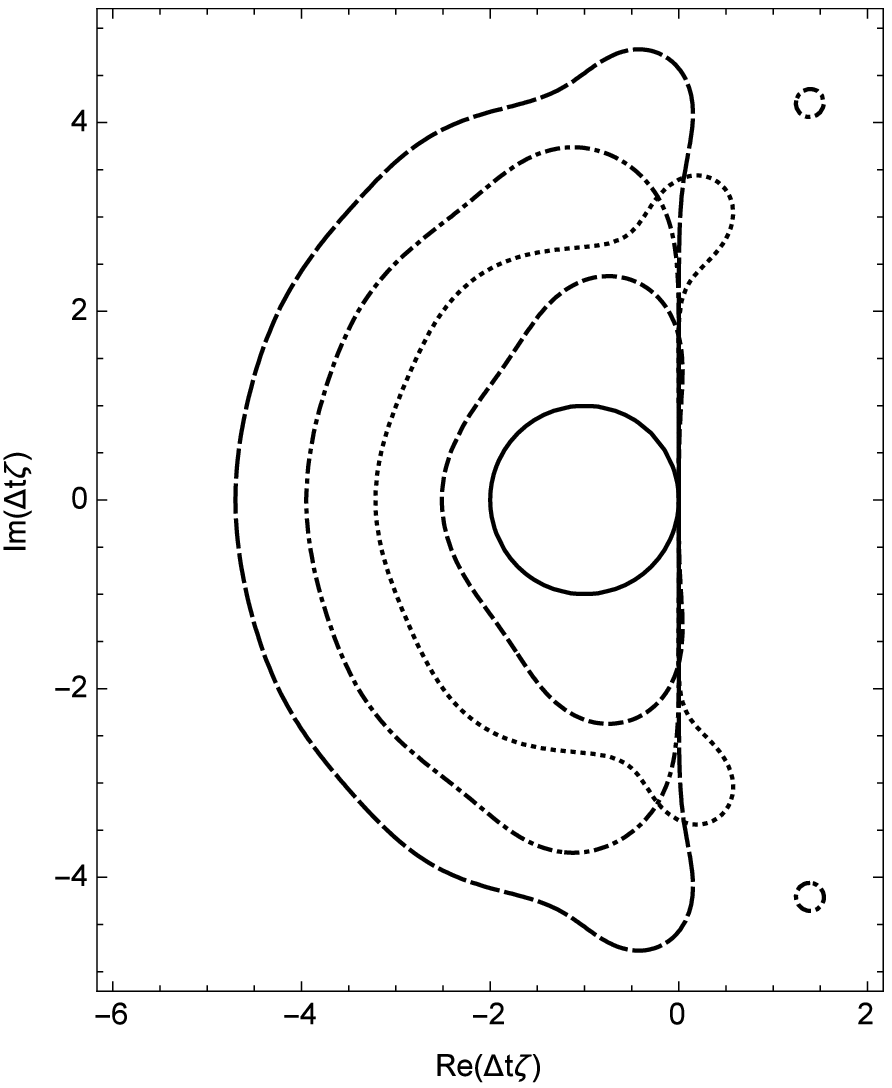}
\hspace*{3em}
\includegraphics[height=0.4\textheight]{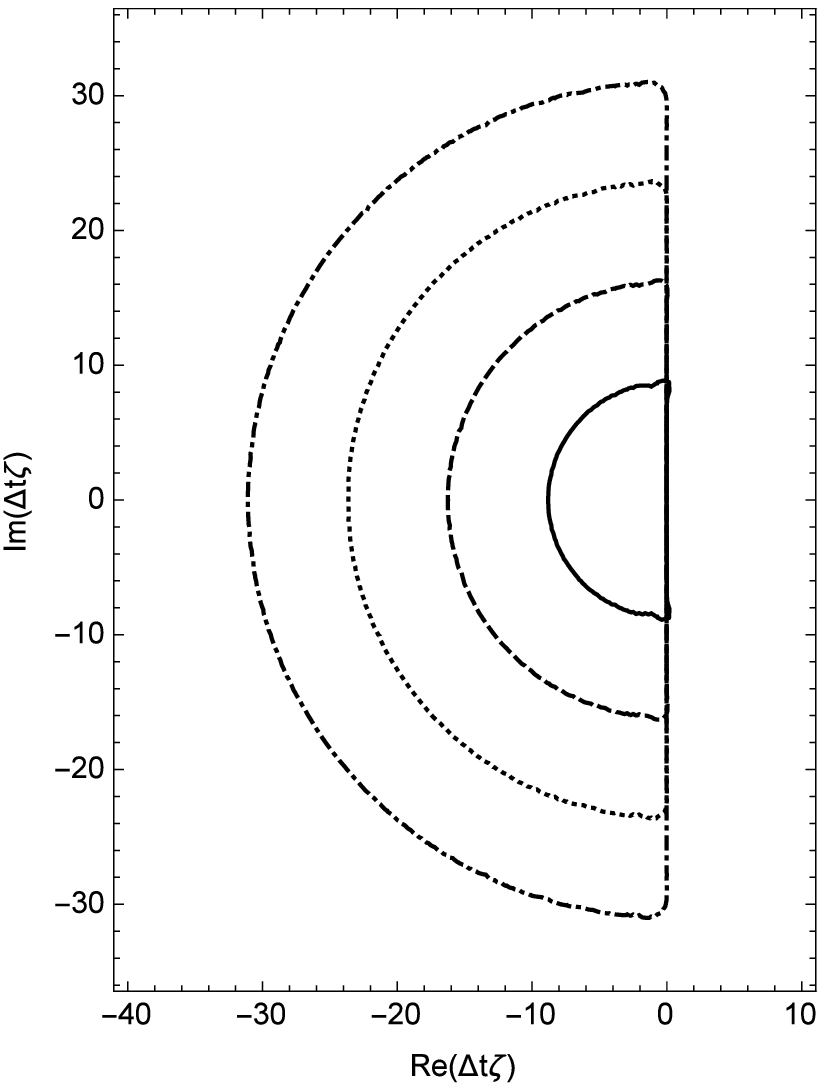}
\caption{Stability analysis of the scalar equation~\eqref{model} with
  complex $\zeta$.  The $m$-dependence of the stability region in the
  $ \Delta t \zeta$ complex plane. Left: $m=1$, 3, 5, 7, 9; Right:
  $m=20$, 40, 60, 80.  The $m$ value increases from the inner curves
  to the outer ones for the both figures.}
\label{sreigion}
\end{figure}

For complex $\zeta$, the boundary of the stability region is defined
in the complex plane of $\Delta t \zeta$. As shown in
Fig.~\ref{sreigion}, the stability region is approximately a
semicircle on the left half of the complex plane $\zeta$, and the
radius of the semicircle grows linearly as $m$ increases.

The preceding analysis shows the effect of the number of the terms in
the Taylor expansion of the propagator, $\exp(\Delta t \zeta)$, on the
stability of the exponential scheme. Next, we analyze the
stability of the PCEXP scheme \eqref{eqn:PCEXP} by using the model
problem \eqref{model}.  Applying the PCEXP scheme of \eqref{eqn:PCEXP}
to \eqref{model}, we have:
\begin{align}
  r
  = 
  &
  \frac{u_{n+1}}{u_n} 
  = 
  \frac{\lambda \,(\lambda-J) + e^{2 J \Delta t} \, 
    (J+\lambda) \, \lambda + 2 e^{J \Delta t} \, (J^2 - \lambda^2)}
       {2 J^2}
  \nonumber
  \\
  = 
  &
  \frac {a (a - b) + e^{2 b} (a+b) a + 2 e^{b} (b^2-a^2)}{2 b^2}
  ,
\label{grow}
\end{align}
where $a := \Delta t \,\lambda$ and $b := \Delta t \, J$.
The Taylor expansion of $r$ in terms of $b$ is
\begin{equation}   
  r 
  = 
  \left(1+a+\frac{a^2}{2} \right) 
  + 
  \left(1+a+\frac{a^2}{2}\right)\, b
  +
  \left(\frac{1}{2}+\frac{2 a}{3}+\frac{7a^2}{24} \right) \, b^2 
  + O(b^3) 
  .
\label{y_poly}
\end{equation}
In the limit that $b \rightarrow {0}$, the growth factor $r$ converges
to the leading term of \eqref{y_poly}
\begin{equation}
\label{rk2}
r \rightarrow 1+a+\frac{a^2}{2},
\end{equation}
which is identical to the second-order Runge-Kutta scheme.

{ The above analysis directly shows the following
  two important features of the PCEXP scheme in the limits of $N
  \rightarrow 0$ and $J \rightarrow 0$}:
\begin{enumerate}

\item { The PCEXP scheme reduces to the exact solution
  of the linear equation in the limit of vanishing nonlinear term $N
  \rightarrow 0$, as shown by \eqref{eqn:growthfactor};}

\item { The PCEXP scheme converges to the second-order
  Runge-Kutta scheme in the limit of vanishing linear term $J
  \rightarrow {0}$, as shown by \eqref{rk2}.}

\end{enumerate}

\begin{figure}[htb!]
\centering
\includegraphics[width=0.45\textwidth]{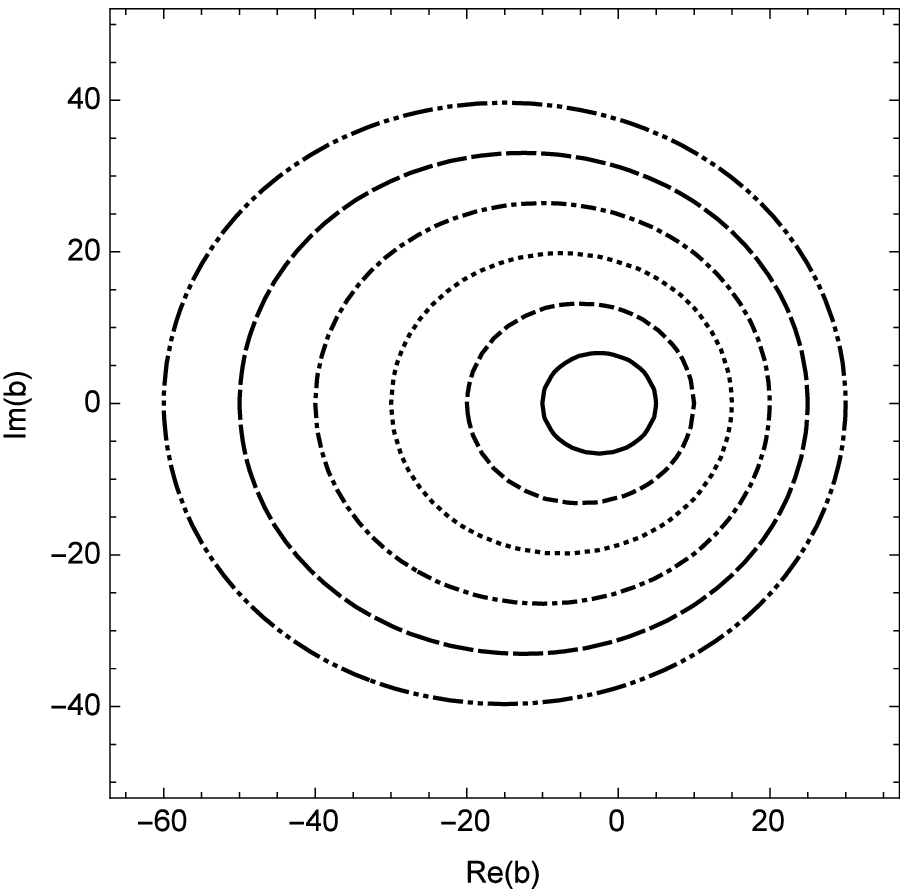}%
\includegraphics[width=0.45\textwidth]{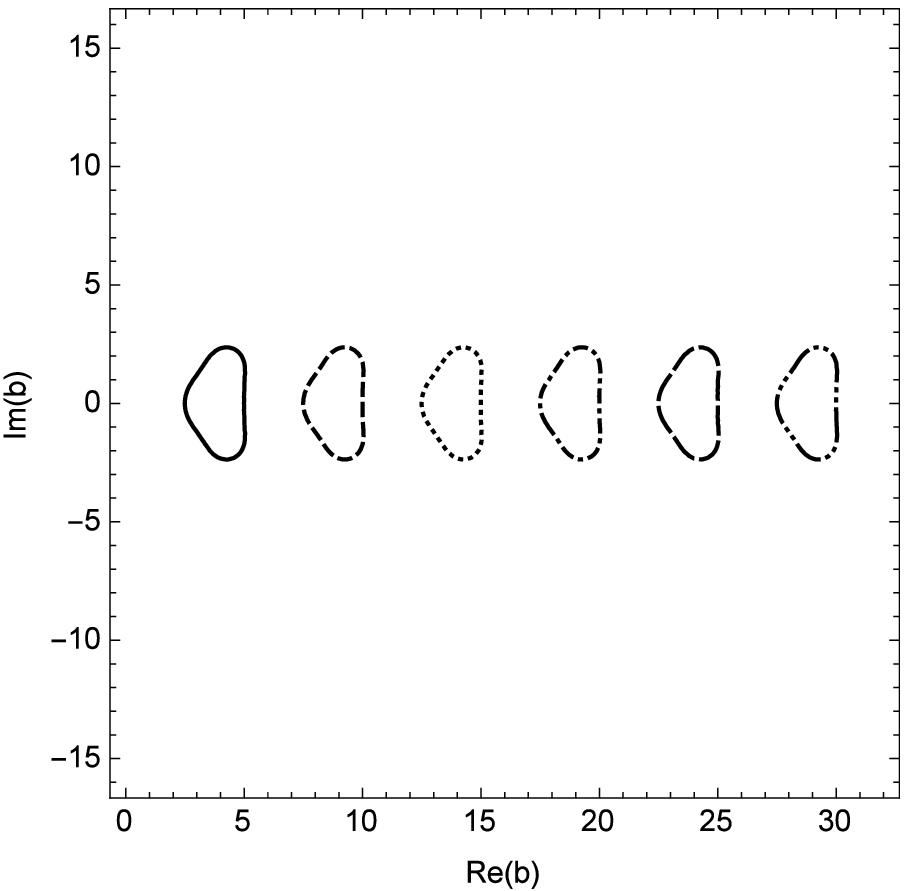}
\caption{The stability region of the PCEXP scheme (left) and the
  TVDRK3 scheme (right) on the complex plan of $b := \Delta t \, J$
  with real $a := \Delta t \, \lambda$ as varying parameter.  $a= 5$,
  10, 15, 20, 25, and 30, corresponding to enlarging elliptical curves
  for the PCEXP scheme (left) { and the curves moving
    away positively from the origin along the real axis on the complex $b$
    plane for the TVDRK3 scheme.  The interior regions of the closed
    curves are the stable regions. Note that the sales of two figures
    are different.}  }
\label{fig:xfix}
\end{figure}

We now consider the dependence of the stability region on the
parameter  $a := \Delta t \,\lambda$ and $b := \Delta t \, J$.
For real $a$ and $b$, the boundary of the stability region is defined
by $|r|=1$. For the PCEXP scheme, $r$ is
given by \eqref{grow}, and $|r| \leq 1$ leads to
\begin{equation}
  a \leq -b, 
  \quad 
  a 
  \geq 
  \frac{2b}{1-e^b} 
  .
\label{pcexp_bd}
\end{equation}
{
For the EXP1 scheme, 
\begin{equation}
  r
  = 
  \frac{u_{n+1}}{u_n} 
  = 
  \frac{a (e^b-1) + b e^b }{b}
  ,
\label{exp1_grow}
\end{equation}
%
then the corresponding stability region is bounded by
\begin{equation}
  a \leq - b , 
  \quad 
  a \geq - b \, \frac{e^b + 1}{e^b - 1}
  = 
  - b \coth \frac{b}{2}
  .
\label{exp1_bd}
\end{equation}
Similarly, 
for the TVDRK3 scheme
\begin{equation}
  r
  = 
  \frac{u_{n+1}}{u_n} 
  = 1+ (a+b) + \frac{1}{2}(a+b)^2 + \frac{1}{6}(a+b)^3
  ,
\label{rkgrow}
\end{equation}
so its stability region is a strip bounded by two parallel lines:
\begin{equation}
  a \leq -b, 
  \quad 
  a 
  \geq 
  -b -1 - \left(4+{\sqrt {17}}\right)^{1/3}
  + \left(4+{\sqrt {17}}\right)^{-1/3} 
  .
\end{equation}
The stability regions of PCEXP, EXP1, and TVDRK3 are illustrated in
Fig.~\ref{fig:bdexp}.

\begin{figure}[htb!]
\centering
\includegraphics[width=0.33\textwidth]{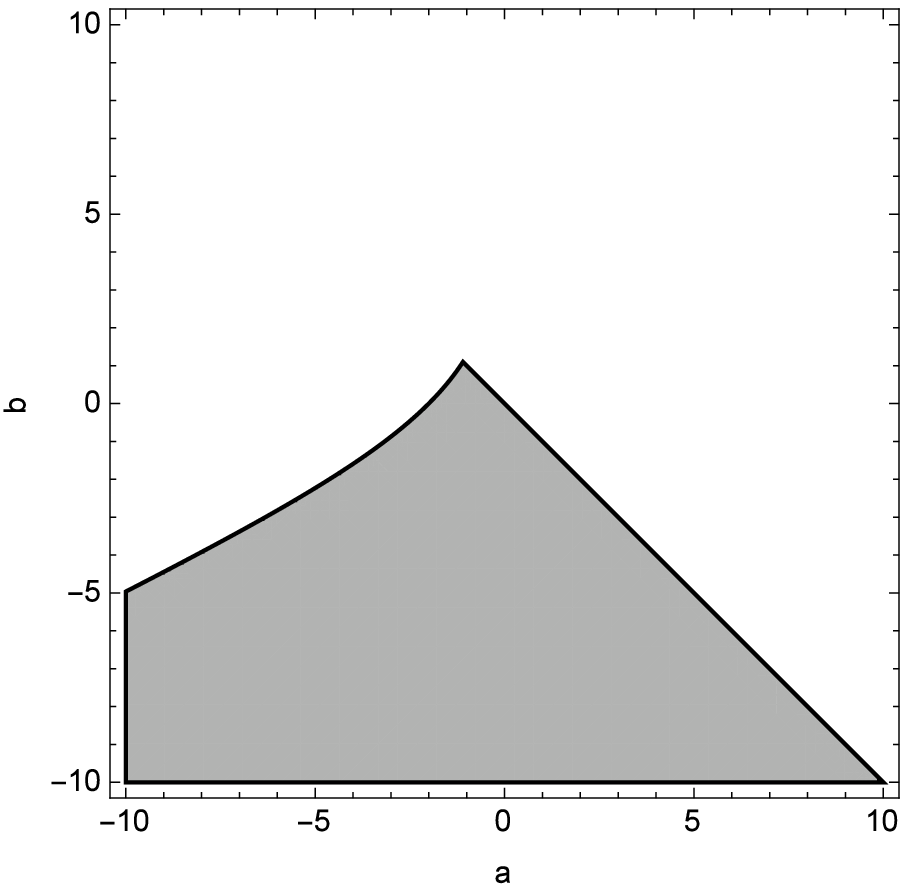}%
\includegraphics[width=0.33\textwidth]{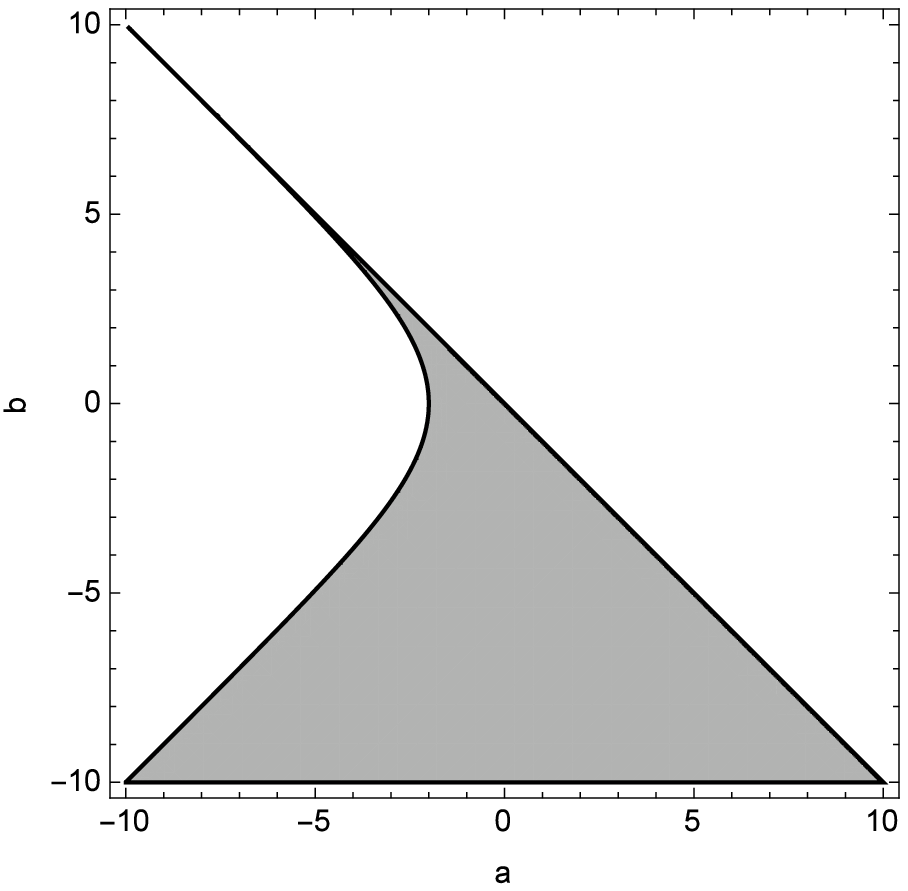}%
\includegraphics[width=0.33\textwidth]{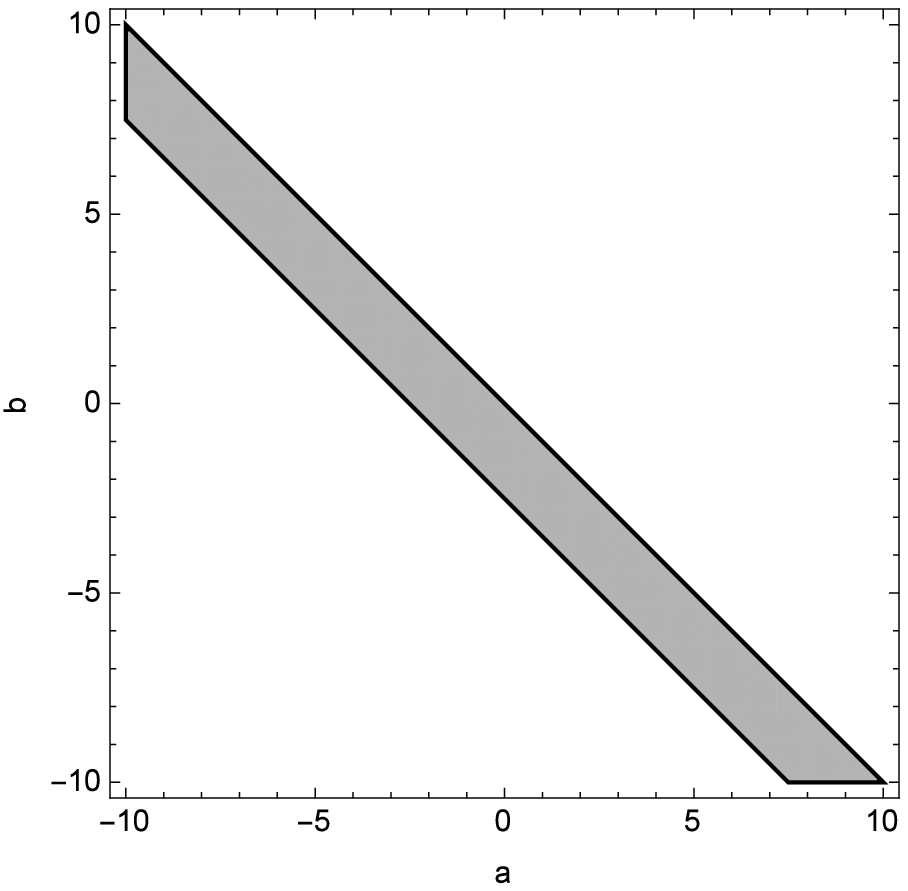}%
\caption{Stability analysis of exponential and TVDRK3 schemes applied
  to the scalar equation~\eqref{model}. The stability region (shaded
  area) within the square $[-10,\, 10] \times [-10,\, 10]$ on $(a,\,
  b)$ plane.  From left to right: PCEXP, EXP1, and {
    TVDRK3}.}
\label{fig:bdexp}
\end{figure}

The stability region of the PCEXP scheme is larger than that
of the EXP1 scheme, therefore the PCEXP scheme is not only more
accurate but also is better in terms of stability.  Note that the
stability regions of EXP1 and PCEXP schemes are infinitely large
fan-shaped areas without a lower bound, while the TVDRK3 scheme in
Fig.~\ref{fig:bdexp} (right) is a narrow strip. They imply that given a
negative $J$ and a fixed $\lambda$ inside the regions, both EXP1 and
PCEXP allow an infinity large $\Delta t$, while TVDRK3 does not.  This
stability feature distinguishes the exponential schemes from the TVDRK3
scheme.

We can also consider the stability of \eqref{grow} with a complex $b =
\Delta t \, J$ and a real $a = \Delta t \, \lambda$, as shown in
Fig.~\ref{fig:xfix}. The stability region of the PCEXP scheme
increases as $|\lambda|$ increases, while that of the TVDRK3 scheme
only shifts a distance along the real axis on the complex $b$-plane
but without changing its area. Thus, the stability region of the
TVDRK3 scheme does not expand under the constraint of $b$.
Clearly, the PCEXP scheme is far more superior than the TVDRK3 scheme
in terms of stability, as expected.
} %

\subsection{Local truncation error}

We study the local truncation error of the PCEXP scheme by using the model
scalar equation \eqref{model}. The Taylor expansion of the solution
\eqref{eqn:growthfactor} of the linearized equation \eqref{model},
$u_{n+1}$, at $t = t_n$ is
\begin{align}
  u_{\mbox{\scriptsize Taylor}} 
  = 
  &
  u_n 
  + \Delta t \, u_n^{(1)} 
  + \frac{1}{2} \Delta t^2 u_n^{(2)} 
  + \frac{1}{6} \Delta t^3 u_n^{(3)} 
  +  O(\Delta t^4)
  \nonumber
  \\
  = 
  &
  \left[
    1 + \Delta t (J+\lambda)  
    + \frac{1}{2} \Delta t^2 (J+\lambda)^2  
    + \frac{1}{6} \Delta t^3 (J+\lambda)^3 + \cdots 
    \right] 
  u_n
  ,
\label{taylor}
\end{align}
where $u_n^{(k)}$ is the $k$-th order derivative of $u(t)$ 
evaluated at $t=t_n$, and we have used the fact that for the
linearized equation of \eqref{model},
\begin{equation}  
  u_n^{(k)} 
  = 
  J \, u_n^{(k-1)} + \lambda \,u_n^{(k-1)}  
  = 
  (J + \lambda )^{k} \, u_n 
  .
\label{ut3}
\end{equation}  
By substitution of the Taylor expansion of the exponential term
$e^{\Delta t \zeta}$ in the solution \eqref{eqn:growthfactor}
obtained by the PCEXP scheme \eqref{eqn:PCEXP} yields
\begin{equation}
  u_{n+1} 
  = 
  \left[
    1 + \Delta t (J+\lambda)  
    + \frac{1}{2} \Delta t^2 (J+\lambda)^2  
    + \frac {1}{2} \Delta t^3 J \lambda (J+\lambda)  
    + \cdots 
    \right] u_n .
\label{eqn:u_n+1}
\end{equation}
The difference between the Taylor expansion \eqref{taylor}
and the approximated solution \eqref{eqn:u_n+1} for $u_{n+1}$ yields
the local truncation error:
\begin{equation}
  \frac {1}{6} \Delta t^3 
  ( J + \lambda )(J^2 - J \lambda + \lambda^2)  \,u_n.
\label{trun1}
\end{equation}
Similarly, the leading-order local truncation error of the EXP1 scheme
is
\begin{equation}
  \frac {1}{2} \Delta t^2 \lambda (J+\lambda) \,u_n.
\label{trun2}
\end{equation}
Obviously, the accuracy of the PCEXP scheme is one-order higher than the ETD1 scheme
in the frame of classical truncation error analysis.

\section{Spatial discretization}
\label{sec:DG-Euler}

In this section, we apply the PCEXP scheme to the Euler equations
discretized with the discontinuous Galerkin (DG) method in space.

\subsection{Governing equations }

Consider the Euler equations in a rotating frame of reference in $d$
dimensional space:
\begin{equation}
  \frac{\partial \mathbf{U}}{\partial t} 
  + \bm{\nabla} \cdot \tensor{F} 
  = 
  \mathbf{S},
\label{euler}
\end{equation}
where $\mathbf{U} \in \mathbb{R}^{d+2}$ stands for the vector of
conservative variables, $\tensor{F} \in \mathbb{R}^{(d+2) \times d}$
the convective flux, and $\mathbf{S} \in \mathbb{R}^{d+2}$ the source
term:
\begin{equation}
  \mathbf{U}
  = 
  \left( 
  \begin{array}{c} 
    \rho \\ 
    \rho \bm{v} \\
    \rho E
  \end{array} 
  \right) 
  ,
  \quad 
  \tensor{F}
  =
  \left(
  \begin{array}{c}
    \rho\, (\bm{v}-\bm{v}_r)^{\tensor{T}} 
    \\ 
    \rho\, (\bm{v}-\bm{v}_r)
    \bm{v}^{\tensor{T}} 
    +
    p \, \tensor{I}
    \\ 
    \rho\, H\, (\bm{v}-\bm{v}_r)^{\tensor{T}} 
  \end{array}
  \right) 
   ,
   \quad
  \mathbf{S}
  = 
  \left( 
  \begin{array}{c} 
    0 
    \\ 
    - \rho \, \bm{\omega} \times \bm{v} 
    \\
    0 
  \end{array} 
  \right) ,
\end{equation} 
where $\bm{v}: = (u,\, v,\, w)^{\tensor{T}}$ is the absolute velocity,
$\bm{\omega} := (\omega_x,\, \omega_y,\, \omega_z)^{\tensor{T}}$ is
the angular velocity of the rotating frame of reference, $\bm{v}_r :=
\bm{\omega} \times \bm{x}$; $\rho$, $p$, and $e$ denote the flow
density, pressure, and the specific internal energy; $E = e +
\frac{1}{2}||\bm{v}||^2$ and $H = E + p/\rho$ denote the total energy
and total enthalpy, respectively; $\tensor{I}$ denotes the $d \times
d$ unit matrix;
and the pressure $p$ is given by the equation of state for a perfect
gas:
\begin{equation}
  p=\rho \left(\gamma-1\right) e,
\end{equation} 
where $\gamma=7/5$ is the ratio of specific heats for perfect
gas.

\subsection{Discontinuous Galerkin discretization}
\label{sec:DG}

The computational domain $\Omega$ is divided into a set of
non-overlapping elements of arbitrary shape. 
We seek an approximation $\mathbf{U}_h$ in each element $E \in \Omega$
with finite dimensional space of polynomial $P^{ p}$ of
order { $p$} in the discontinuous finite element space
\begin{equation}
  \mathbb{V}_h 
  :=
  \{ \psi_i \in L^{2}(\Omega): 
  \psi_i |_E \in {P}^{{ p}}(\Omega),\, 
  \forall \, E \in \Omega \}
  .
\end{equation}
The numerical solution of $\mathbf{U}_h$ can be approximated in the
finite element space $\mathbb{V}_h$
\begin{equation}
  \mathbf{U}_h(\bm{x},\, t )
  =  
  \sum_{j = 1}^n 
  \mathbf{u}_j(t) \psi_j( \bm{x} )
  .
\label{uexp}
\end{equation}
In the weak formulation, the Euler equations \eqref{euler} in an
element $E$ becomes:
\begin{equation}
  \int_{E} 
  \psi_i \psi_j \diff \bm{x}
  \frac{\diff \mathbf{u}_j}{\diff t}  
  = 
  - \int_{\partial E} \psi_i \widetilde{\tensor{F}} 
  \cdot \hat{\bm{n}}
  \, \diff \bm\sigma
  +
  \int_{E} 
  ( \tensor{F} \cdot \bm{\nabla} \psi_i  + \psi_i  \mathbf{S} ) 
  \diff \bm{x}
  :=
  \mathbf{R}_i
  ,
\label{euler3}
\end{equation}
where $\hat{\bm{n}}$ is the out-normal unit vector of the surface
element $\bm \sigma$ with respect to the element $E$,
$\widetilde{\tensor{F}}$ is the Riemann flux \cite{Toro1999},
which will be approximated by Roe's flux \cite{Roe1981jcp},
and the Einstein summation convention is used.
For an orthonormal basis $\{ \psi_i \}$, the term on the left-hand
side of Eq.~\eqref{euler3} becomes diagonal, so the system is in the
standard ODE form of Eq.~\eqref{eqn:starteq}, thus avoiding solving a
linear system as required for a non-orthogonal basis.
More importantly, the use of orthogonal basis would yield more
accurate solutions, especially for high-order methods with $p \gg 2$.

\subsection{Orthogonal basis in the Cartesian coordinates}

In this paper, the basis function $\psi _i(\bm{x})$ is defined on the
global Cartesian coordinate $\bm{x}:=(x,\, y,\, z)$ rather than on the
cell-wise, local reference coordinates.  The variable values on the
{ Gaussian quadrature} points for computing the surface
fluxes can be easily accessed without the Jacobian mapping between the
local reference coordinates to the global Cartesian ones
\cite{DengS2006anacm, Bergot2013nmpde}, and it also makes the
discontinuous Galerkin method feasible on arbitrary polyhedral grids
\cite{Botti2012siamjsc}.

A simple choice of the basis function in \eqref{uexp} may be the
monomials \cite{Wolkov2011} or Taylor basis \cite{LuoH2012aiaa0461}.
However, in the case of distorted meshes, the non-orthogonality of
these basis functions may yield an ill-conditioned mass matrix,
resulting in degradation of accuracy and even loss of numerical
stability.  In this work, to construct an orthonormal basis set $\{
\psi_i(\bm{x}) \}$, we start with the normalized monomials $\{
\chi_i(\bm{x}) \}$:
\begin{subequations}
\begin{align}
  {\{\chi_i(\bm{x})\}}
  &
  := 
  \left\{
  \left.
  \frac{ ( x - x_c )^{p_1} ( y - y_c)^{p_2} ( z - z_c)^{p_3}}
       {{L_x}^{p_1} {L_y}^{p_2} {L_z}^{p_3}}
        \right| 0 \le p_1,\, p_2,\, p_3 \, ;
       \ p_1 + p_2 + p_3 \le i-1 
  \right\}
  ,
\label{basis2}
\\
  {\{\psi_i(\bm{x}) \}}
  &
  :=  
  \left\{
  s_{i} 
  \left[
  \left.
    \chi_i(\bm{x}) 
    +
    \sum_{j = 1}^{i - 1} 
    c_{ij} \chi_j(\bm{x}) 
    \right]  \right|
  1 \le i \le N \right\},
\label{basis}
\end{align}
\end{subequations}
where 
\begin{align*}
  &
  x_c 
  := 
  \frac{1}{| E |} \int_{E} x \, \diff \bm{x} 
  , 
  &&
  y_c 
  := 
  \frac{1}{| E |} \int_E y \, \diff \bm{x} 
  , 
  &&
  z_c 
  := 
  \frac{1}{| E |} \int_E z \, \diff \bm{x} 
  , 
  \\
  &
  L_x 
  :=
  \frac{1}{2}\left(x_{\max} - x_{\min}\right)
  ,
  &&
  L_y
  :=
  \frac{1}{2}\left(y_{\max} - y_{\min}\right)
  ,
  &&
  L_z
  :=
  \frac{1}{2}\left(z_{\max} - z_{\min}\right)
  ,
\end{align*}
and the total number of basis functions $N = (p + 1)(p + 2)(p + 3)/6$
for the $p$-th order DG approximation in 3D space.  With the following
definition of the inner product on an element $E$:
$
\left\langle f,\, g \right\rangle_E 
:= \int_E f \left(\bm{x}\right) g\left(\bm{x}\right) \, \diff \bm{x}
,
$
the coefficients $\{ s_{i} \}$ and $\{c_{ij}\}$ can be computed with
the modified Gram-Schmidt (MGS) orthogonalization described in the
Algorithm~\ref{alg:alg2}.

\begin{algorithm}
\caption{Basis orthonormalization}               
\label{alg:alg2}                         
\begin{algorithmic}
\FOR {$i = 1$ to $\rm{N}$}          
    \FOR {$j = 1$ to $i-1$}          
    \STATE$w_{ij} = \left \langle {\psi}_i,\chi_j \right \rangle$\\
      \STATE ${\psi}_i={\psi}_i-w_{ij} \chi_j$\\
      \STATE $w_{ii} = \sqrt{\left \langle {\psi}_i,{\psi}_i \right \rangle}$\\
      \STATE ${\psi}_i={\psi}_i/w_{ii}$
    \ENDFOR   
    \ENDFOR
\end{algorithmic}
\end{algorithm}

\subsection{Exact Jacobian matrix for the exponential schemes}

The convergence rate and stability of the PCEXP scheme rely on the
accuracy to approximate the Jacobian matrix $\tensor{J}$, which is
directly determined by the local truncation error of \eqref{trun1}.
In the PCEXP scheme, the broadcasting of global information is
achieved through the exact Jacobian matrix which accurately includes
the information of both the interior and the boundary of the
elements. The details of computing the exact Jacobian is discussed
next.

The diagonal Jacobian can be obtained by taking the derivative of
\eqref{euler3} with respect to the $\mathbf{u}_j$ of the host cell
with the label ``L'':
\begin{align}
  \frac{\partial \mathbf{R}_i}{\partial \mathbf{u}_j^{\rm L} } 
  & 
  = 
  - \int_{\partial E} \psi_i 
  \frac{\partial \widetilde{\tensor{F}} (\mathbf{U}_{\rm L},\,
    \mathbf{U}_{\rm R} )} 
       {\partial \mathbf{U}_{\rm L}} 
  \frac{\partial \mathbf{U}_{\rm L}} 
       {\partial \mathbf{u}_j^{\rm L}}   
  \, \diff \bm{\sigma} 
  + 
  \int_{E} 
  \left( 
  \bm{\nabla} \psi_i   
  \frac{\partial \tensor{F}}{\partial \mathbf{U}} 
  \frac{\partial \mathbf{U}}{\partial \mathbf{u}_j} 
  + 
  \psi_i  
  \frac{\partial \mathbf{S}}{\partial \mathbf{U}} 
  \frac{\partial \mathbf{U}}{\partial \mathbf{u}_j}
  \right) 
  \diff \bm{x} 
  \nonumber
  \\
  & 
  = 
  - \int_{\partial E} 
  \psi_i^{\rm L} \, 
  \psi_j^{\rm L} \, 
  \frac{\partial \widetilde{\tensor{F}}(\mathbf{U}_{\rm L},\,
    \mathbf{U}_{\rm R}) }
       {\partial \mathbf{U}_{\rm L}}  
  \, \diff \bm{\sigma} 
  + 
  \int_E 
  \left( 
  \psi_j \, \bm{\nabla} \psi_i \, 
  \frac{\partial \tensor{F}}{\partial \mathbf U}  
  + \psi_i \, \psi_j\, 
  \frac{\partial \mathbf{S}}{\partial \mathbf{U}} \,
  \right) 
  \diff \bm{x} 
  . 
\label{jac1}
\end{align}
Similarly, the off-diagonal Jacobian can be obtained by taking the
derivative of \eqref{euler3} with respect to the $\mathbf{u}_j$ of the
neighboring cells around the host cell ``L'', which is marked with the
label ``R'':
\begin{equation}
  \frac{\partial \mathbf{R}_i}{\partial \mathbf{u}_j^{\rm R}}
  = 
  - \int_{\partial E} 
  \psi_i \, 
  \frac{\partial \widetilde{\tensor{F}}(\mathbf{U}_{\rm L},\, 
    \mathbf{U}_{\rm R})}
       {\partial \mathbf{U}_{\rm R}} 
  \frac{\partial \mathbf{U}_{\rm R}} 
       {\partial \mathbf{u}_j^{\rm R}}  
  \, \diff \bm{\sigma}
  = 
  - \int_{\partial E} 
  \psi_i^{\rm L} \, \psi_j^{\rm R} \, 
  \frac{\partial \widetilde{\tensor{F}}(\mathbf{U}_{\rm L},\, 
    \mathbf{U}_{\rm R})} 
       {\partial \mathbf{U}_{\rm R}}
  \, \diff \bm{\sigma}
  .
\label{jac2}
\end{equation}
The Riemann flux Jacobian matrices 
$\partial \widetilde{\tensor{F}}(\mathbf{U}_{\rm L},\, \mathbf{U}_{\rm
  R}) / \partial \mathbf{U}_{\rm L}$, $\partial \widetilde{\tensor{F}}
(\mathbf{U}_{\rm L},\, \mathbf{U}_{\rm R}) / \partial \mathbf{U}_{\rm
  R}$ in \eqref{jac1} and \eqref{jac2} are evaluated exactly through
 the automatic differentiation (AD), others can be derived
easily.

The global Jacobian matrix $\tensor{J}$ is made of the diagonal and
off-diagonal matrices above.  When ${\bm{\sigma}}$ is an interior
face, the flux $\widetilde{\tensor{F}}(\mathbf{U}_{\rm L},\,
\mathbf{U}_{\rm R})$ is calculated with Roe's Riemann solver
\cite{Roe1981jcp}.  When $\bm{\sigma}$ is a boundary face with a
appropriate boundary condition, one has
\begin{equation}
  \widetilde{\tensor{F}}_{\mbox{\scriptsize bc}} 
  = 
  \widetilde{\tensor{F}}(\mathbf{U}_{\rm L},\, \mathbf{U}_{\rm ghost}),
\end{equation}
where $\mathbf{U}_{\rm{ghost}}$ is a function of $\mathbf{U}_{\rm L}$
corresponding the boundary condition, and $\widetilde{\tensor{F}}$ is also
consistently computed by the same Roe's Riemann solver used on the
interior faces.  Then,  the boundary flux Jacobian matrix can be expressed as
\begin{equation}
  \frac{\partial \widetilde{\tensor{F}}_{\mbox{\scriptsize bc}} }
       {\partial \mathbf{U}_{\rm L}} 
  =
  \frac{\partial \widetilde{\tensor{F}}}{\partial \mathbf{U}_{\rm L}}
  +
  \frac{\partial \widetilde{\tensor{F}}}{\partial \mathbf{U}_{\rm
      ghost}} 
  \frac{\partial \mathbf{U}_{\rm ghost}}{\partial \mathbf{U}_{\rm L}}
    .
\end{equation}
The Jacobian matrix $\bm{\nabla} \psi_j \, {\partial \tensor{F}} /
{\partial \mathbf{U}}$ in the volume integration and the source-term
Jacobian matrix ${\partial \mathbf{S}} / {\partial \mathbf {U}}$ are
given by \eqref{dfdu} and \eqref{dsdu}, respectively, in the Appendix.
The Jacobians $\partial \widetilde{\tensor{F}} / \partial
\mathbf{U}_{\rm L}$, $\partial \widetilde{\tensor{F}} / \partial
\mathbf{U}_{\rm ghost}$ and $\partial \mathbf{U}_{\rm ghost} / \partial
\mathbf{U}_{\rm L}$ are obtained exactly by the automatic
differentiation (AD).

\section{Numerical Results} 
\label{sec:results}

{ The PCEXP scheme is tested for the time marching of the
  Euler equations for both steady and unsteady flows.  
Its accuracy and efficiency are investigated in the unsteady flow case
and compared with two widely-used explicit and implicit schemes: the
third-order TVD Runge-Kutta (TVDRK3) scheme and the second-order
backward difference formula (BDF2).  In the steady flow case, the
performance of the PCEXP scheme is also investigated by comparing with the
implicit backward Euler (BE) and the implicit BDF2 schemes given
below:
\begin{align}     
  &\mbox{BE:}   
  &\mathbf{u}_{n+1}-\mathbf{u}_n
  & = {\Delta t_n}\,  \mathbf{R}(\mathbf{u}_{n+1}), &\label{be} \\
  &\mbox{BDF2:}
  & \frac{1+2 r_n}{1+r_n}  {\mathbf{u}_{n+1}}
  -(1+r_n) {\mathbf{u}_{n}}
  +\frac{r_n^2}{1+r_n} {\mathbf{u}_{n-1}}
  & =
  {\Delta t_n} \, \mathbf{R}(\mathbf{u}_{n+1}), &
  \label{bdf2}
\end{align}
where $r_n:=\Delta t_n/\Delta t_{n-1}$ so that a variable time-step
size is allowed in the BDF2 context \cite{VBDF2}.  The resulting linear systems are
solved by an ILU preconditioned GMRES method.  In both the exponential
and the implicit methods, the dimension of the Krylov basis
$m=30$. The convergence tolerance of the Krylov subspace is set to
$1.0 \times 10^{-5}$.

} 

For all the schemes used in this work, the time-step size $\Delta t$
is determined by
\begin{equation}
 \Delta t 
 =   
 \frac{\mbox{CFL} \, h_{\rm 3D}}
      {\left(2 p+ 1\right) \left(\|\bm{v}\|+ c \right)} ,
 \quad
 h_{\rm 3D}
 :=  
 2 d 
 \frac{| E |}{| \partial E |},
\label{x3d}
\end{equation}
where $\mbox{CFL}$ is the { global Courant-Friedrichs-Lewy
  (CFL) number},
$p$ the order of polynomial in DG,
$\bm{v}$ the velocity vector at the cell center, 
$c$ the speed of sound,
$d$ the spatial dimension,
$| E |$ and $| \partial E |$ are the volume and the surface area
of the boundary of $E$, respectively;
and $h_{\rm 3D}$ represents a characteristic size of a cell in 3D
defined by the ratio of its volume and surface area.
{
The CFL number is a constant for unsteady flows and a variable for
steady flows (cf. Eq.~\eqref{cfl} and related discussion later).}

For the quasi-2D problems, we extrude a 2D mesh to a 3D (quasi-2D)
mesh by one layer of grids and use $h_{\rm 2D}$ instead of $h_{\rm
  3d}$ to eliminate the effect of the $z$ dimension on obtaining the
truly 2D time step.  Given the cell size $\Delta z$ in the $z$
direction, ${h}_{\rm 2D}$ is determined by
\begin{equation}
  \frac{2}{h_{\rm 2D}} 
  = 
  \frac{3}{{h}_{\rm 3D}} - \frac{1}{\Delta z}.
\label{x2d}
\end{equation}

\subsection{Temporal accuracy test for an unsteady problem}

The vortex transportation by a uniform flow of velocity $(U_\infty,\, 0)$
\cite{WangJZ2013ijnmf} is computed to test the temporal accuracy
of PCEXP. The initial conditions of the flow are
\begin{equation}
\begin{aligned}
  U_0 
  &
  = 
  U_{\infty} - \beta U_{\infty} \frac{y-y_c}{R} \, 
  \exp \left({-\frac {r^2}{2}} \right) 
  ,
  \\
  V_0 
  &
  =  
  \beta U_{\infty} \frac{x-x_c}{R} \, 
  \exp \left( -\frac {r^2}{2} \right ) 
  ,
  \\
  T_0 
  &
  =  
  T_{\infty} 
  - \frac{\beta U_{\infty}^2}{2 C_p} \, 
  \exp \left( -\frac {r^2}{2} \right) 
  ,
\end{aligned}
\end{equation}
where $r := \sqrt{(x-x_c)^2 + (y-y_c)^2}/R$, $R= 0.05$, and 
$(x_c,\, y_c) = (0.05,\, 0.05)$ is the initial position of the vortex
center. The Mach number is set to 0.5, $\beta=0.2$, 
$T_{\infty} = 300\, ^\circ$K, $P_\infty = 10^5$\,N/m$^2$, $C_p = R_{\mbox{\scriptsize
    gas}}\gamma / (\gamma-1)$, $R_{\mbox{\scriptsize gas}}$ is the
ideal gas constant and $\gamma=7/5$.  
The reminding variables, $\rho$ and $e$, are determined by the
equation of state for perfect gas.  
Periodic boundary conditions are used in all dimensions.  
On a finite domain with periodic boundary conditions, the motion of
the vortex is periodic with the period $T=L_x/U_{\infty}$, where $L_x$
is the domain size in $x$ direction.
A uniform mesh of size $[N_x,\, N_y] = [24, \, 24]$ is used on a
computational domain of size $[0, L_x] \times [0,L_y]$ and
$L_x=L_y=0.1$.

Evaluation of the temporal order of accuracy requires the time-exact
solution, which may be approximated by a solution obtained with a
sufficiently small time-step size (\ie, we use $\mbox{CFL}=0.1$ ) so that the
temporal error is negligible.
Specifically, the time step size $\Delta t$ is decreased until the
following entropy error becomes a constant
\begin{equation}
  E_{L_2 \left(\Omega \right)} \left(s\right)
  :=  
  \sqrt{ \frac{1}{|\Omega|}
    {\displaystyle \int_\Omega \left( \frac{s}{ s_0} \right)^2 
      \diff \bm{x} - 1 }}
      \label{entropy_error}
\end{equation}
where $s = p / \rho^\gamma$ and $s_0$ is the entropy of the free
stream. 

{
We use $\mbox{CFL}=0.1 \times 2^n$ with $0 \leq n \leq 5$, to measure
the order of convergence with respect to the time-exact solution.  The
order of convergence for the TVDRK3, BDF2, and PCEXP schemes are all
shown in Fig.~\ref{torder1} with the order of polynomials $p=0$ to
$p=3$.  The formal orders of accuracy are verified for all the cases,
which validate our implementation.  It is also apparent that the
temporal error of BDF2 is much larger than that of PCEXP with an equal
time-step size.  Overall, the error of PCEXP is one order of magnitude
smaller than that of BDF2, although both schemes are second-order
accurate.

\begin{figure}[htbp!]
\centering
\subfigure[Convergence order]{
\label{torder1}
\begin{minipage}[b]{0.4\textwidth}
\includegraphics[width=1\textwidth]{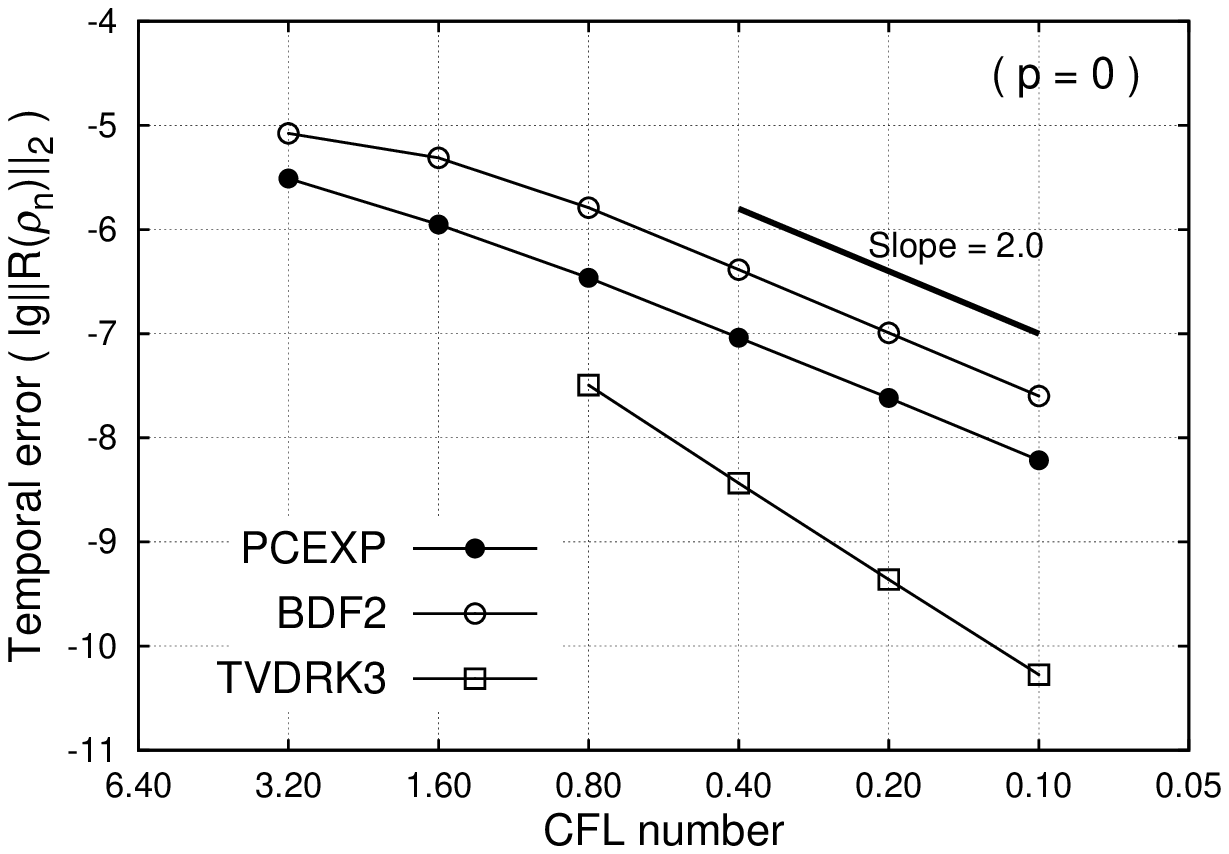}\\
\includegraphics[width=1\textwidth]{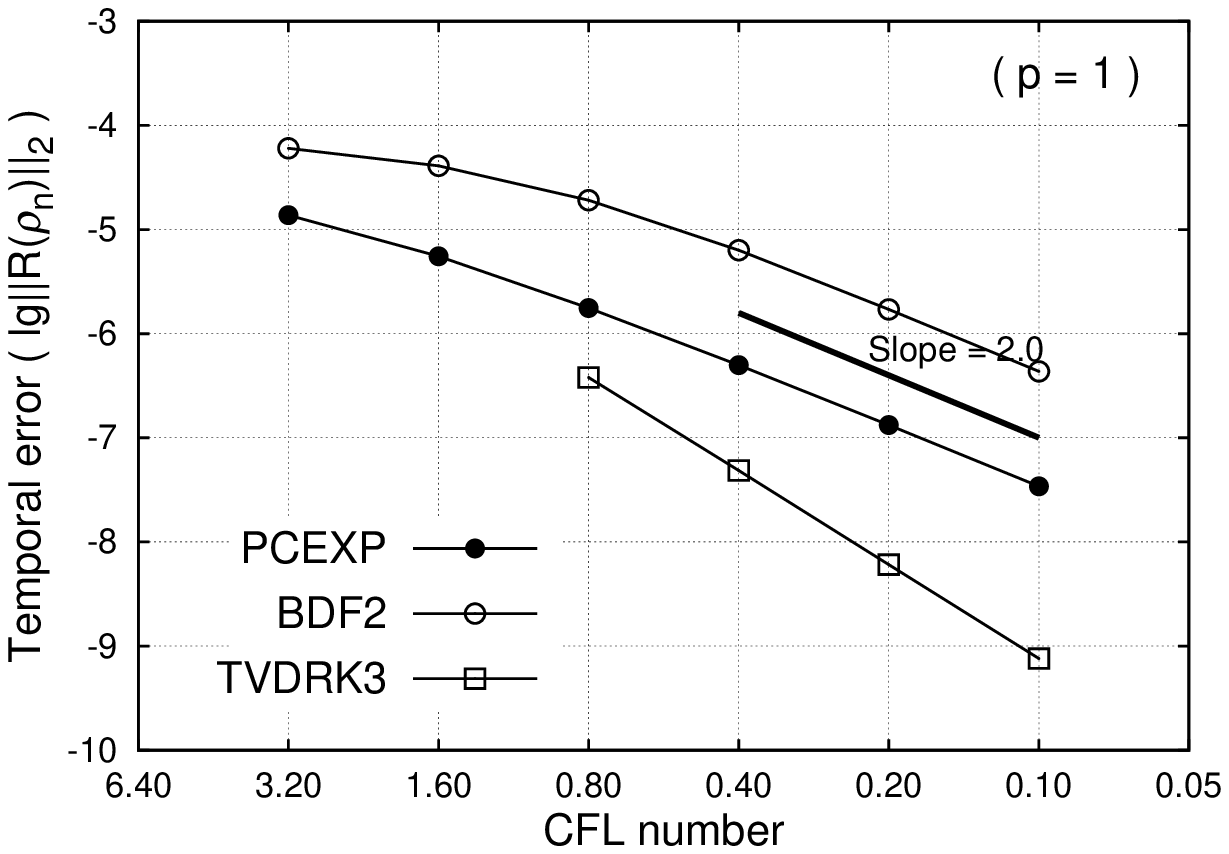}\\
\includegraphics[width=1\textwidth]{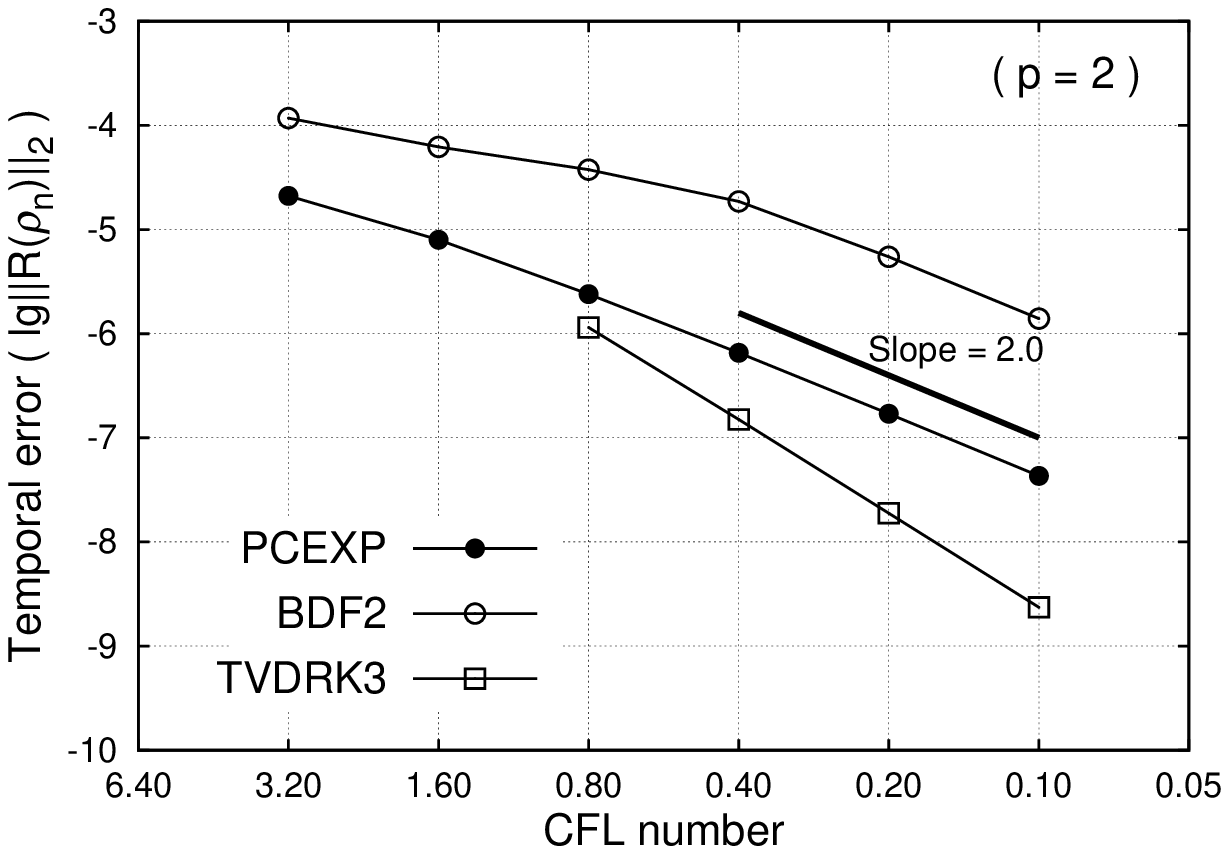}\\
\includegraphics[width=1\textwidth]{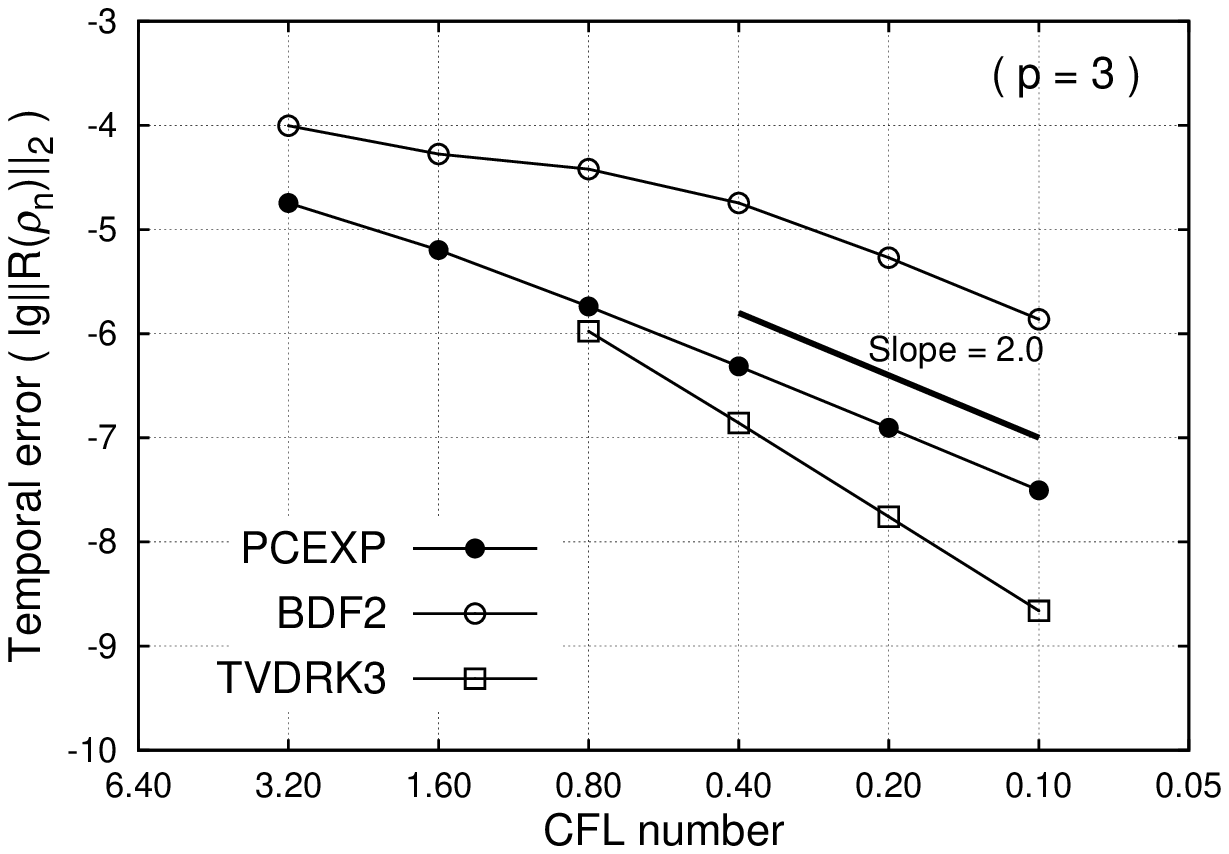}%
\end{minipage}
}
\subfigure[CPU time]{
\label{torder2}
\begin{minipage}[b]{0.4\textwidth}
\includegraphics[width=1\textwidth]{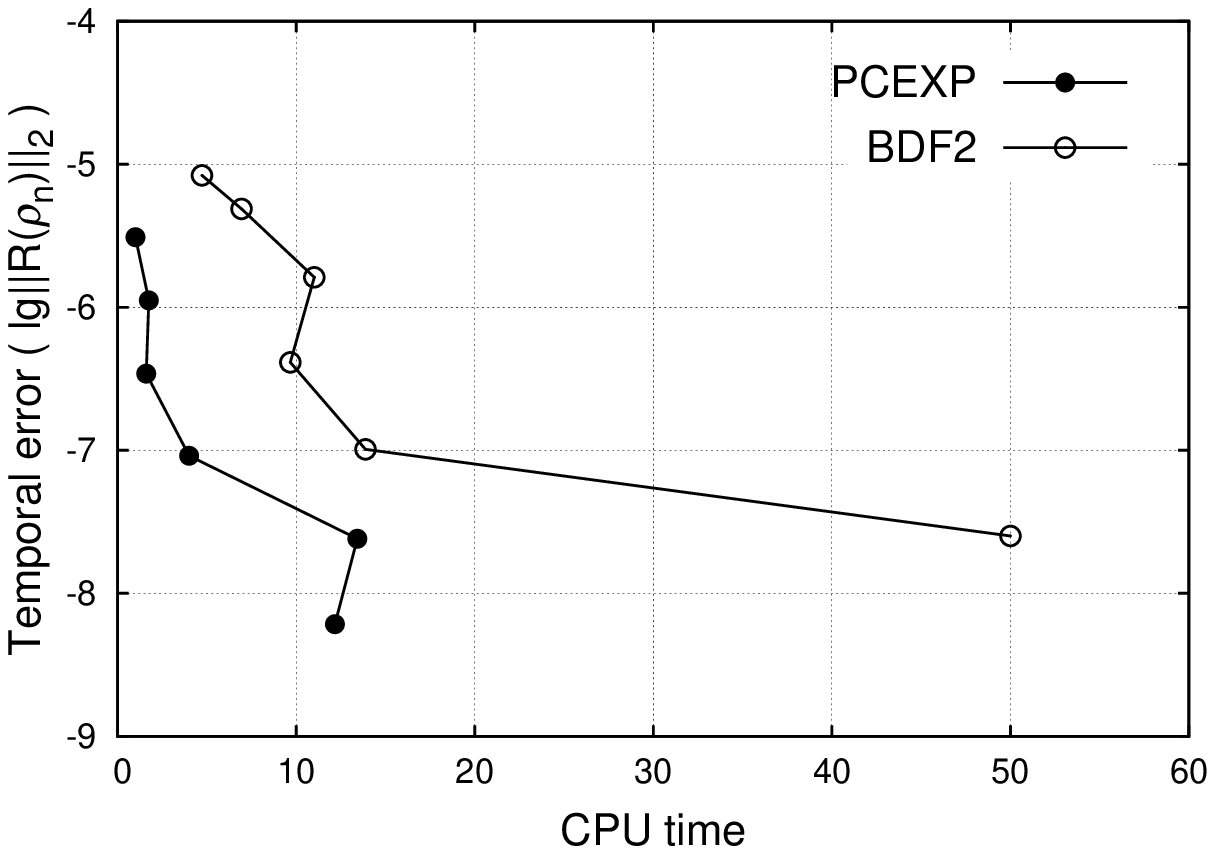}\\
\includegraphics[width=1\textwidth]{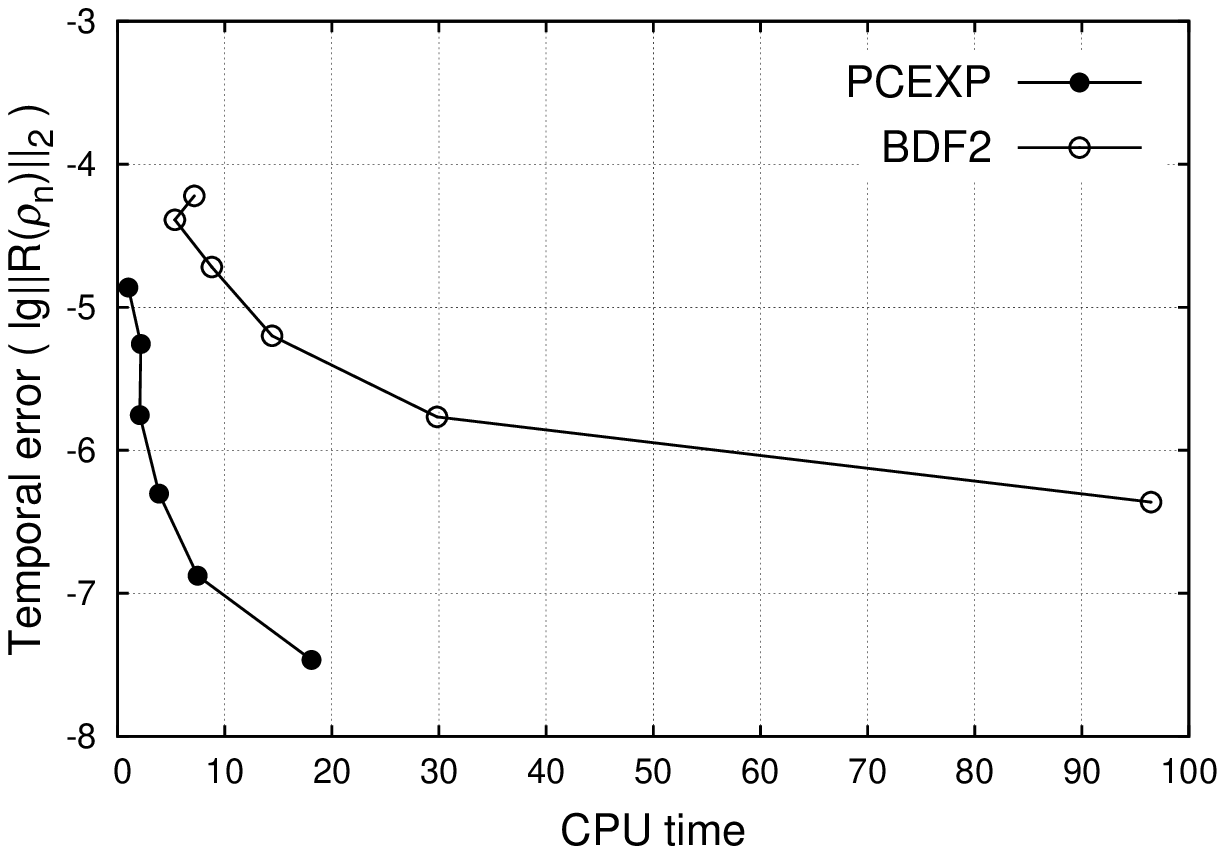}\\
\includegraphics[width=1\textwidth]{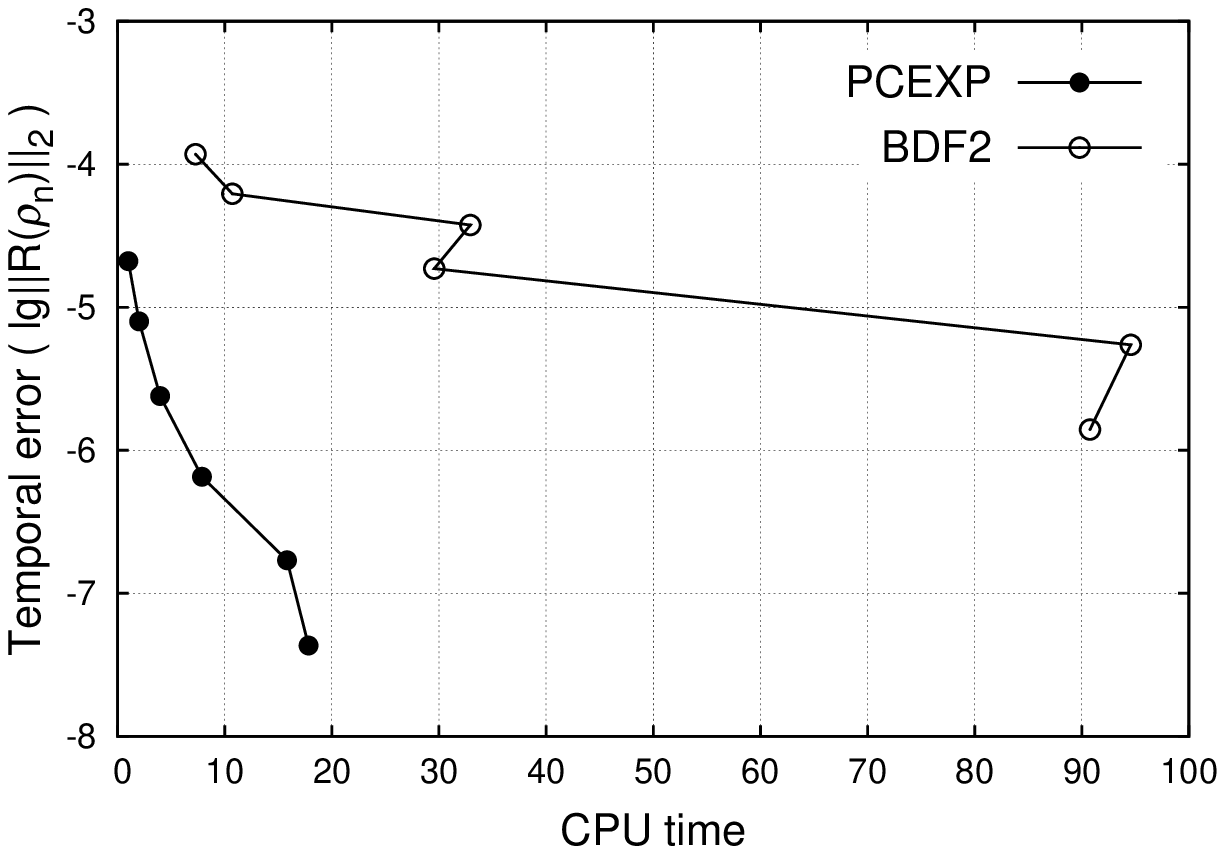}\\
\includegraphics[width=1\textwidth]{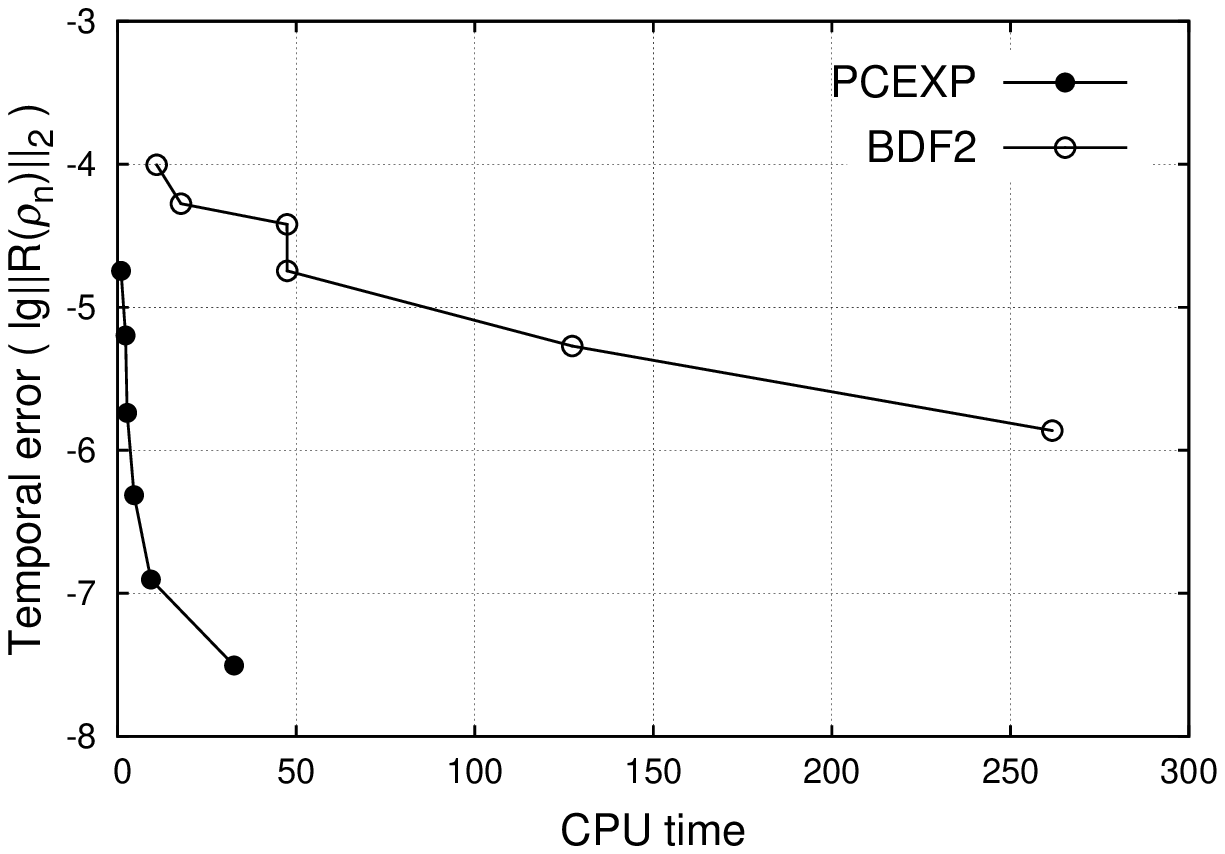}%
\end{minipage}
}
\caption{ Vortex transportation on a $24 \times 24$ uniform mesh
  with $\mbox{CFL} = 0.1 \times 2^n$, $0 \leq n \leq 5$.
Left: Temporal convergence of PCEXP, BDF2, and TVDRK3 schemes.  Right:
Evolution of the errors versus CPU time for PCEXP and BDF2
schemes. From top to bottom: $0 \le p \le 3$.}
\label{fig:torder}
\end{figure}

Besides the accuracy, the computational efficiency is also
investigated by showing the evolution of the error with respect to CPU
time.  We focus on two cases, one with uniform grids, and the other
with highly stretched grids.

With uniform meshes, small time-steps are used to test the order of
accuracy.  Obviously, when the time step size is sufficiently small,
both the implicit and exponential schemes are not as efficient as the
explicit TVDRK3 scheme because of its simplicity. Therefore, we only
compare the CPU times of the exponential and implicit schemes in
Fig.~\ref{torder2}.  The results show that the temporal error of the
PCEXP scheme not only is one order of magnitude smaller than that of
the BDF2 scheme, as shown in Fig.~\ref{torder1}, but also decays much
faster than that of the BDF2 scheme, as shown in
Fig.~\ref{torder2}. This shows that the PCEXP scheme is more accurate
and efficient than the BDF2 scheme.

For a stiff case, the following highly stretched non-uniform mesh is
used:
\begin{equation}
  x_j 
  = 
  \frac{1}{2} \left(1-\xi_j^3 \right) x_{\rm L} 
  + 
  \frac{1}{2} \left(1+\xi_j^3 \right) x_{\rm R}
  , 
  \quad 
  \xi_j = \frac{2}{N} \left(j-1\right)-1,
  \quad
  1 \leq j \leq N,
\end{equation}
where $x_{\rm L} = 0$, $x_{\rm R} = 0.1$ and $N = N_x = N_y = 24$.
The mesh in $y$ direction is also clustered in the same manner.
The grids are concentrated about the cube center, with a minimal grid
size of $3 \times 10^{-5}$ and a maximal one of 0.0115 for inducing
the mesh stiffness.

The solution computed by using the third-order TVDRK3 scheme with
$\mbox{CFL}=1.2$ is used as the reference solution.  The results
computed by using both second-order schemes BDF2 and PCEXP with
$\mbox{CFL}=1000$ are compared with each other since both permit large
time steps and should be more efficient in this case.  First, the
accuracy of the solutions obtained by using both schemes at the end of
one period are shown in Fig.~\ref{fig:cfl1000_accuracy}.  The result
of BDF2 exhibits a visible phase delay caused by a large temporal
error, which is evident in Fig.~\ref{torder1}, while the result of
PCEXP shows very little phase error and agrees well with the result of
TVDRK3. This validates the fact that the PCEXP scheme generates a
temporal error much smaller than what the BDF2 scheme does.

\begin{figure}[htb!]
\centering 
\includegraphics[width=0.3\textwidth]{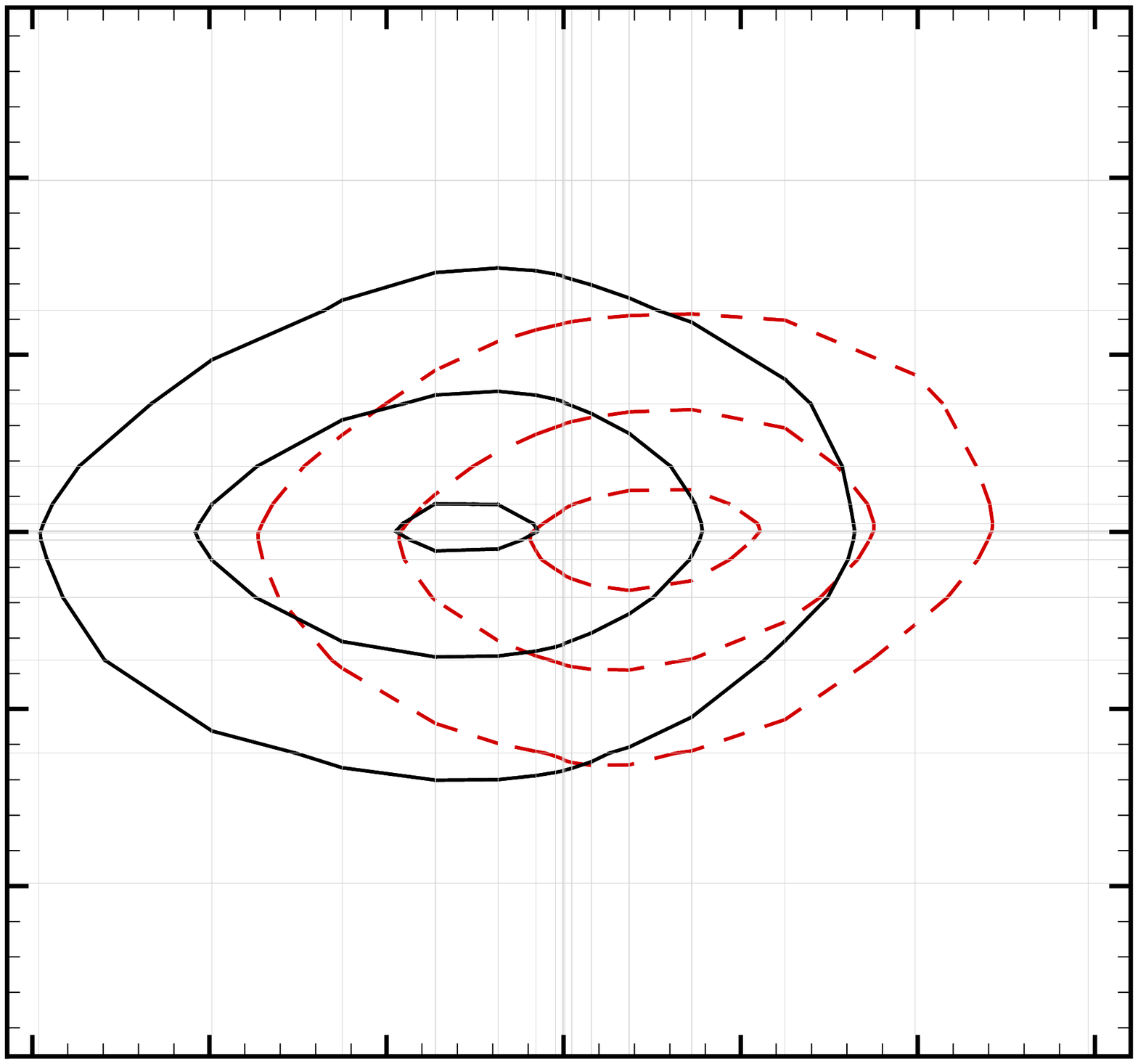}%
\includegraphics[width=0.3\textwidth]{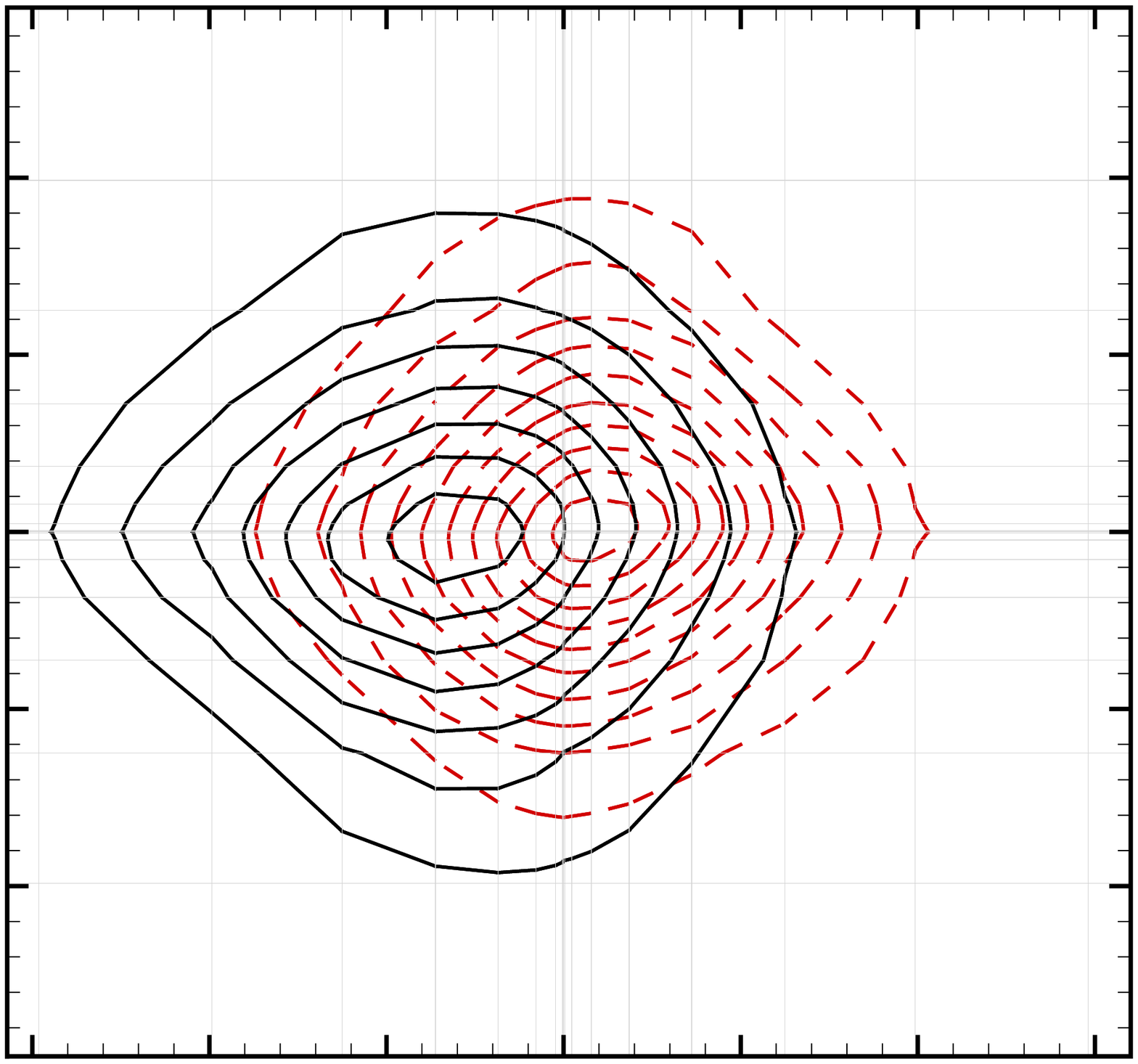}%
\includegraphics[width=0.3\textwidth]{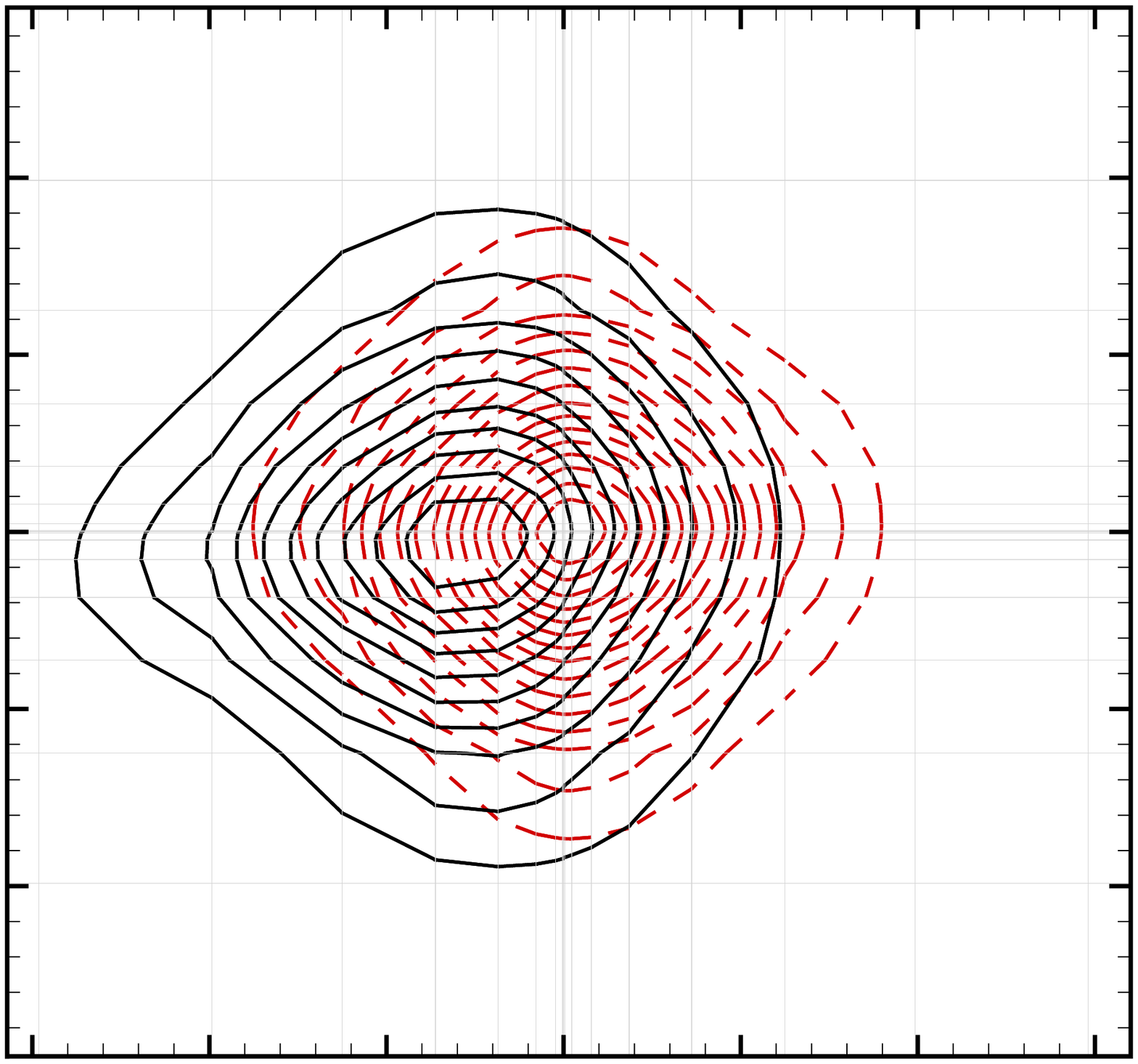}\\
\includegraphics[width=0.3\textwidth]{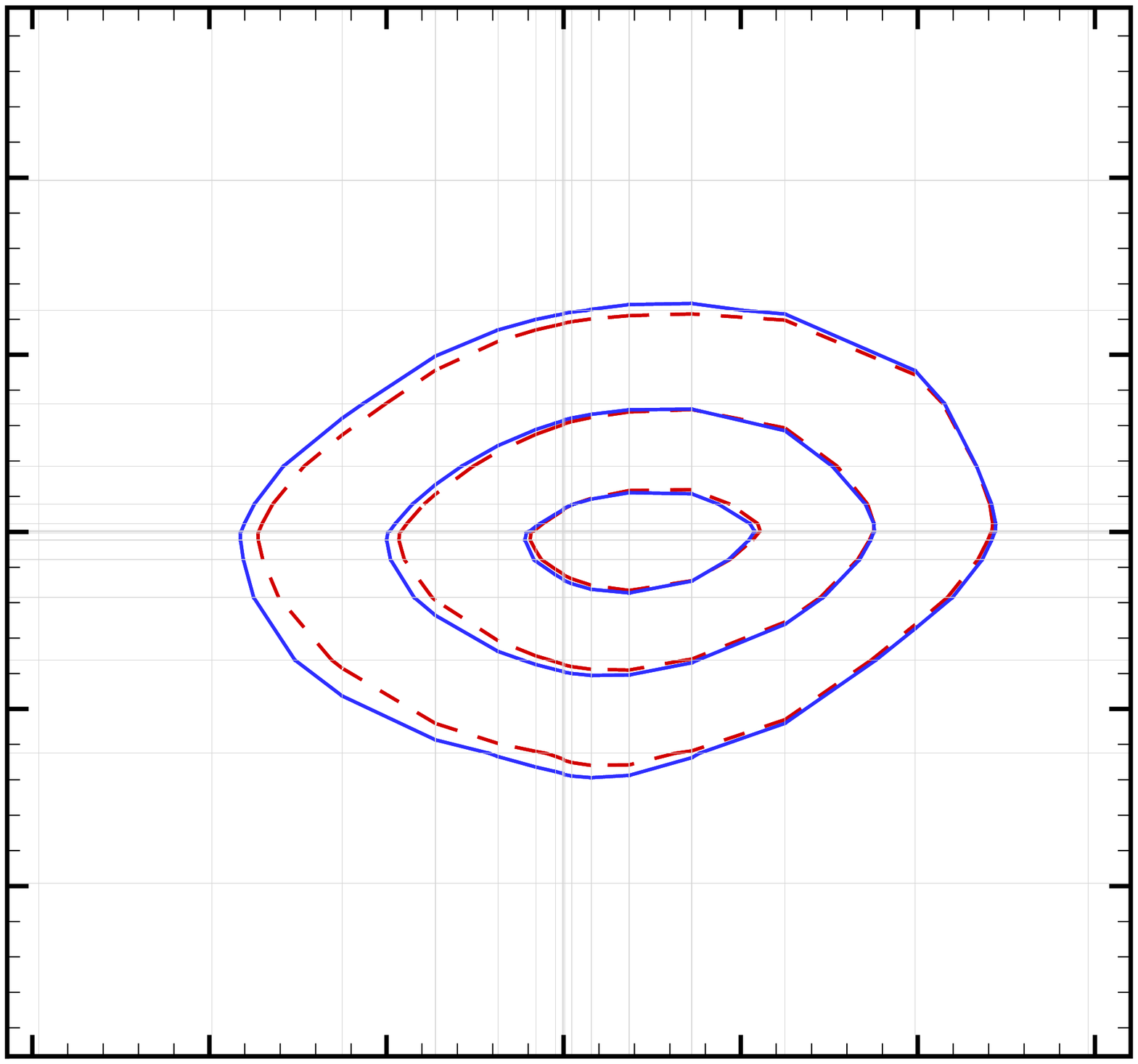}%
\includegraphics[width=0.3\textwidth]{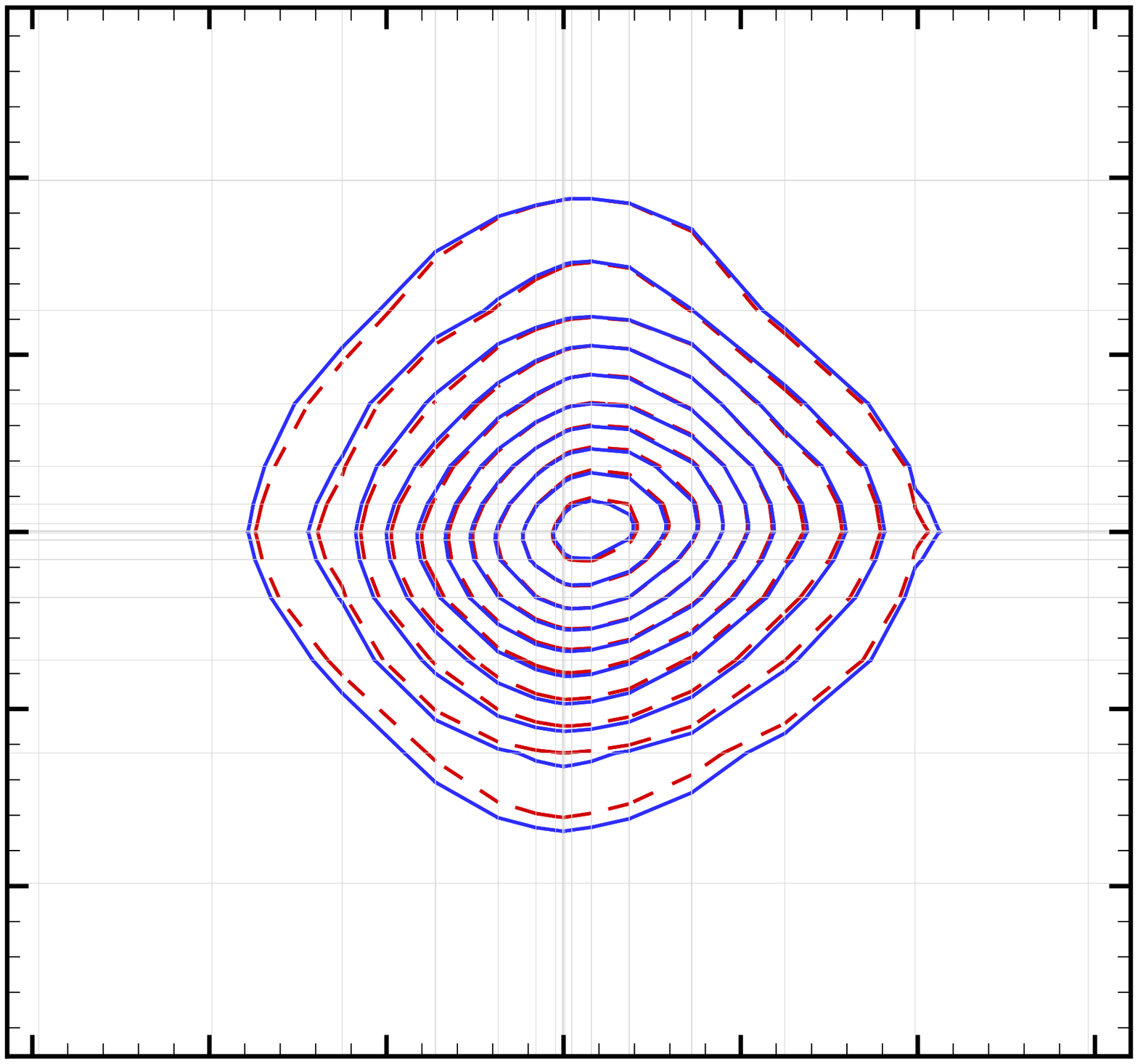}%
\includegraphics[width=0.3\textwidth]{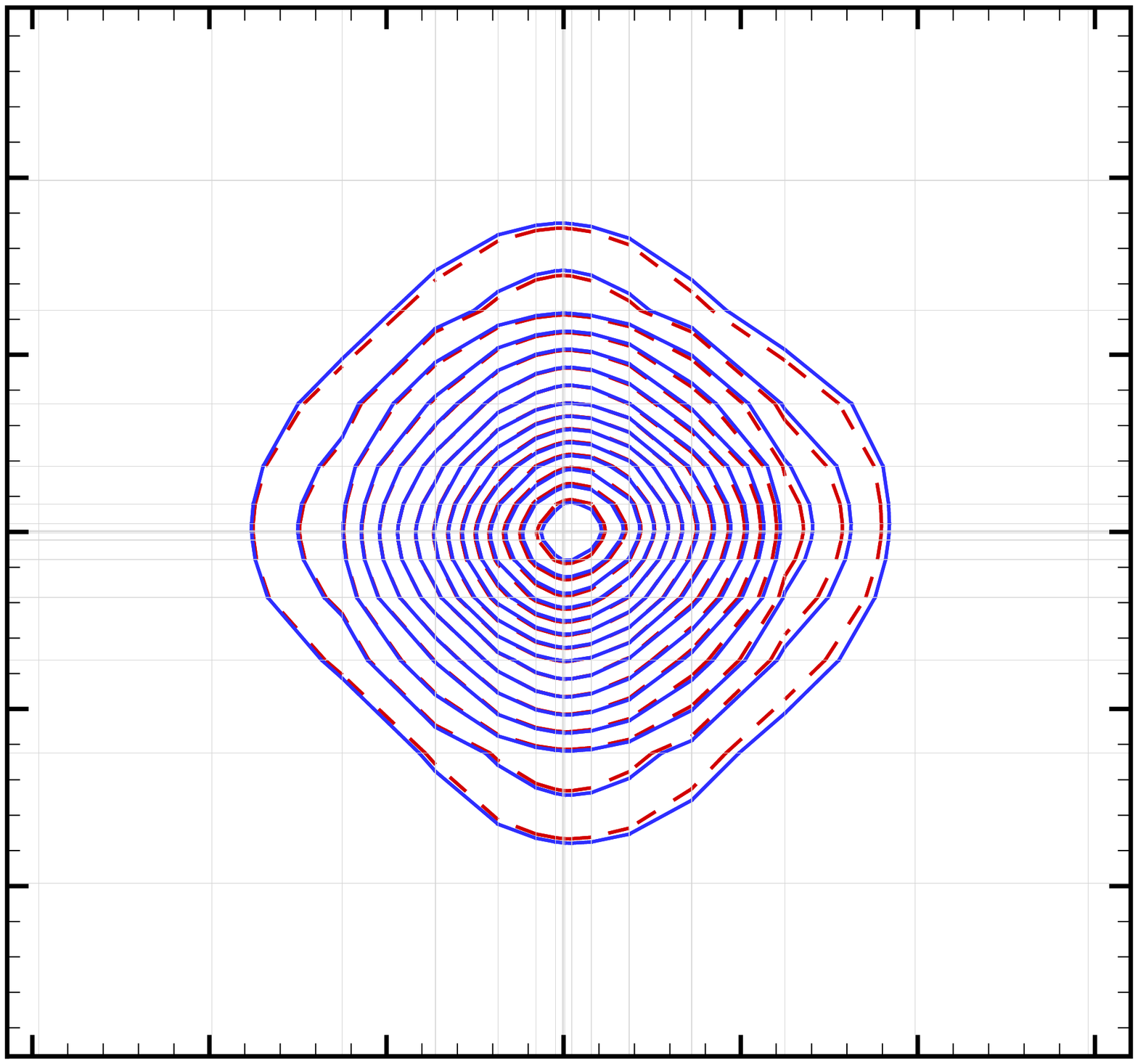}%
\caption{ Vortex transport on a $24 \times 24$ stretched
  mesh. Density Contours of 15 iso-lines in the range of $[0.99691,\,
    0.99974 ]$.  Top: BDF2 (black solid lines) \textit{versus} TVDRK3
  (red dashed lines). Bottom: PCEXP (blue solid lines) \textit{versus}
  TVDRK3 (red dashed lines). From the left to the right: $p=1$, 2, and
  3.}
\label{fig:cfl1000_accuracy}
\end{figure}

The total error, which includes both temporal and spatial errors, is
an important factor which should be considered.  The behavior of the
total error versus time is shown in Fig.~\ref{fig:time_slot}.  The
errors in the solutions obtained by using the third-order TVDRK3
scheme with $\mbox{CFL}=1.2$ and 0.1 are also included, and the latter
is used as the approximated time-exact solution.
Two observations can be made. First, the total errors of the TVDRK3
scheme with $\mbox{CFL}=1.2$ and 0.1 are almost indistinguishable with
a fixed polynomial order $p$. 
This suggests that the total error is indeed dominated by spatial
error, and the temporal error has little effect, if any. Therefore,
the total error will not be reduced by either decreasing the CFL number or using
even higher-order time discretizations.
Second, the errors of the PCEXP scheme with $\mbox{CFL}=1000$ are
rather close to that of the TVDRK3 scheme for all cases of $0 \leq p
\leq 3$; and the PCEXP scheme is certainly more accurate than the BDF2
scheme with the same CFL number.

\begin{figure}[htb!]
\centering 
\includegraphics[width=0.45\columnwidth]{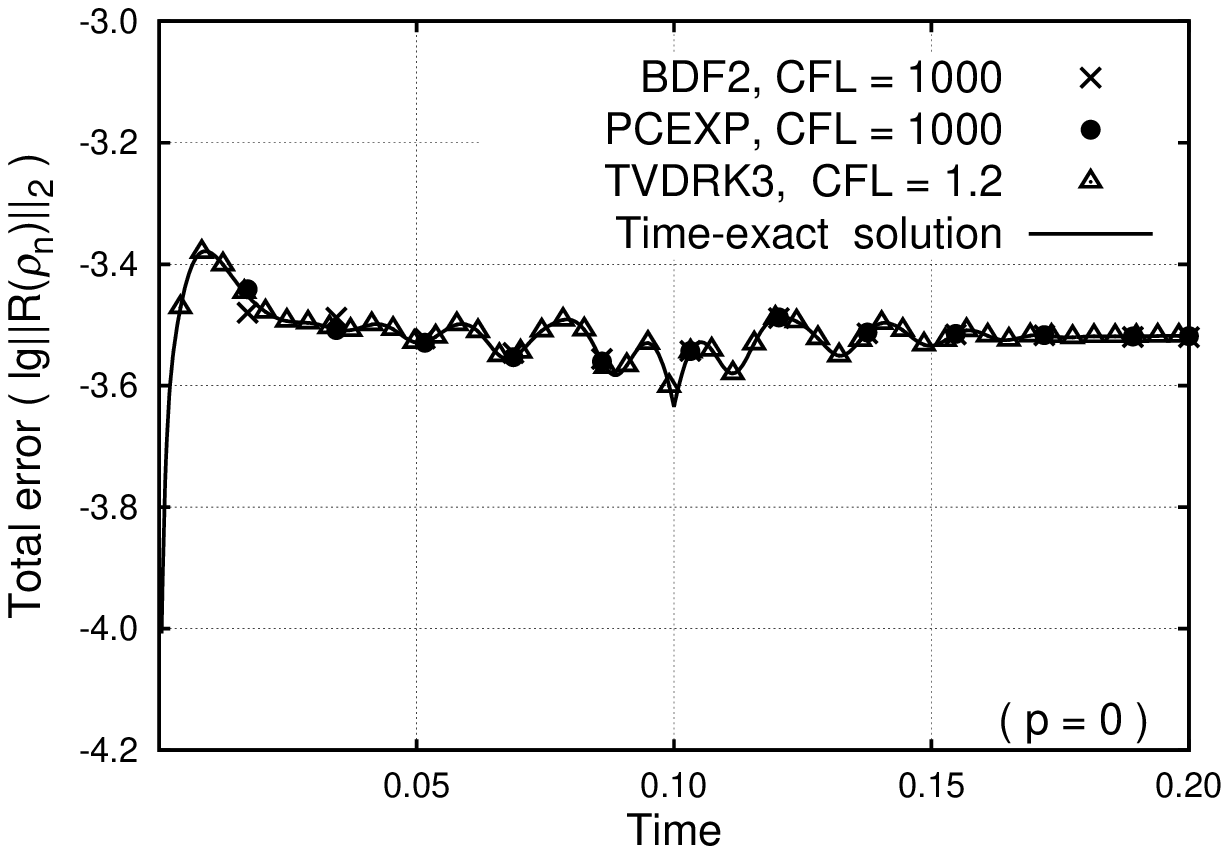}%
\includegraphics[width=0.45\columnwidth]{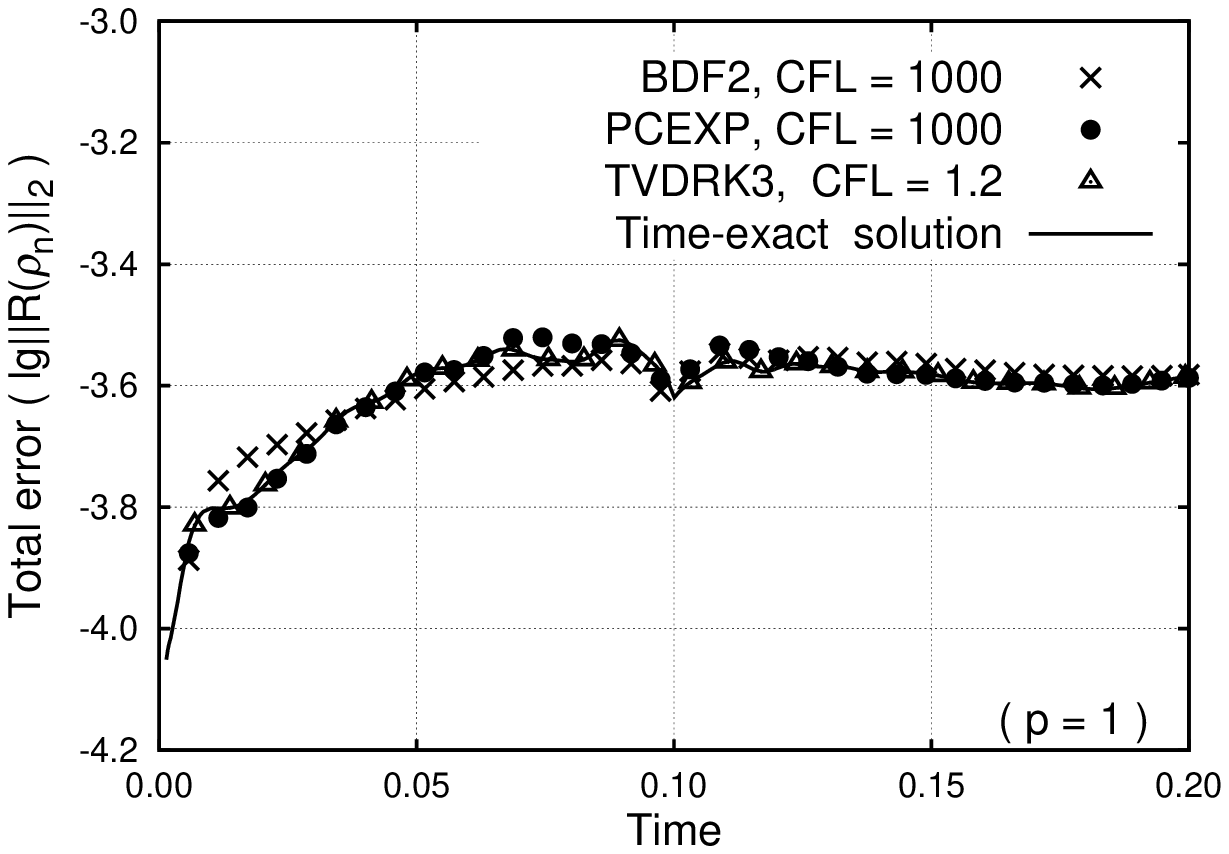}\\  
\includegraphics[width=0.45\columnwidth]{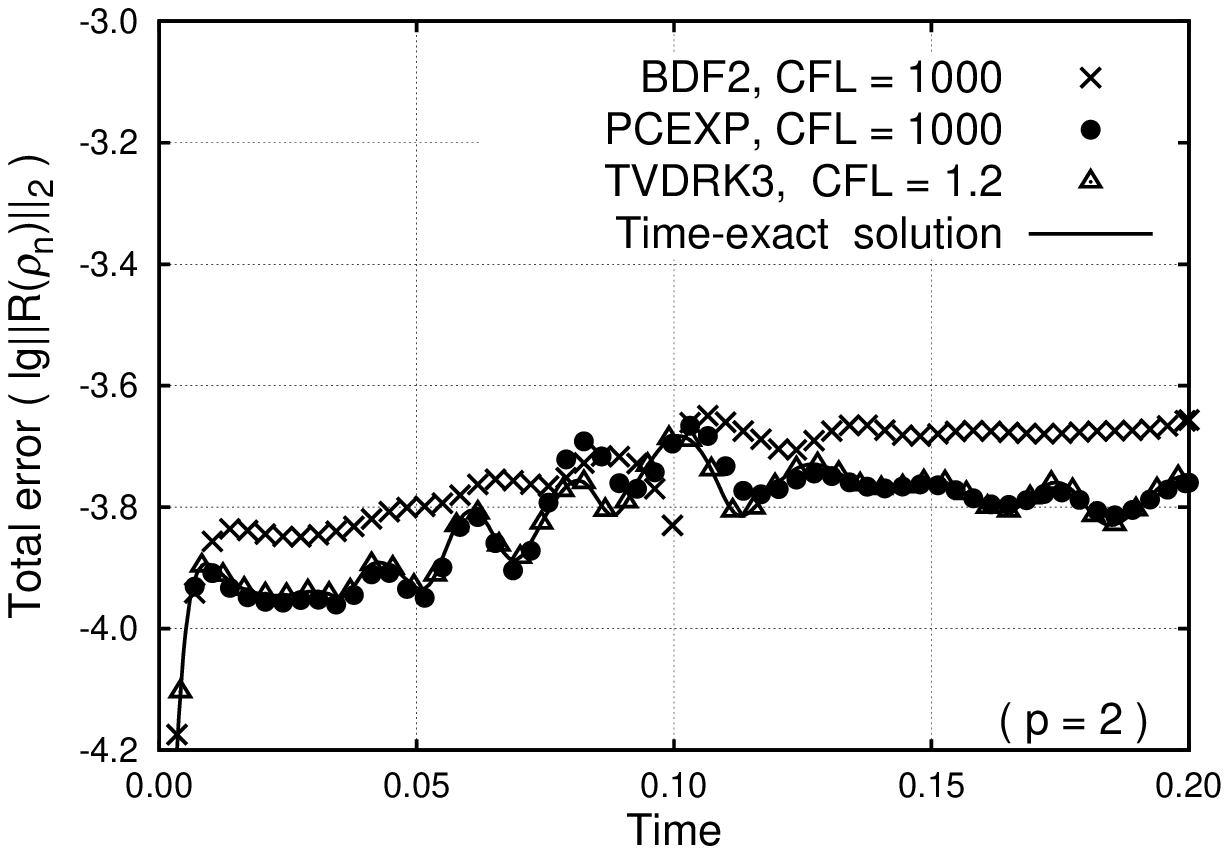}%
\includegraphics[width=0.45\columnwidth]{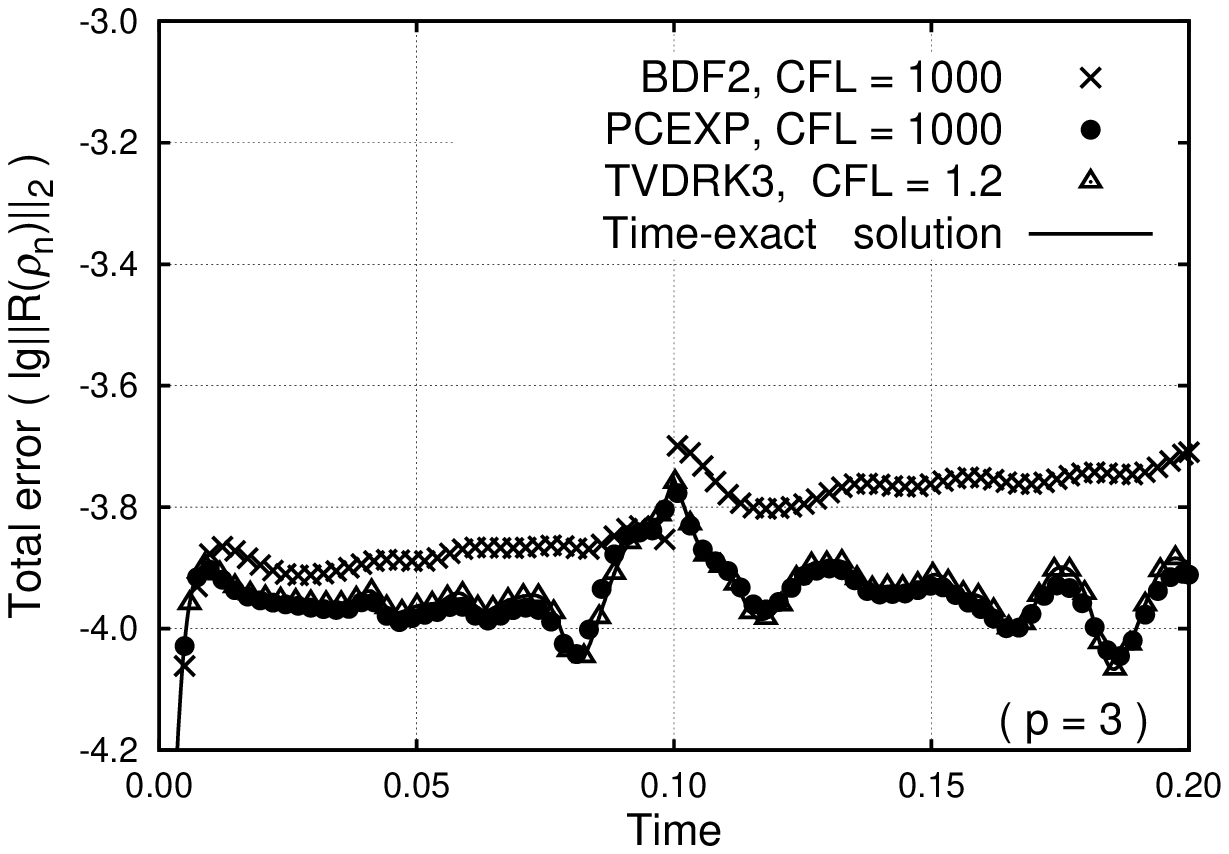}
\caption{ Vortex transport on a $24 \times 24$ uniform
  mesh.  The evolution of the total error of density.  The time-exact
  solution is obtained by using the TVDRK3 scheme with
  $\mbox{CFL}=0.1$.}
\label{fig:time_slot}
\end{figure}

Finally, the computational efficiency of three schemes, PCEXP, BDF2,
and TVDRK3 are compared by measuring the CPU time, and the results are
summarized in Table~\ref{vort_speedup}.  For $p=0$, the PCEXP scheme
is about 3.9 times faster than BDF2, and both the PCEXP and BDF2
schemes are much faster than TVDRK3 while the accuracy is maintained.

\begin{table}[htb!]
\caption{ Vortex transportation on $24 \times 24$ stretched
  mesh in one period.  The $t_{\mbox{\scriptsize {P}}}$,
  $t_{\mbox{\scriptsize {B}}}$, and $t_{\mbox{\scriptsize {T}}}$ are
  the CPU times (in minutes) corresponding to the PCEXP, BDF2, and
  TVDRK3 schemes, respectively.}
\vskip 0.5cm
\label{vort_speedup}
\centering
\begin{tabular}{crcr|rcr|rcr}
\toprule
 \multicolumn{4}{c}{PCEXP} &
  \multicolumn{3}{c}{BDF2} &
  \multicolumn{3}{c}{TVDRK3} 
\\
\hline 
\multicolumn{1}{c}{$p$ order}  &
\multicolumn{1}{r}{Steps} & \multicolumn{1}{c}{CFL}  & 
\multicolumn{1}{c}{$t_{\mbox{\scriptsize {P}}}$} &
\multicolumn{1}{r}{Steps} & \multicolumn{1}{c}{CFL}  & 
\multicolumn{1}{c}{$t_{\mbox{\scriptsize {B}}}/t_{\mbox{\scriptsize {P}}}$} &
\multicolumn{1}{r}{Steps} & \multicolumn{1}{c}{CFL}  & 
\multicolumn{1}{c}{$t_{\mbox{\scriptsize {T}}}/t_{\mbox{\scriptsize {P}}}$}
\\
\midrule
$p=0$  
& 12 & 1000.0 & 0.03
& 12 & 1000.0 & 3.90 
& 9696  & 1.2 & 48.97 
\\
$p=1$  
& 35 & 1000.0 & 1.10
& 35 & 1000.0 & 1.13
& 29106  &1.2 & 10.70
\\
$p=2$  
& 52 & 1000.0 & 12.10
& 52 & 1000.0 & 1.00 
& 48507  &1.2 & 4.98 
\\
$p=3$  
& 84 & 1000.0 & 76.50
& 84 & 1000.0 & 1.60 
& 67893  &1.2 & 3.80 
\\
\bottomrule
\end{tabular}
\end{table}

}

\subsection{Performance assessments for steady problems}

{ Unconditional stable implicit methods are highly
  efficient for solving steady problems, often achieving orders of
  magnitude speedup relative to explicit methods such as the
  Runge-Kutta types.  In this section, the exponential schemes are
  compared with two implicit methods including the backward Euler (BE)
  and the second-order backward difference formula (BDF2). Both
  schemes use an ILU preconditioned GMRES linear solver.  Two
  exponential schemes, the first-order PCEXP scheme, \ie, the EXP1
  scheme, which skips the second stage evaluation in \eqref{eqn:PCEXP}
  and the second-order PCEXP scheme are applied to steady flow
  problems in both 2D and 3D.  To enhance the computational efficiency
  and maintain stability for steady problems, the \mbox{CFL} number
  for all the exponential and implicit schemes is dynamically
  determined by the following formula:
\begin{subequations}
\begin{align}
  &
  \mbox{CFL}(n)
  = 
  \min 
  \left\{ 
  \mbox{CFL}_{\max}, \, 
  \max
  \left[ \| R(\rho_n) \|_2^{-3},\, 1
    + \frac{( n - 1 )}{ (2p+1) } 
    \right] 
  \right\} 
  ,
\label{cfl}
  \\
  &
  \| R(\rho_n) \|_2 
  := 
  \frac{1}{|\Omega|}
  \left[ {\displaystyle \int_\Omega R(\rho_n)^2 
      \diff \bm{x}  } \right]^{1/2}
  ,
\end{align}
\end{subequations}
{ where , $R(\rho_n)$ denotes the residual of density},
$\mbox{CFL}_{\max}$ is the user-defined maximal CFL number, $n$ is the
number of iterations, and $p$ is the spatial order of accuracy.  Thus,
the value of \mbox{CFL} starts at unity initially ($n=1$)
{ when $\| R(\rho_n) \|_2$ is large}, and gradually
increases to its maximum $\CFLmax$ { as $\| R(\rho_n)
  \|_2$ diminishes.

}

\subsubsection{Subsonic flow over a \mbox{NACA0012} airfoil in 2D}

In this Section, we consider a subsonic flow over the \mbox{NACA0012}
airfoil with the Mach number $\Ma = 0.63$ and the angle of attack
$\alpha = 2^\circ$.  The computational domain is a circular disc with
the radius of $5$ in the unit of the chord length equal to 1, as shown
in Fig.~\ref{naca_mesh}.  The mesh is a quasi-2D one consisting of
1322 quadratic curved wedge elements.
The minimal and maximal grid sizes are about $0.01$ and 2.0,
respectively. The CFL number is determined by \eqref{cfl} with
$\CFLmax=1000$.

\begin{figure}[htb!]
\centering
\includegraphics[width=0.6\textwidth]{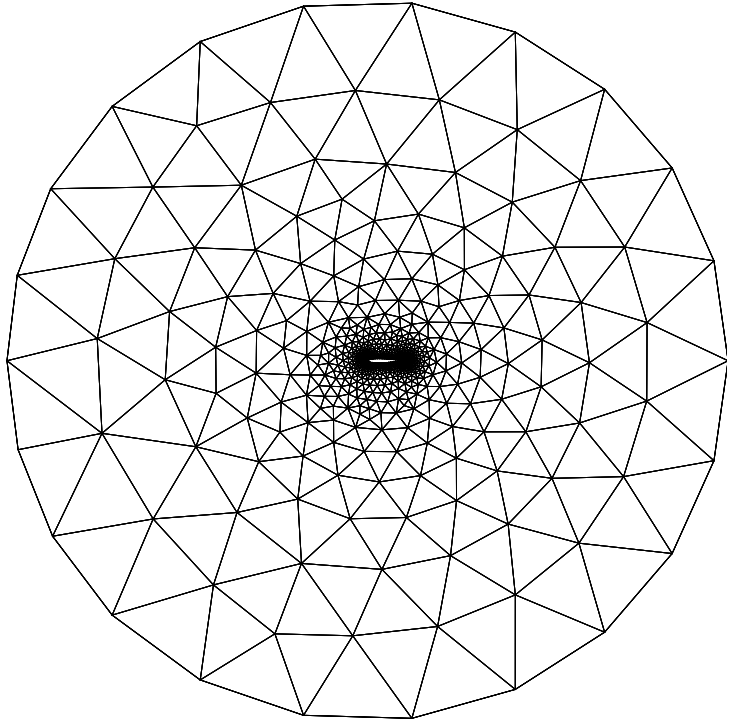}
\caption{The subsonic flow over a \mbox{NACA0012} airfoil with $\Ma
  =0.63$ and $\alpha = 2^\circ$. An illustration of a typical mesh.}
\label{naca_mesh}
\end{figure}

Figure~\ref{naca_res} shows the convergence behaviors of the density
residual $R(\rho_n)$ with the $L_2$ norm for all the time-marching
schemes with different order $p$ in terms of the number of iterations
and CPU time.
For $p=0$, the PCEXP scheme requires the least number of iterations to
converge to the steady state, while the BE scheme
requires the shortest CPU time. For $1 \leq p \leq 3$, the BE scheme
is the most efficient one in terms of both the number of iterations
and CPU time.  While the number of iterations to attain convergence is
rather similar for all the schemes, the required CPU time is rather
different; the BE scheme is by far the fastest one in this case.

\begin{figure}[htbp!]
\centering 
\includegraphics[width=0.425\columnwidth]{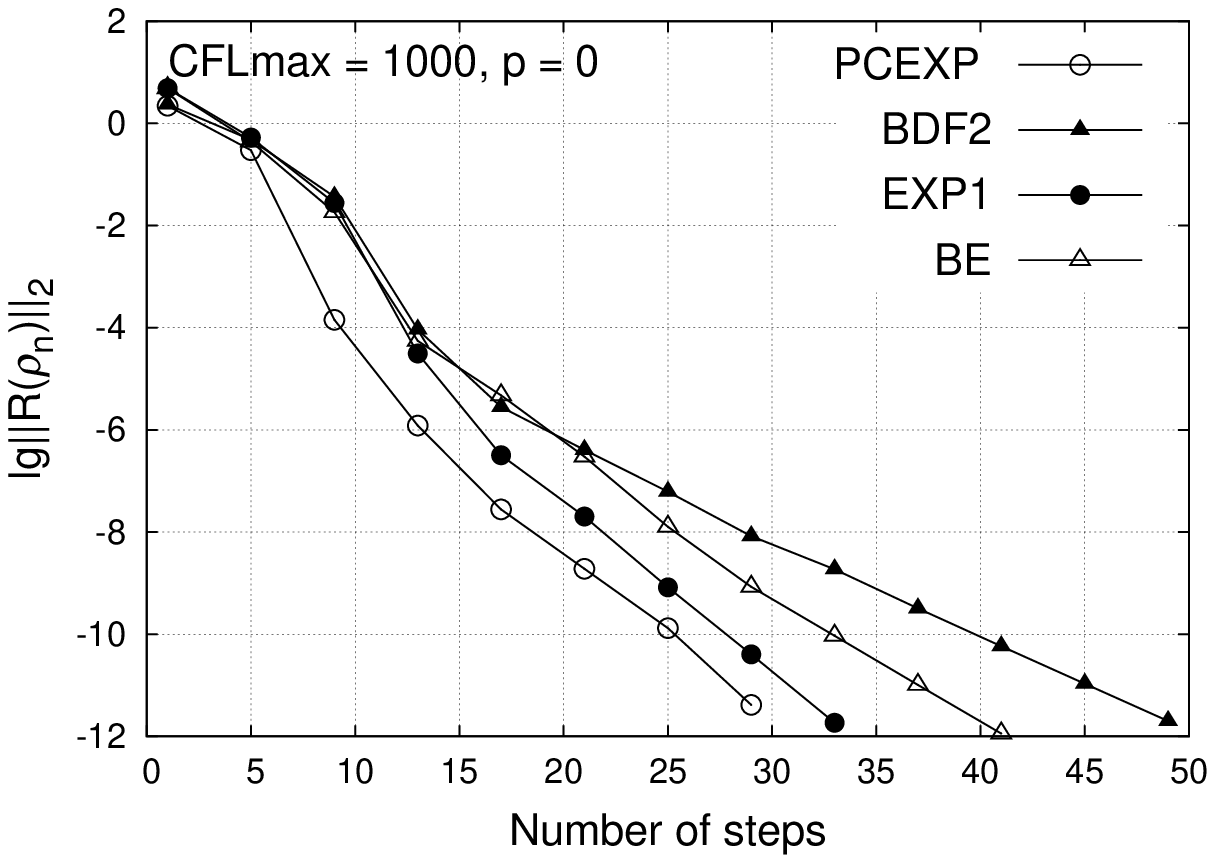}%
\includegraphics[width=0.425\columnwidth]{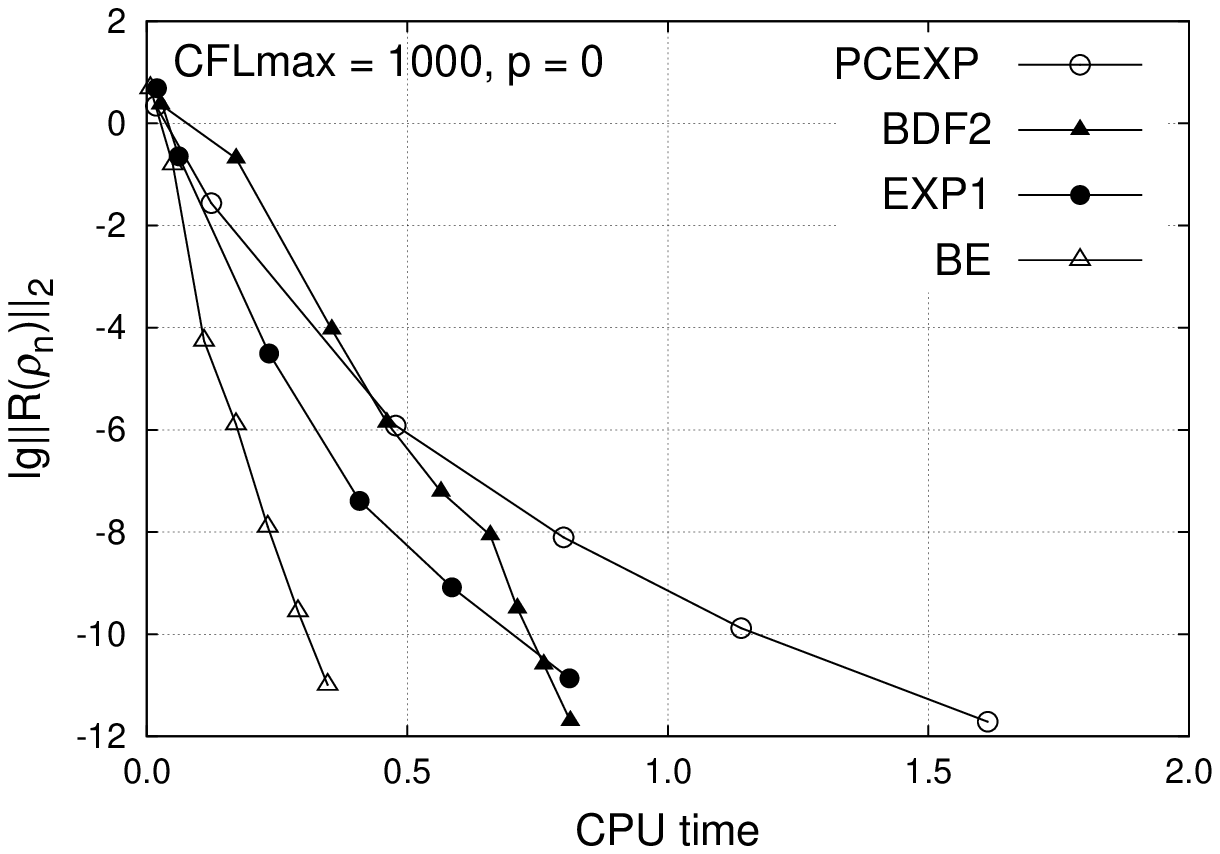}\\
\includegraphics[width=0.425\columnwidth]{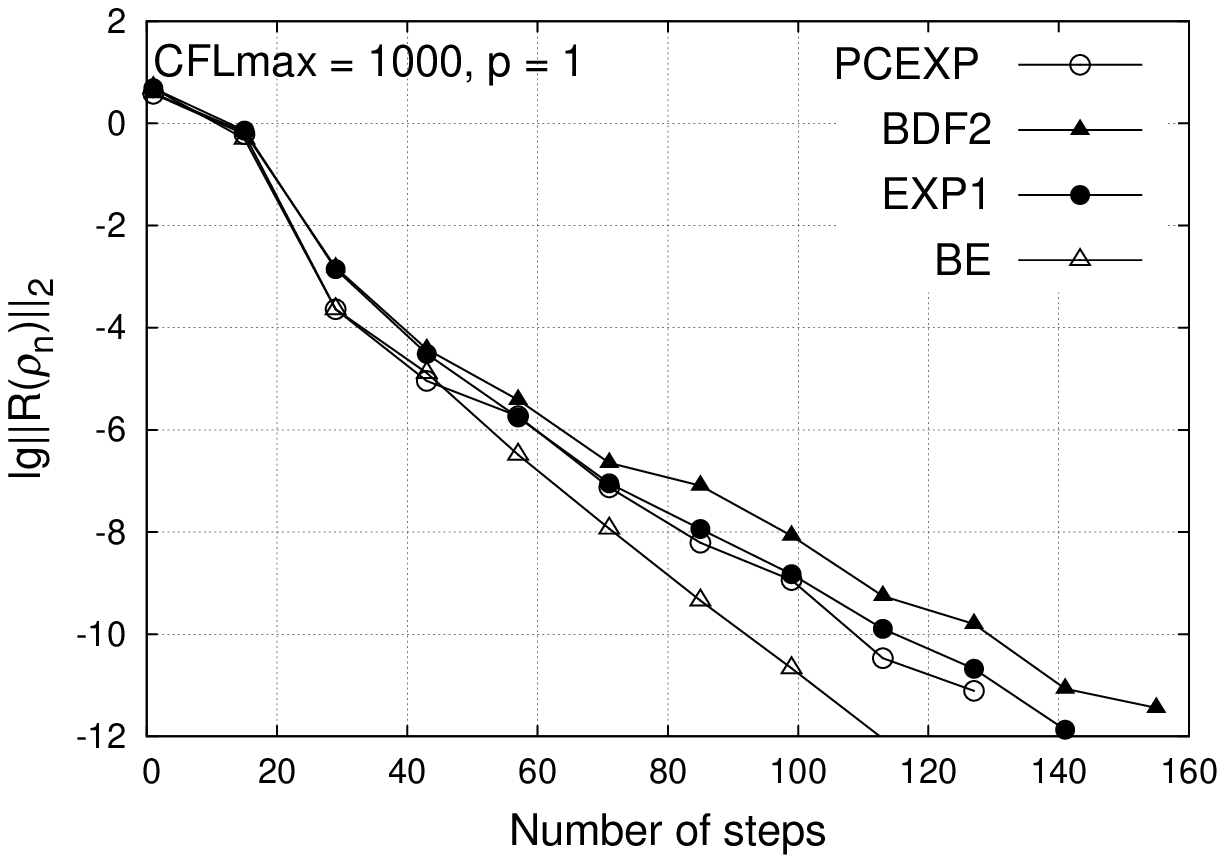}%
\includegraphics[width=0.425\columnwidth]{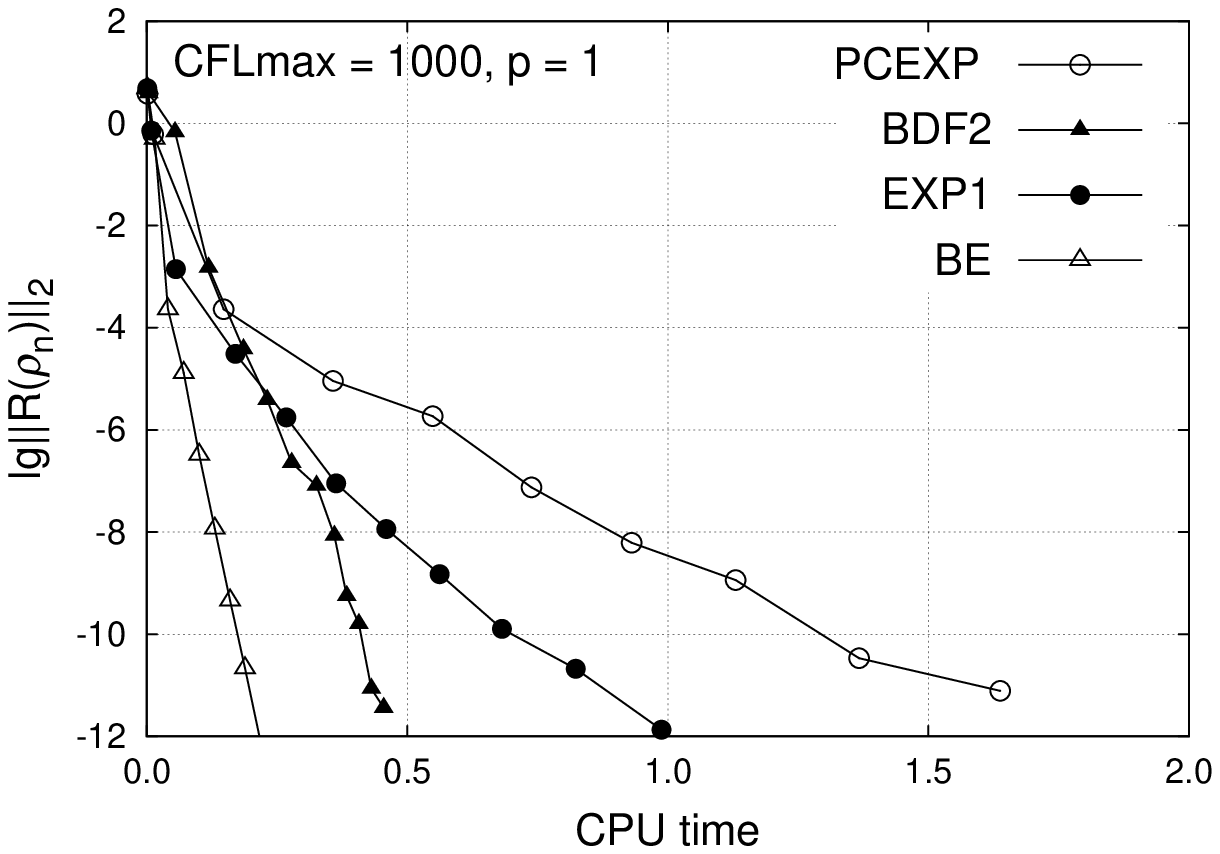}\\
\includegraphics[width=0.425\columnwidth]{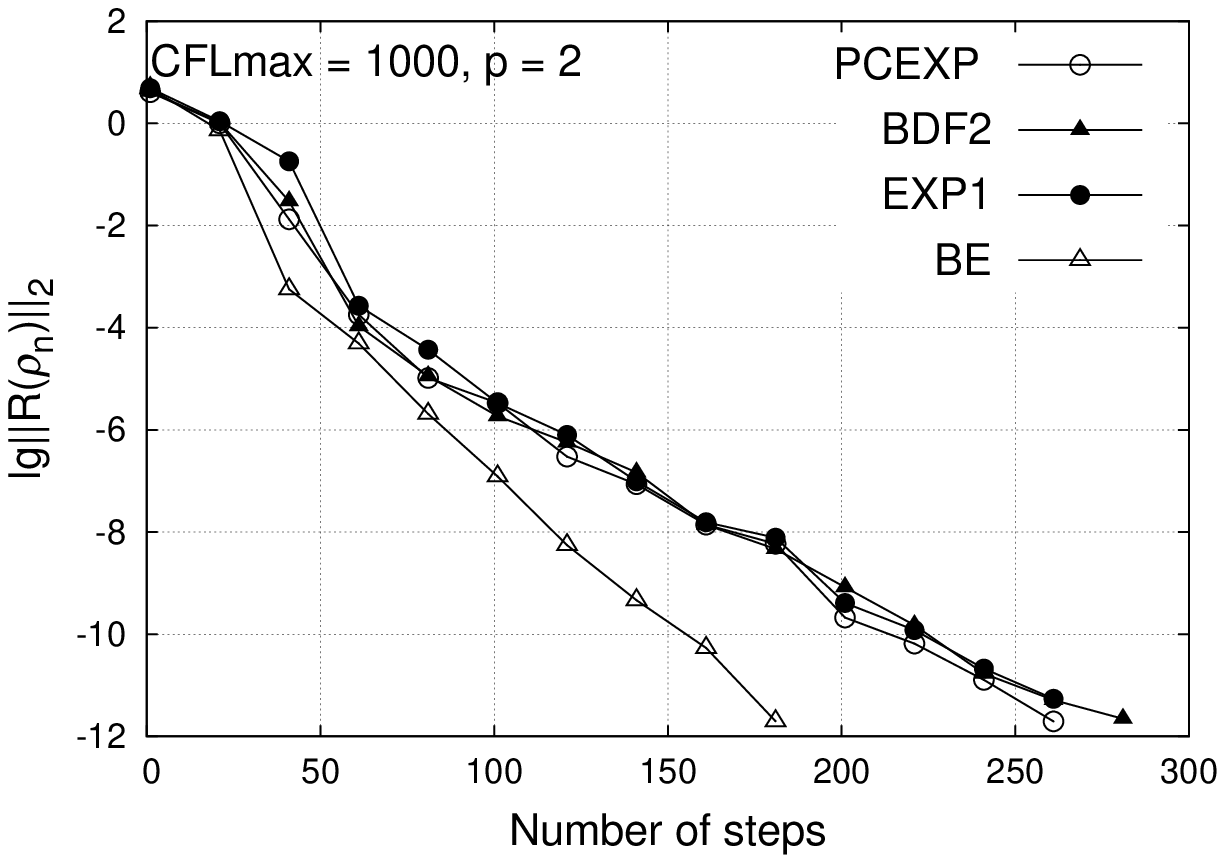}%
\includegraphics[width=0.425\columnwidth]{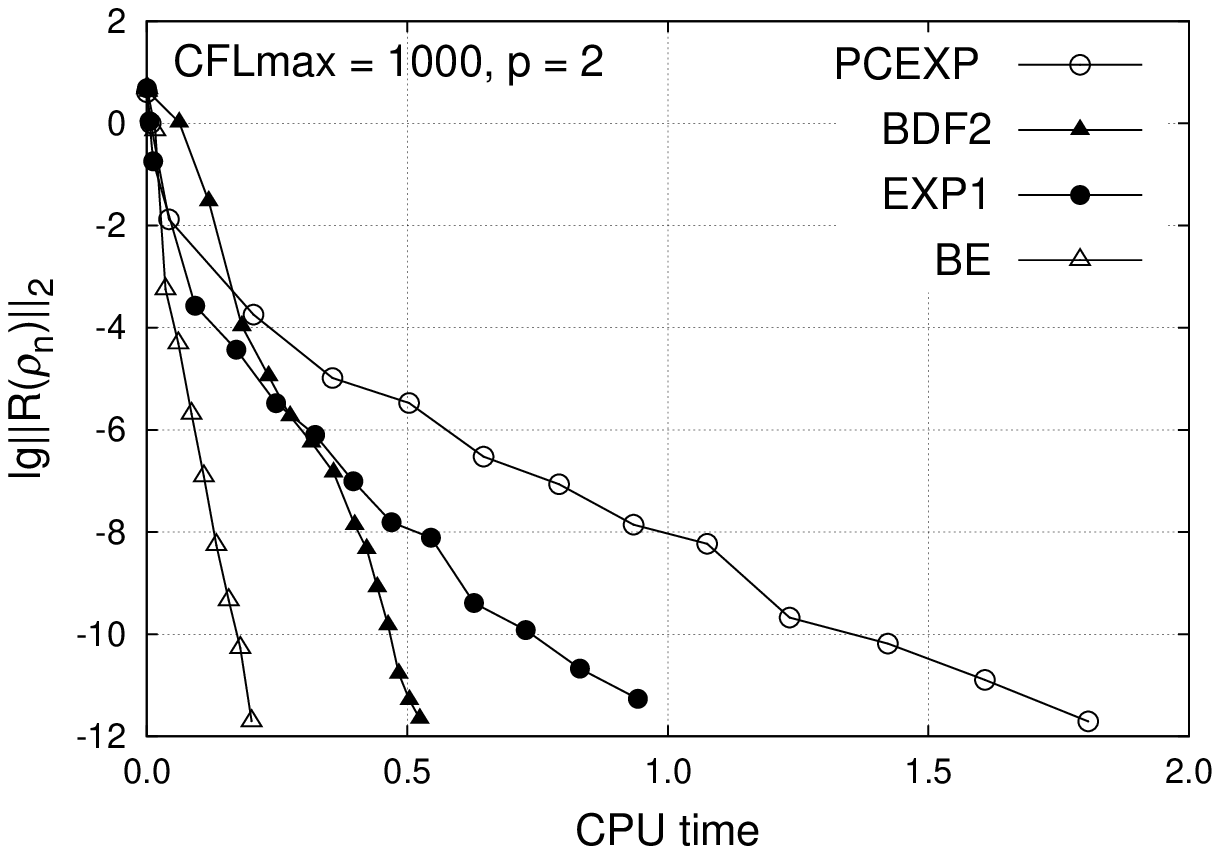}\\
\includegraphics[width=0.425\columnwidth]{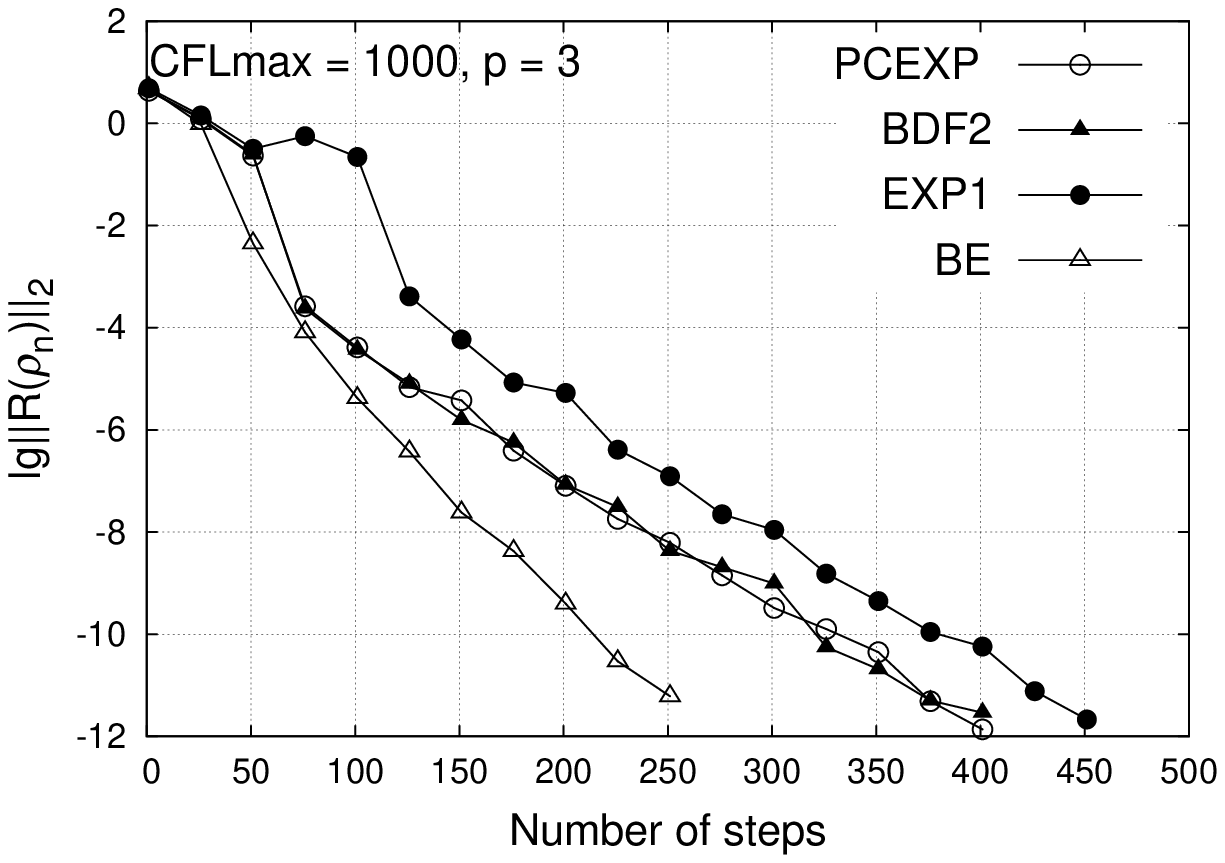}%
\includegraphics[width=0.425\columnwidth]{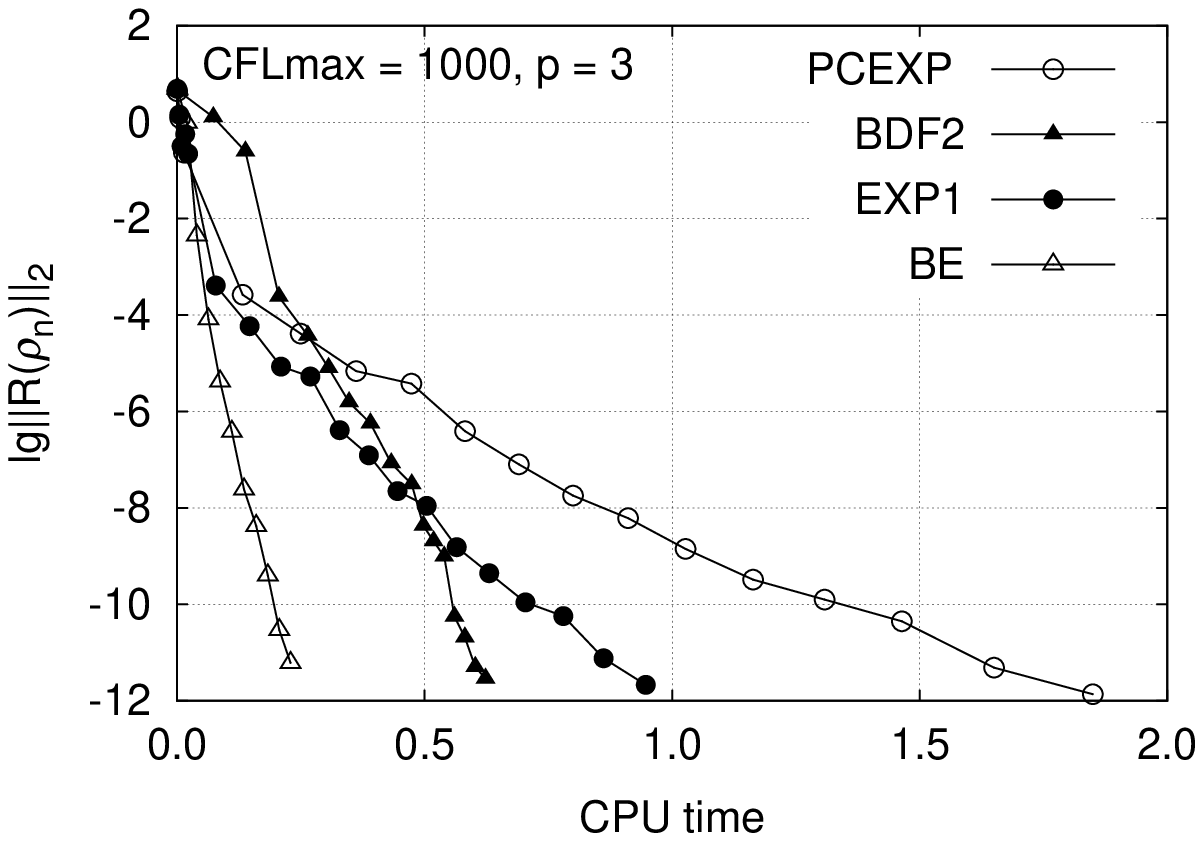}%
\vskip -0.5cm
\caption{Flow past a \mbox{NACA0012} airfoil at $\Ma=0.63$.
  Convergence behaviors of the density residual $\lg
  \|R(\rho_n) \|_2$ for the exponential and implicit schemes in
  terms of the number of iterations (left) and the elapsed CPU time
  (right), with DG discretizations of $p=0$ to $p=3$ (from top to
  bottom). }
\label{naca_res}
\end{figure}

\subsubsection{Subsonic flow over a sphere in 3D}

In this Section, we evaluate the computational efficiency of the
exponential schemes for a three-dimensional flow past a
sphere with the Mach number $\Ma = 0.3$.  The radius of the sphere is
set to 1. The computational domain is the spherical shell with the
inner and outer radius of 1 and 5, respectively.
The inner boundaries of the computational domain are the slip wall,
and the outer ones are the far-field characteristic boundaries defined
by Riemann invariants.
The CFL number of all the schemes is determined by \eqref{cfl} with
$\mbox{CFL}_{\max} = 1000$.

The mesh respects the flow symmetries of the horizontal and vertical
planes, on which the symmetry boundary condition is imposed.
The curved mesh consists of 9778 tetrahedrons and 4248 prisms.
A close-up view of the mesh about the sphere and the pressure field
computed with the PCEXP scheme of $p=2$ discretization is illustrated
in Fig.~\ref{ball_mesh}.

\begin{figure}[htb!]
\centering
\includegraphics[width=0.7\textwidth]{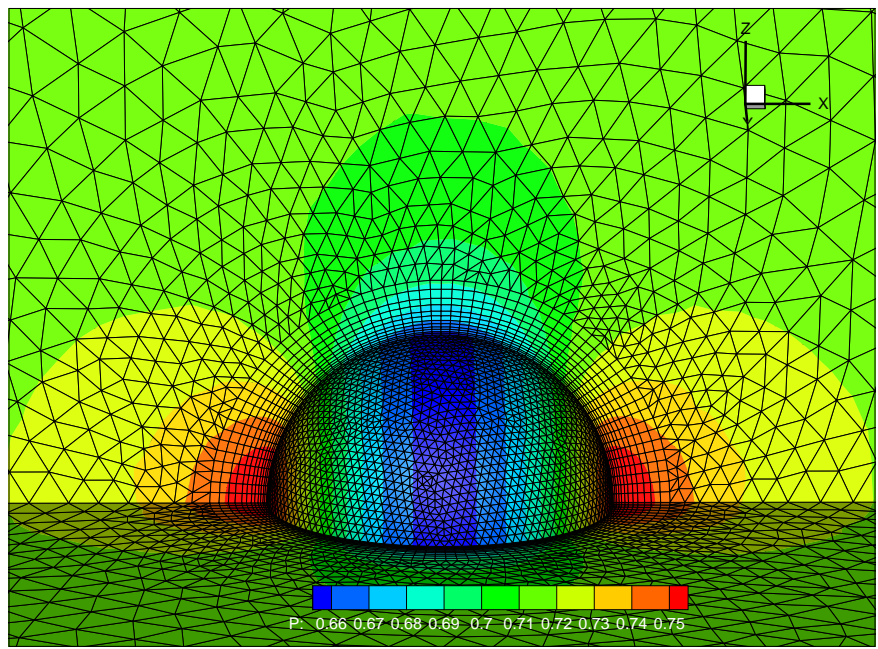}
\caption{Flow past a sphere at $\Ma = 0.3$.  A close-up view of the
    mesh of hybrid quadratic curved elements about the sphere along
    with the pressure computed with $p=2$ discretization.}
\label{ball_mesh}
\end{figure}

Figure~\ref{fig:ball} shows the convergence histories of the
exponential schemes, EXP1 and PCEXP, and the implicit schemes, BE and
BDF2, with the spatial orders $0 \leq p \leq 2$.  

\begin{figure}[htbp!]
\centering 
\includegraphics[width=0.45\columnwidth]{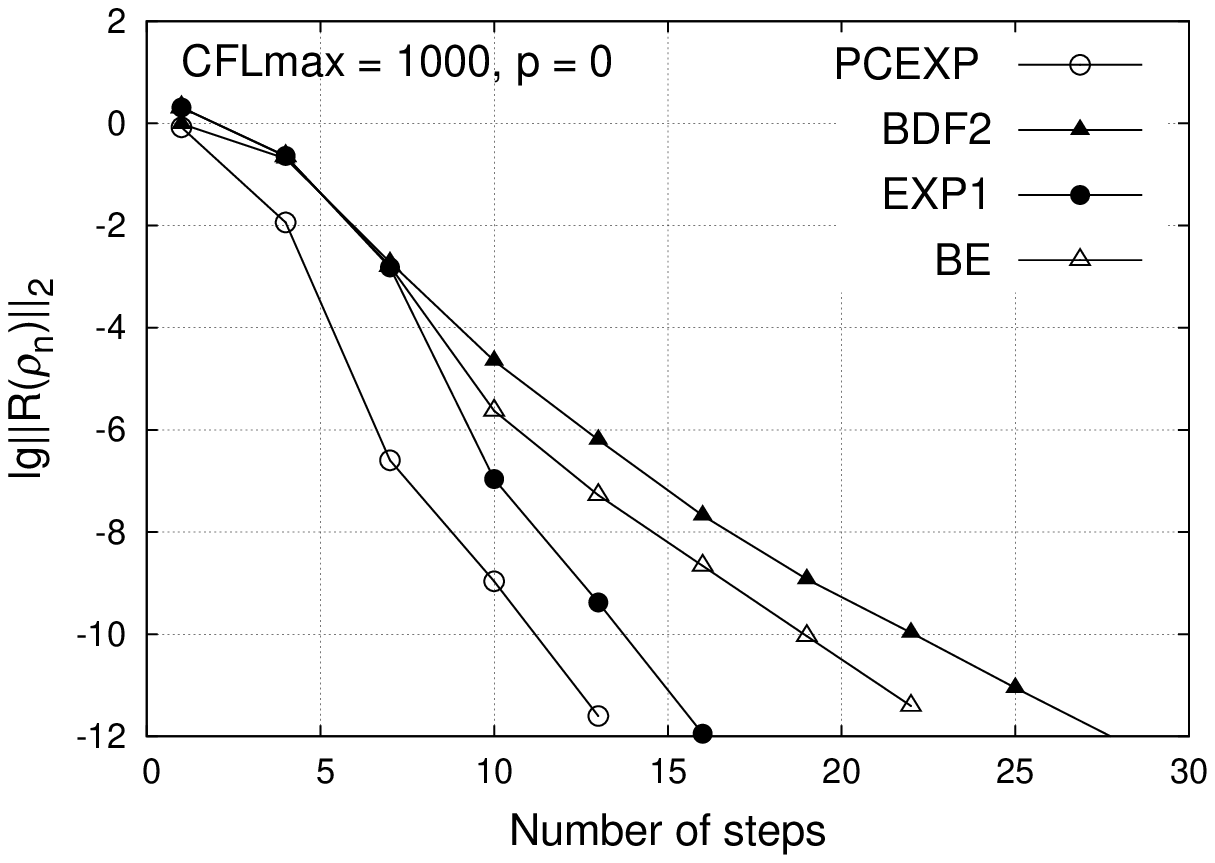}%
\includegraphics[width=0.45\columnwidth]{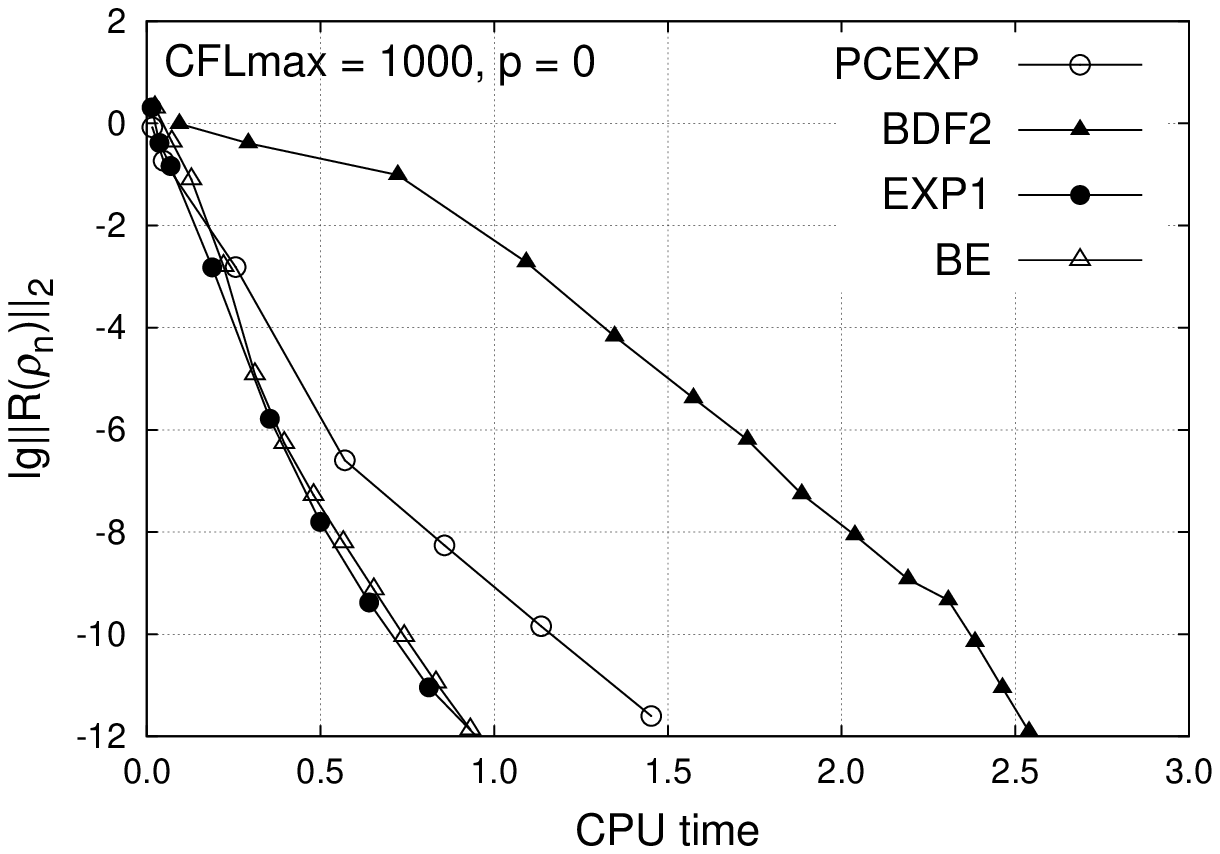}%
\\
\includegraphics[width=0.45\columnwidth]{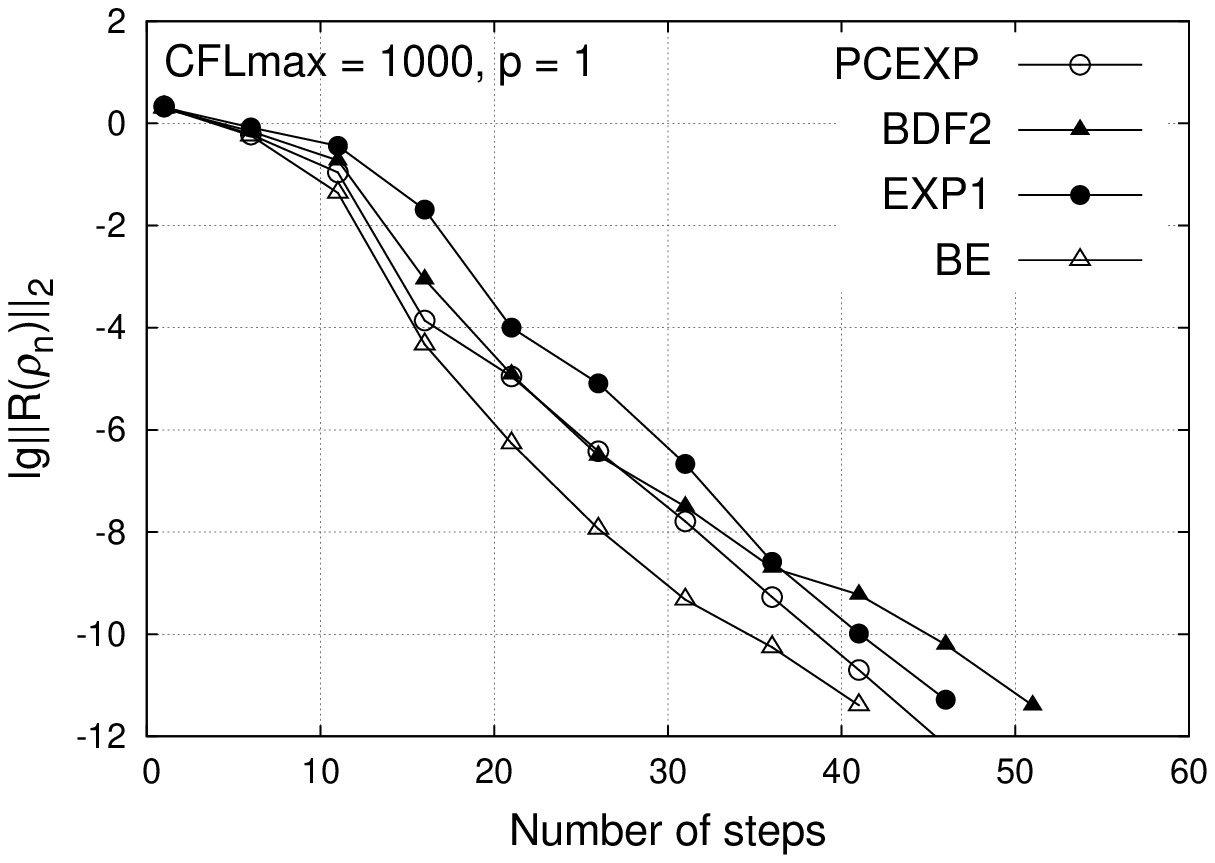}%
\includegraphics[width=0.45\columnwidth]{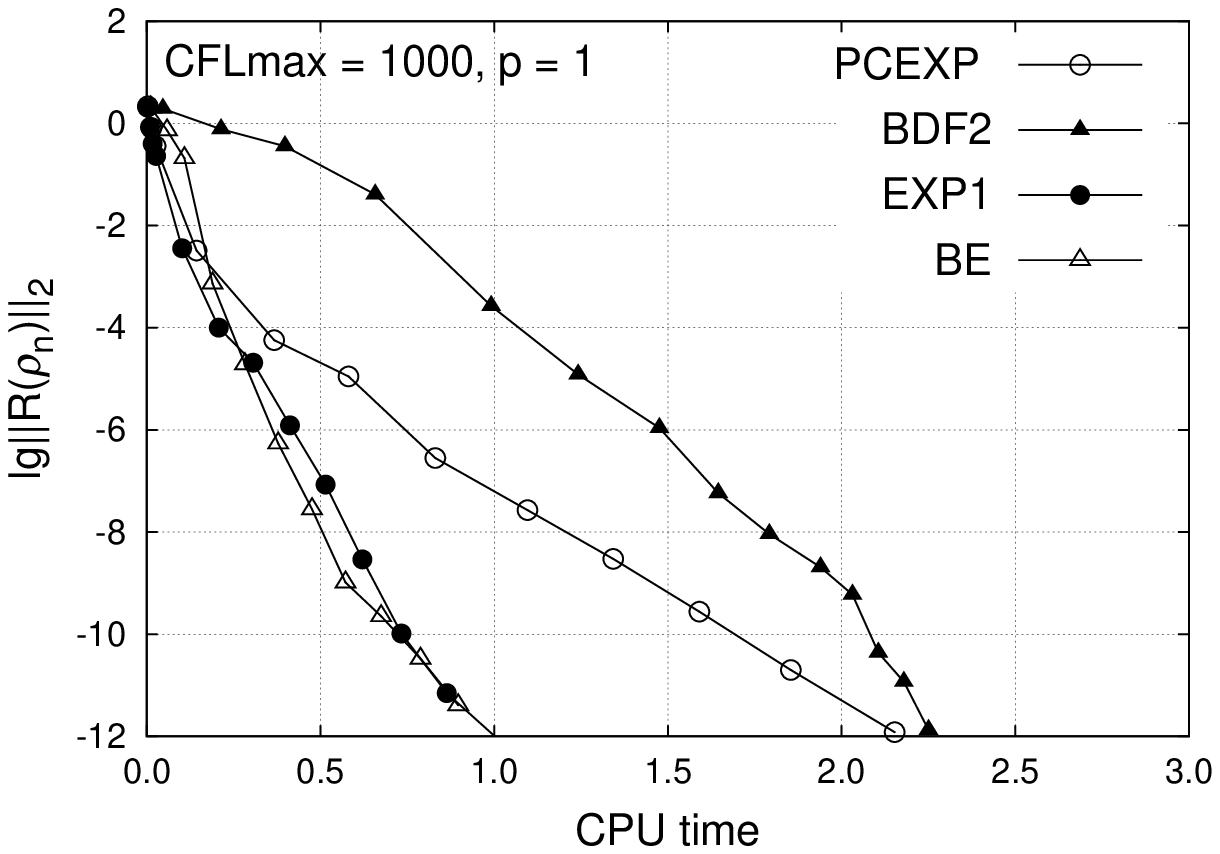}%
\\
\includegraphics[width=0.45\columnwidth]{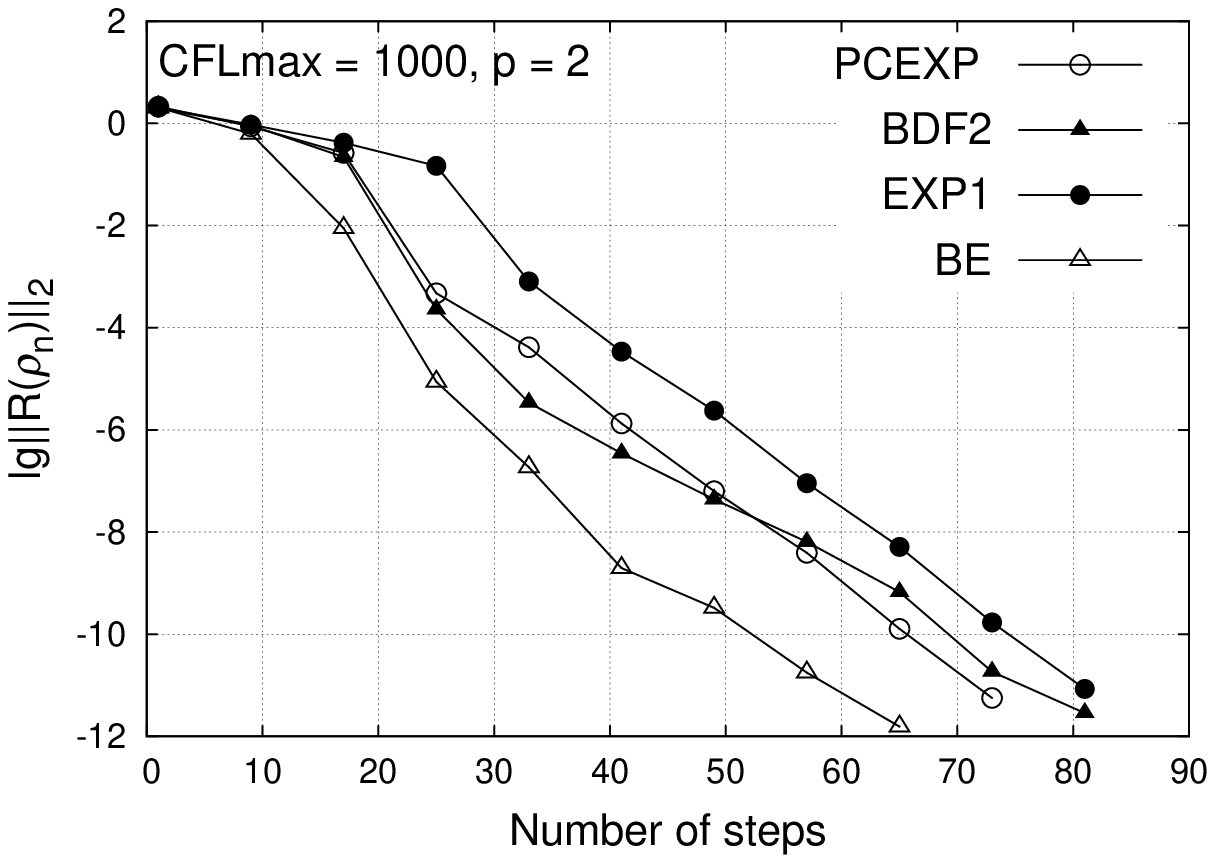}%
\includegraphics[width=0.45\columnwidth]{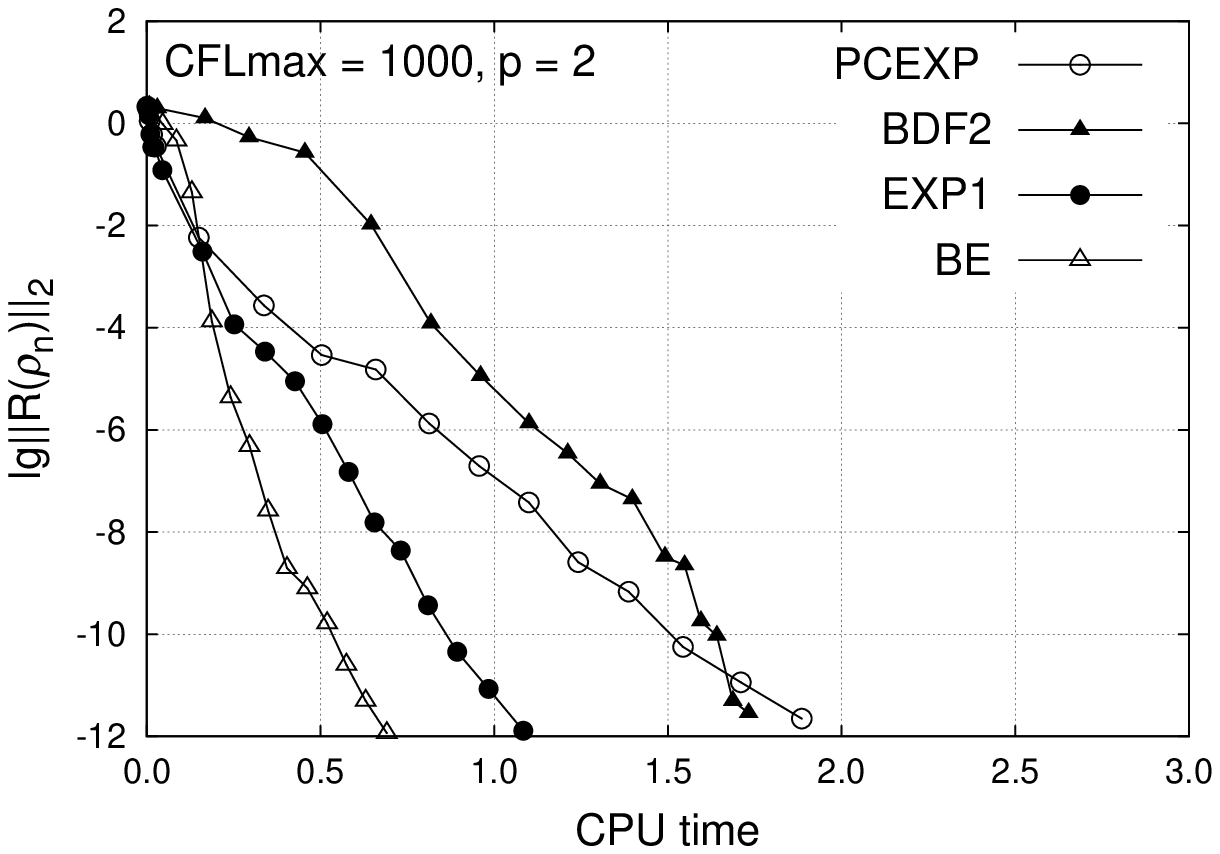}%
\caption{The subsonic flow over a sphere in 3D with $\Ma = 0.3$.
  Convergence behaviors of the density residual $\lg \| R(\rho_n)
  \|_2$ for the exponential and implicit schemes in terms of the
  number of iterations or steps (left) and the elapsed CPU time
  (right), with DG discretizations of $p=0$ to $p=2$ (from top to
  bottom). }
\label{fig:ball}
\end{figure}

The rates of convergence for the exponential and the implicit schemes
in terms of the number of iterations are similar, as shown in
Fig.~\ref{fig:ball} (left), because both types of schemes are of
global nature.
The PCEXP scheme requires the least number of iterations to converge
when $p=0$, while the BE scheme does so when $p = 1$ and $2$.
We note that the implicit schemes with the GMRES linear solver
preconditioned by an ILU is among the fastest solvers.  Using other
less efficient linear solvers would degrade the efficiency of an implicit solver.

In Fig.~\ref{fig:ball} (right), the computational efficiency is
measured by the elapsed CPU time to achieve convergence.  Both
first-order schemes, EXP1 and BE, converge faster than their
second-order counterparts, PCEXP and BDF2, respectively.
Again the BE scheme is the fastest in terms of CPU time, followed
closely by the EXP1 scheme. Interestingly, the BDF2 scheme is the
slowest in terms of CPU time in this case. 
It is also observed that the EXP1 scheme converges roughly twice as
fast as its second-order counterpart, the PCEXP scheme. 
This suggests that first-order schemes are the most efficient for
steady-state calculations, and higher-order temporal accuracy is  
inefficient.
It can also be seen that the EXP1 scheme performs better than the BDF2
scheme in all cases of $0 \leq p \leq 3$, as opposed to the previous
case of the flow past a NACA0012 airfoil in 2D, in which the BDF2
scheme performs better (cf. Fig.~\ref{naca_res}).  For steady problems
in 3D, the computational efficiency and performance of the exponential
schemes are comparable to those of the implicit schemes in terms of
either the number of iterations or CPU time; and the first-order
schemes perform better than their second-order counterparts.

\section{Conclusions}
\label{sec:finale}

{ An exponential scheme, PCEXP, has been developed for the
  time marching of steady and unsteady inviscid flows in both 2D and
  3D. The PCEXP scheme, a one-step scheme based on the
  predictor-corrector methodology, allows large time-step size while
  maintaining second-order accuracy in time. The temporal accuracy of
  the PCEXP scheme is verified; its convergence behavior and
  computational efficiency are validated for both steady and unsteady
  test cases.

  The unsteady flow of a vortex moving with a constant velocity in 2D
  is used to verify the temporal accuracy of the proposed PCEXP
  scheme. The comparisons are carried out with uniform (nonstiff
  case) and highly clustered non-uniform (stiff case) meshes.  On the
  uniform meshes, the order of temporal accuracy of the PCEXP and BDF2
  schemes are verified. In addition, the PCEXP scheme is shown to be
  much more accurate than the BDF2 scheme. Specifically for the vortex
  transportation in 2D, the magnitude of total error in the solution of the
  PCEXP scheme can be more than one order of magnitude smaller than
  that of the BDF2 scheme with the same time-step size.  Thus, the
  PCEXP scheme is far more effective and more efficient than the BDF2
  scheme. The above conclusion applies to both nonstiff and stiff
  cases.

}

For steady-state problems, the PCEXP scheme allows large time-step
sizes thus can achieve a rapid convergence. Both the EXP1 and PCEXP
schemes enjoy the rates of convergence comparable to their implicit
counterparts, the BE and BDF2 schemes, respectively, in terms of the
number of iterations. Also, the first-order exponential scheme, EXP1,
is more efficient than its second-order counterpart, PCEXP, as
expected.

In conclusion, we have successfully demonstrated the effectiveness and
efficiency of the proposed PCEXP scheme for accelerating computations
of unsteady flows, especially for stiff problems. Comparing to the BDF2
scheme, the PCEXP scheme generates a much smaller temporal error,
although both schemes are second-order accurate. To enhance
the efficiency of  exponential schemes including the PCEXP scheme, the computational cost per iteration 
of the matrix exponential  worth a further investigation.

}

\section*{Acknowledgments}

This work is funded by the National Natural Science Foundation of
China (NSFC) under the Grant U1530401.  The computational resources
are provided by the Special Program for Applied Research on Super
Computing from the NSFC-Guangdong Joint Fund (Phase~2) under
Grant U1501501 and Beijing Computational Science Research Center
(CSRC).  The third author would like to acknowledge the support from the US
National Science Foundation under the Grant DMS-1521965 and the US
Department of Energy under the Grant DE-SC0016540.  The authors would
like to thank Dr. Ken C.Y. Loh for his careful proof-reading of the
manuscript.  { The authors would also like to thank the
  anonymous referees whose comments helped us improve the paper
  significantly.}

\appendix

\section{The Jacobian matrices}


The matrix $\bm{\nabla} \psi \, \partial \tensor{F} /
\partial \mathbf{U}$ in \eqref{jac1} is
%
\begin{equation}       
\left(                 
\begin{array}{ccccc}       
  -B_2 & \psi_x & \psi_y & \psi_z & 0
  \\ 
  a_0 \psi_x - B_1 u &  B_1 - B_2 - a_3 u \psi_x & u \psi_y - a_2 v
  \psi_x &  u \psi_z - a_2 w \psi_x & a_2 \psi_x
  \\  
  a_0 \psi_y - B_1 v &  v \psi_x - a_2 u \psi_y &  B_1 - B_2 - a_3 v
  \psi_y &  v \psi_z - a_2 w \psi_y & a_2 \psi_y
  \\  
  a_0 \psi_z - B_1 w & w \psi_x - a_2 u \psi_z & w \psi_y - a_2 v
  \psi_z &  B_1 -B_2  - a_3 w \psi_z & a_2 \psi_z
  \\  
  (a_0-a_1)  B_1 & a_1 \psi_x - a_2 B_1 u & a_1 \psi_y - a_2 v  B_1 &
  a_1 \psi_z - a_2 B_1 w & \gamma B_1 - B_2 \\
  \end{array}
\right),
\label{dfdu}
\end{equation}
where $\bm{v} := (u,\, v,\, w)$, $\bm{\omega} := (\omega_x,\,
\omega_y,\, \omega_z)$, $ \bm{\nabla} \psi
:= (\psi_x,\, \psi_y,\,  \psi_z)$, 
%
\begin{equation}
\begin{aligned}
  &
  a_0 
  := 
  \frac{1}{2} (\gamma -1) (u^2 + v^2 + w^2) 
  ,
  \quad
  a_1 
  := 
  \gamma e - a_0 ,
  \quad
  a_2 
  := 
  \gamma - 1,
  \quad
  a_3 
  := 
  \gamma - 2, 
  \\
  &
  B_1 
  := 
  \bm{v} \cdot \bm{\nabla} \psi
  =
  u \psi_x  + v \psi_y  + w \psi_z ,
  \quad
 B_2
  :=
  (\bm{\omega} \times \bm{x} ) \cdot \bm{\nabla}{\psi} .
\end{aligned}
\end{equation}
%

The source-term Jacobian matrix $\partial \mathbf{S} /\partial
\mathbf{U}$ in \eqref{jac1} is 
%
\begin{equation}      
  \frac{\partial \mathbf{S}}{\partial \mathbf{U}} =
  \left(                 
  \begin{array}{ccccc}       
    0& 0& 0& 0& 0\\ 
    0 & 0& -\omega_z & \omega_y & 0\\  
    0 & \omega_z & 0 & -\omega_x & 0 \\
    0 & -\omega_y & \omega_x & 0 & 0 \\
    0 & 0 &0 &0 & 0 \\
  \end{array}
  \right)
  .
\label{dsdu}   
\end{equation}

%



\bibliography{shujie}

\end{document}